\documentclass[12pt]{article}
\usepackage{amsmath,amsfonts,amssymb,amsthm,epsfig, graphicx, hyperref}
\usepackage[dvipsnames,table,dvipsnames*, svgnames*, hyperref]{xcolor}
\usepackage[linesnumbered,ruled]{algorithm2e}
\usepackage{tcolorbox}
\usepackage{natbib}
\usepackage{empheq}
\usepackage{caption}
\usepackage{subcaption}
\usepackage{babel,blindtext}
\usepackage{mwe}
\usepackage{float}
\usepackage{mathrsfs,enumitem,FACE_notations}
\usepackage{booktabs}
\usepackage{lscape}
\DeclareMathSizes{12}{12}{5}{3}

\newcommand{\blind}{1}

\definecolor{jcolor}{RGB}{041,122,000}
\definecolor{darkred}{RGB}{100,000,000}
\definecolor{purple}{RGB}{200,000,200}

\newcommand{\cut}[1]{}
\newcommand{\commentout}[1]{}

\usepackage{xr}

\makeatletter
\newcommand*{\addFileDependency}[1]{% argument=file name and extension
  \typeout{(#1)}
  \@addtofilelist{#1}
  \IfFileExists{#1}{}{\typeout{No file #1.}}
}

\newtheorem*{assumption*}{\assumptionnumber}
\providecommand{\assumptionnumber}{}
\newenvironment{assumption}[2]
 {%
  \renewcommand{\assumptionnumber}{Assumption #1#2}%
  \begin{assumption*}%
  \protected@edef\@currentlabel{#1#2}%
 }
 {%
  \end{assumption*}
 }

\makeatother

\myexternaldocument{_Supplement_Sep2023}

\begin{document}

\sloppy

\def\spacingset#1{\renewcommand{\baselinestretch}%
{#1}\small\normalsize} \spacingset{1}

\allowdisplaybreaks

%%%%%%%%%%%%%%%%%%%%%%%%%%%%%%%%%%%%%%%%%%%%%%%%%%%%%%%%%%%%%%%%%%%%%%%%%%%%%%

\if1\blind
{
  \title{\bf Federated Adaptive Causal Estimation (FACE) of Target Treatment Effects}
  \author{Larry Han$^{1,2}$, Jue Hou$^{3}$, Kelly Cho$^{4}$, Rui Duan$^{1\dag}$, Tianxi Cai$^{1,5\dag}$ \\ \\
    1 Department of Biostatistics, Harvard University \\ 
    2 Department of Health Sciences, Northeastern University \\ 
    3 Division of Biostatistics, University of Minnesota \\ 
    4 Massachusetts Veterans Epidemiology Research and \\ Information Center,
    US Department of Veteran Affairs \\ 
    5 Department of Biomedical Informatics, Harvard Medical School \\ 
    $\dag$ Co-corresponding authors}
  \maketitle
} \fi

\if0\blind
{
  \bigskip
  \bigskip
  \bigskip
  \begin{center}
    {\LARGE \bf Federated Adaptive Causal Estimation (FACE) of Target Treatment Effects}
\end{center}
  \medskip
} \fi

\bigskip

% JASA limit is 200 words.
\begin{abstract}
Federated learning of causal estimands may greatly improve estimation efficiency by leveraging data from multiple study sites, but robustness to heterogeneity and model misspecifications is vital for ensuring validity. We develop a Federated Adaptive Causal Estimation (FACE) framework to incorporate heterogeneous data from multiple sites to provide treatment effect estimation and inference for a flexibly specified target population of interest. FACE accounts for site-level heterogeneity in the distribution of covariates through density ratio weighting. To safely incorporate source sites and avoid negative transfer, we introduce an adaptive weighting procedure via a penalized regression, which achieves both consistency and optimal efficiency. Our strategy is communication-efficient and privacy-preserving, allowing participating sites to share summary statistics only once with other sites. We conduct both theoretical and numerical evaluations of FACE and apply it to conduct a comparative effectiveness study of BNT162b2 (Pfizer) and mRNA-1273 (Moderna) vaccines on COVID-19 outcomes in U.S. veterans using electronic health records from five VA regional sites. We show that compared to traditional methods, FACE meaningfully increases the precision of treatment effect estimates, with reductions in standard errors ranging from $26\%$ to $67\%$.
\end{abstract}

\noindent%
\vfill

\newpage
\spacingset{1.53} 
\section{Introduction}
\label{sec:intro}

Multi-center, federated causal inference is of great interest, particularly when studying novel treatments, rare diseases, or in times of urgent health crises. For example, the COVID-19 pandemic has highlighted the need for novel approaches to efficiently and safely evaluate the effectiveness of novel therapies and vaccines, while leveraging data from multiple healthcare systems to ensure the generalizability of findings. Over the past few years, many research networks and data consortia have been built to facilitate multi-site studies and have been actively contributing to COVID-19 studies, including the Observational Health Data Sciences and Informatics (OHDSI) consortium \citep{hripcsak2016characterizing} and the Consortium for Clinical Characterization of COVID-19 by Electronic Health Records (EHR) \citep{brat2020international}.

Analyzing data collected from multiple healthcare systems, however, is highly challenging for several reasons. Various sources of heterogeneity exist in terms of (i) differences in the underlying population of each dataset and (ii) policy-level variations of treatment assignment. Since treatment effects may differ across different patient populations, it would be of interest to infer the average treatment effect (ATE) for specific target populations. However, the presence of heterogeneity and potential model misspecification poses great difficulty in ensuring valid estimates for the target average treatment effect (TATE). Furthermore, patient-level data typically cannot be shared across healthcare centers, which brings additional practical challenges. To overcome these challenges, we propose a  Federated Adaptive Causal Estimation (FACE) framework that aims to incorporate heterogeneous data from multiple sites to make inferences about the TATE, while accounting for heterogeneity and data-sharing constraints.

Most existing literature on federated learning has focused on regression and classification models \citep{chen2006regression, li2013statistical,chen2014split,lee2017communication, lian2017divide,wang2019distributed, duan2020learninga}. Limited federated learning methods currently exist to make causal inferences with multiple heterogeneous studies. Recently, \cite{xiong2021federated} proposed federated inverse probability weighted (IPW) estimation of the ATE specifically for an entire study population. Although \cite{xiong2021federated} provided multiple methods for point estimation and variance estimation, the choice of the proper method depends on prior knowledge about model homogeneity and specification, which are difficult to verify in practice. No empirical study in \cite{xiong2021federated} was provided to test the robustness of the approach to the covariate shift assumption.  In addition, their methods cannot be used to estimate the ATE of a target population that differs from the full study population. \cite{vo2021federated} proposed a Bayesian approach that models potential outcomes as random functions distributed by Gaussian processes. Their focus is also on the population ATE rather than any particular target population, and their approach requires specifying parameters and hyperparameters of Gaussian processes and modeling between-site covariate correlations through kernel functions, which can be numerically intensive. Compared to these approaches, our approach estimates the TATE in a particular target population and accounts for the heterogeneity across populations without requiring prior information on the source data distribution or the validity of model specifications. Our approach further safeguards against incorporating source datasets that may introduce bias to the TATE estimate, known as negative transfer \citep{pan2009survey,weiss2016survey}.

Another related strand of literature concerns the generalizability and transportability of randomized clinical trials to EHR studies. For example, \citet{stuart2011use,stuart2015assessing,stuart2018generalizability} assessed the generalizability of results from randomized trials to target populations of interest. \cite{dahabreh2020extending, josey2022calibration, lee2023improving} all focused on extending inferences about treatments from a randomized trial to a new target population by using different weighting schemes. For a comprehensive review of statistical methods for generalizability and transportability, see \citet{degtiar2023review}. However, to date, no literature in generalizability and transportability has sought to leverage observational data from a potentially large number of source sites in a data-adaptive manner to obtain unbiased, efficient, and robust estimation of target treatment effects. 

The major contributions of FACE can be summarized as follows. First, FACE allows for flexibility in the specification of the target population. For example, the target population in a research network can be defined as the underlying population of a given healthcare center, or multiple healthcare centers that share certain properties (e.g., geographic location), or the overall population combining all sites. This flexibility provides stakeholders and policymakers at different levels with information on their respective target populations.
Second, using a semiparametric density ratio weighting approach, FACE allows the distribution of covariates to be heterogeneous across sites. Third, FACE protects against negative transfer through an adaptive integration strategy that anchors on the target data and computes data-adaptive weights for source sites. In the context of statistical inference, negative transfer occurs when incorporating a source dataset increases the bias or asymptotic variance of the estimator as compared to not including it. In doing so, FACE can achieve optimal efficiency while maintaining consistency, and it is robust to the distribution of data and potential model misspecifications in the source sites. Moreover, FACE is a communication-efficient federated algorithm that allows each participating site to keep its data stored locally and only share summary statistics once with other sites.

The remainder of the paper is organized as follows. In Section \ref{sec:setting}, we introduce the problem setting, notation, and assumptions required for identification of the TATE.
In Section \ref{sec:method}, we describe the proposed FACE
framework for estimating the TATE.
We introduce the in-site estimators based on the target population and source populations separately in Sections \ref{ssec:method-target} and \ref{ssec:method-source} and present the adaptive and distributed integration in Section \ref{ssec:method-aggre}. In Section \ref{sec:theory}, we provide the theoretical guarantees of FACE, including double robustness, asymptotic normality, and relative efficiency. In Section \ref{sec:sim}, we conduct extensive simulations for various data generating mechanisms and show robustness to misspecification of different models. In Section \ref{sec:data}, we apply FACE to conduct a comparative effectiveness study of COVID-19 vaccines using the EHRs from five geographic regions of the Department of Veterans Affairs (VA). We conclude in Section \ref{sec:discuss} with key takeaways and directions for future research.

\section{Setting and Notation}\label{sec:setting}

For the $i$-th observation, we denote the outcome as $Y_{i}\in \R$, the $p$-dimensional baseline covariate vector as $\bX_{i} = (X_{i1},...,X_{ip})^\top \in \Xcal \subset \R^p$, and the indicator for binary treatment as  $A_{i} \in \{0,1\}$. There are $J \geq 1$ target sites and another $K \geq 0$ source sites. Let $\tgt\subseteq [J+K]$ indicate sites that are in the target population and $\tgs\subset [J+K]$ indicate sites that are in the source population, where $[K]=\{1, ..., K\}$ for any integer $K$. Under the federated learning setting, a total of $N$ observations are stored at $J+K$ study sites, where the $k$-th site has sample size $n_k$, and $N = \sum_{k=1}^{J+K} n_k$. Let $R_{i}$ be a site indicator such that $R_{i} = k$ indicates the $i$-th patient in the $k$-th site.
Indexing the site by a single integer $R_i$, we assume that each observation may only belong to one site. We summarize the observed data at each site $k$ as $\Dscr_k = \{(Y_{i}, \bX_{i}^{^\top} , A_{i},R_{i})^\top, R_i = k\},$ and consider a federated data setting where each site has access to its own patient-level data but can share only summary statistics with other sites.
We denote the index set for each site as $\Ical_k = \{i: R_i = k\}$. The data included in the target sites are denoted by $\Dscr_{\tgt}$.
For simplicity of notation, we use $(Y, \bX, A, R)$ without subscripts to state general assumptions and conclusions.

Under the potential outcomes framework \citep{neyman1923application, rubin1974estimating}, we denote  $Y^{(a)}$ as the potential outcome of  patients under treatment $A=a$, $a=0,1$.
Our goal is to estimate the TATE for a specified target population $\tgt$,
\begin{equation}
    \Delta_{\tgt} = \mu^{(1)}_{\tgt} - \mu^{(0)}_{\tgt}, \quad \mu^{(a)}_{\tgt} = \E(Y^{(a)} \mid R \in \tgt),
\end{equation}
where the expectation is taken over the distribution in the target population. The target population can be specified at multiple levels (e.g., single site, multiple sites, all sites) corresponding to different targets of real-world interest.  
This distinction between target and source sites also distinguishes our setting from that of \cite{xiong2021federated}, in which the target population always contains all participating sites. 

To identify the TATE, we make the following standard assumptions \citep{imbens2015causal,hernan2020causal}
throughout the paper:
\begin{assumption}{1}{}\label{assume:causal}
For a positive constant $\varepsilon > 0$, $a \in \{0,1\}$,
 and $\bx \in \Xcal$,
\begin{enumerate}[label = (\alph*), ref = \ref{assume:causal}(\alph*)]
    \item\label{assume:causal-cons} Consistency: $Y = Y^{(A)}$.
    \item\label{assume:causal-posA} Overlapping of treatment arms: $\P(A=a \mid \bX=\bx, R=k) \in (\varepsilon,1-\varepsilon)$, $k \in [J+K]$.
    \item\label{assume:causal-posR} Overlapping of site populations: $\P(R=k \mid \bX=\bx) > \varepsilon$, $k \in [J+K]$.
    \item\label{assume:causal-unconfA} Ignorability: $\left(Y^{(1)},Y^{(0)}\right) \indep (A,R) \mid \bX$ for $R \in \{\tgt,\tgs^*\}$ for some $\tgs^* \subseteq \tgs$. 
\end{enumerate}
\end{assumption}

\begin{remark}\label{remark:source}
Assumption \ref{assume:causal-unconfA} implies that 
the underlying true treatment response pattern is shared across target sites and an unspecified subset of source sites $\tgs^* \subseteq \tgs$ so that the treatment effect estimates from $\tgt$ and $\tgs^*$ can be safely combined to estimate the TATE.  
Our adaptive selection and aggregation step in FACE, as detailed in Section \ref{ssec:method-aggre}, is designed to incorporate these informative source sites $\tgs^*$ for precision gain while preventing negative transfer from non-informative source sites $\tgs\setminus\tgs^*$.
Assumption \ref{assume:causal-unconfA} assumes that controlling for observed confounders is sufficient and is similar to assumption C1 made in \cite{dehejia2021local}.
Assumption \ref{assume:causal-unconfA} may be violated, for example, when the target and source populations differ along unobserved features. \cite{nie2021covariate} considered such a setting by assuming that the distribution of potential outcomes across target and source populations are the same conditioning on observed confounders $\bX$ and unmeasured effect modifiers $\bU$ and derive bounds for the TATE by assuming a sensitivity model that directly implies a bound on the unobserved distribution shift ratio. Since violations of the transportability assumption are in general untestable, many works have also proposed sensitivity analysis for how much violation of the assumption can result in transportability bias \citep{andrews2017weighting, nguyen2017sensitivity}.
\end{remark}

We denote the specified models for the site-specific propensity score (PS) and outcome regression (OR) as:
\begin{alignat}{2}
&\text{PS :} &\quad& \P(A=a \mid R=k, \bX)   = \pi_{k} (a,\bX; \bga_k), \label{model-PS} \\
& \text{OR : } &\quad& \E(Y \mid R=k, A=a, \bX)  = m(a,\bX; \bgb_{a,k}). \label{model-OR} 
\end{alignat}
For the target sites, we require $E(Y^{(a)} \mid R=k, \bX)$ to be shared but do not require $\bga_k$ to be the same across $\tgt$. Under possible model misspecifications, we allow either (i) the outcome models in (\ref{model-OR}) to be correctly specified with  $\bgb_{a,k}=\bgb_a$, or (ii) the PS models in (\ref{model-PS}) to be correctly specified, for $k \in \tgt$.  

Since the distribution of the covariates $\bX$ can be heterogeneous across sites,  we characterize the difference in covariate distributions between a target site $k_t \in \tgt$ and 
a source site $k_s \in \tgs$ through a density ratio
$$
\omega_{k_t,k_s}(\bx) = \frac{f(\bX \mid R = k_t)}{f(\bX \mid R = k_{s})} = \frac{\P(R = k_t \mid \bX=\bx) \P(R = k_{s})}{\P(R = k_{s} \mid \bX=\bx)\P(R = k_t)}.
$$
We choose flexible semiparametric models for the density ratio 
\begin{equation}\label{def:drw}
\omega_{k_t,k_s}(\bX; \bgg_{k_t,k_s}) = \exp\{\bgg_{k_t,k_s}\trans \bgps(\bX)\},
\end{equation}
where $\bgps: \R^p \mapsto \R^q$ is a vector-valued basis function with an intercept term. One may specify a range of basis functions to capture potential non-linearity in the density ratio model to improve the robustness of the estimation for $\omega_{k_t,k_s}(\bx)$.

\begin{remark}\label{remark:drw}
The exponential tilt density ratio model \eqref{def:drw} is widely used to account for heterogeneity between two distributions \citep{qin1998inferences, qin11, duan20201fast}. 
By including higher-order terms of $\bx$ in $\bgps(\bx)$, higher-order differences such as variance and skewness can be captured. We propose in Section \ref{sec:method} a communication-efficient approach to estimate $\bgg_{k_t,k_s}$ in covariate distributions between a target site and source site without sharing individual-level data.
In the simulation study and real-data example, we have selected the exponential tilt model with $\bgps(\bx)= \bx$, which recovers the whole class of natural exponential family distributions, including the normal distribution with mean shift, Bernoulli distribution for binary covariates, etc. More flexible choices for $\bgps(\cdot)$ could help calibrate higher-order moments of covariates. However, if the number of covariates $p$ is high, one must consider the trade-off between the amount of information that is shared and the feasibility of balancing covariate distributions. 
\end{remark}

%%%%%%%%%%%%%%%%%%%%%%%%%%%%%%%%%%%%%%%
\section{Method}\label{sec:method}
%%%%%%%%%%%%%%%%%%%%%%%%%%%%%%%%%%%%%%%

In this section, we detail the FACE method. We start with an overview of its main workflow, where a schematic illustration can be found in Figure \ref{fig:flowchart} of the Supplementary Materials.
In step 1,  each target site calculates summary statistics of its covariate distribution, $\bar{\bgps}_{k} = n_{k}^{-1} \sum_{i \in \Ical_{k}}  \bgps(\bX_i)$ for $k \in \tgt$, a key quantity for estimating the density ratio model to balance covariate distributions, and broadcasts them to all source sites, along with its OR parameters $\{\hat{\bgb}_{a, k}, a = 0,1\}$. Each target site also constructs a doubly robust estimator \citep{bang2005doubly} for its site-specific ATE,  obtains additional summary statistics needed for the adaptive aggregation, and shares them with the leading analysis center (AC) (see Section \ref{ssec:method-target}). In Step 2, each source site uses the summary statistics of the target site ($\bar{\bgps}_{k}$ from $k \in \tgt$) to fit its density ratio model and construct an augmentation term $\widehat{\delta}_{\tgt,k}$ for $k \in \tgs$ for the TATE. Each source site shares the augmentation term, together with additional summary statistics needed for the aggregation, to the AC (see Section \ref{ssec:method-source}). In Step 3, the AC performs the aggregation with estimators and parameters from Steps 1 and 2 to obtain the final FACE estimator, $\widehat{\Delta}_{\tgt, \rm{FACE}}$ (see Section \ref{ssec:method-aggre}). Overall, each site is only required to share information one time with other sites. 

We detail each step of FACE in Sections \ref{ssec:method-target}-\ref{ssec:method-aggre} with generic models. Each site will need to fit both the OR model and the PS model using its own local data. Standard regression models such as logistic regression and generalized linear models can be used. Non-linear basis functions can be included to incorporate non-linear effects. For $k \in [J+K]$, we denote the estimated PS as $\pi_{k} (a,\bX; \hat{\bga}_k)$ and the predicted outcome for treatment $a$ as $m(a,\bX; \hat{\bgb}_{a,k})$, where $\hat{\bga}_k$ and $\hat{\bgb}_{a,k}$ can be achieved via classical estimation methods such as maximum likelihood estimation or estimating equations. An example with logistic regression models is given in Section \ref{ssec:theory-real} of the Supplementary Materials.

\subsection{Step 1: Estimation Using Target Data}\label{ssec:method-target}

The initial doubly robust TATE estimator is obtained from the site-specific ATE of the target sites. Within target sites $k \in \Tcal$, we compute the doubly robust TATE \citep{bang2005doubly}, $\widehat{\Delta}_{\tgt,k}=\hat{M}_{k} + \widehat{\delta}_{\tgt,k}$, where $$\hat{M}_{k} = n_{k}^{-1} \sum_{i \in \Ical_{k}}
    \left\{m(1,\bX_i; \hat{\bgb}_{1,k})  - m(0,\bX_i; \hat{\bgb}_{0,k})\right\} \text{ for } k \in \tgt$$ is the OR model based estimate of the TATE, and 
\begin{align}
    \widehat{\delta}_{\tgt, k} &= n_{k}^{-1} \sum_{i \in \Ical_{k}}
    \frac{(-1)^{1-A_i}}{\pi_{k} (A_i,\bX; \hat{\bga}_{k})} \{Y_i - m(A_i,\bX_i; \hat{\bgb}_{A_i,k})\} \text{ for } k \in \tgt \label{aug-target}
  \end{align}
is the augmentation term that guards against misspecification of the OR model.  In addition, we calculate summary statistics for the $k \in \tgt$ target site covariate distribution, $\bar{\bgps}_{k} = n_{k}^{-1} \sum_{i \in \Ical_{k}}  \bgps(\bX_i)$. The AC can construct the initial TATE estimate, $$\widehat{\Delta}_{\tgt, \tgt} = N_{\tgt}^{-1}\sum_{k \in \tgt} n_{k}  \widehat{\Delta}_{\tgt, k} 
%= \hat{M}_{\tgt} -\widehat{\delta}_{\tgt,\tgt},
$$ 
with summary data from target sites, $\{\widehat{\Delta}_{\tgt, k}, n_{k}: k \in \Tcal\}$. The consistency of $\widehat{\Delta}_{\tgt, \tgt}$ is ensured when either the PS or OR is consistently estimated for each $k \in \tgt$.

\begin{remark}
\label{remark:co-train}
    Here, we estimate $\bgb_{a}$ in each target site $k \in \tgt$ as $\hat{\bgb}_{a,k}$. Alternatively, one could estimate $\bgb_{a}$ jointly at the cost of one additional round of communication between target sites. A jointly estimated $\bgb_{a}$ could benefit from efficiency gain under certain model specification conditions. Previous literature have developed distributed methods for aggregating estimates of $\bgb_{a}$ \citep{chen2006regression, huang2019distributed, duan2020odal}. In practice, one should balance the advantage of potential efficiency gain with the cost of additional cross-site communication.
\end{remark}

To facilitate optimal aggregation, we also share the estimators
for the  variance-covariance of scaled estimators, $\sqrt{n_{k}}(\hat{M}_{k}, \widehat{\delta}_{\tgt, k},
\bar{\bgps}_{k}, \hat{\bgb}_{1,k},\hat{\bgb}_{0,k})$,
which we denote as $\hat{\Sigma}_{k}$ for the target sites $k \in \tgt$. Variance estimation $\hat{\Sigma}_{k}$ for $k \in \tgt$ can be conducted through classical influence functions
or bootstrapping within site.
The exact role of the matrix in the aggregation will be unveiled after introducing the optimal combination weights in \eqref{def:ada-aggre}, which is the centerpiece of the adaptive aggregation step.

\subsection{Step 2: Estimation Using Source Data}\label{ssec:method-source}

To safely use source data to assist in estimating $\Delta_{\tgt}$, we further account for the covariate shifts between the source sites and the target sites by tilting the source sites to the target population through the density ratios $\omega_{k_t,k_s}(\bX;\bgg_{k_t,k_s})$. If individual-level data could be shared, estimating $\hat{\bgg}_{k_t,k_s}$ could be achieved by constructing a pseudo-likelihood function as in \cite{qin1998inferences}. However, such an estimator cannot be directly obtained in a federated data setting. Instead, we propose a simple estimating equation approach that can be calculated in each source site $k_s \in \tgs$ using its data, along with summary statistics $\bar{\bgps}_{k_t}$ obtained from the target sites $k_t \in \tgt$. Specifically, we estimate $\bgg_{k_t,k_s}$ as 
\begin{equation}\label{dr_est}
\hat{\bgg}_{k_t,k_s}: \ \mbox{solution to \ }    n_{k_s}^{-1}\sum_{i \in \Ical_{k_s}} \omega_{k_t,k_s} \left(\bgps(\bX_i);\bgg_{k_t,k_s}\right)\bgps(\bX_i) = \bar{\bgps}_{k_t}.
\end{equation}

\begin{remark}
    Our approach is related to recent work that adjusts for observed differences in covariate distributions between a target population and the population that actually receives treatments \citep{hirshberg2019minimax, tan2020model}. \cite{hirshberg2019minimax} construct minimax linear weights that achieve approximate sample balance as in \ref{dr_est} uniformly over an absolutely convex class $\mathcal{M}$. They show that when $\mathcal{M}$ is selected appropriately, the solution to \ref{dr_est} converges in empirical mean square to the functional’s Riesz representer, i.e., the unique square-integrable function that satisfies the corresponding population balance condition for all square-integrable functions \citep{hirshberg2021augmented}. Relatedly, \cite{tan2020model} propose regularized calibrated estimators in the high-dimensional setting under minimal sparsity assumptions.
\end{remark}

For each source site, we construct a site augmentation term similar to the augmentation term in (\ref{aug-target}) for the target sites but with an additional density ratio weight
$$
  \widehat{\delta}_{\tgt,k_s}
  = N_\tgt^{-1}\sum_{k_t \in \tgt} \frac{n_{k_t}}{n_{k_s}} \sum_{i \in \Ical_{k_s}} \omega_{k_t,k_s}(\bX_i;\hat{\bgg}_{k_t,k_s}) \frac{(-1)^{1-A_i}}{\pi_{k_s} (A_i,\bX_i; \hat{\bga}_{k_s})} 
  \{Y_i - m(A_i,\bX_i; \hat{\bgb}_{A_i,k_t})\} \text{ for } k_s \in \tgs.
$$
We use the OR estimates from target sites $\hat{\bgb}_{A_i,k_t}$
to ensure robustness when the OR is misspecified. See Remark \ref{remark:TATE_source} for details.

Then, the site-specific augmentation terms $\widehat{\delta}_{\tgt,k_s}$ are shared back to the
AC, together with (i) $\hat{\sigma}^2_{k_s}$, an estimate for
the scaled conditional variance $n_{k_s}\Var\left(\widehat{\delta}_{\tgt,k_s}\mid \Dscr_{\tgt}\right)$, 
and (ii) $\hat{\bd}_{k_t,k_s}$, an estimate for the partial derivatives of $\widehat{\delta}_{\tgt,k_s}$
with respect to $\bar{\bgps}_{k_t}$, $\hat{\bgb}_{1,k_t}$, and $\hat{\bgb}_{0,k_t}$. The role of $\hat{\bd}_{k_t,k_s}$ in the aggregation will be explained in \eqref{def:ada-aggre}.
Both $\hat{\sigma}^2_{k_s}$ and $\hat{\bd}_{k_t,k_s}$ can be constructed from classical influence functions. Alternatively, $\hat{\sigma}^2_{k_s}$ can be estimated by bootstrapping within site 
and $\hat{\bd}_{k_t,k_s}$ can be estimated by numerical derivatives.

\begin{remark}\label{remark:TATE_source}
Combining the source site augmentation term $\widehat{\delta}_{\tgt,k_s}$ with the initial TATE OR estimator from the target sites $\hat{M}_{\tgt} = N_\tgt^{-1} \sum_{k_t \in \tgt} n_{k_t} \hat{M}_{k_t}$, we obtain the $k_s \in \tgs$ source site  estimators $\widehat{\Delta}_{\tgt,k_s} =  \hat{M}_{\tgt} + \widehat{\delta}_{\tgt,k_s}$ as 
\begin{align*}
   \widehat{\Delta}_{\tgt,k_s} &= N_{\tgt}^{-1}  \sum_{k_t \in \tgt}n_{k_t}\Bigg(n_{k_t}^{-1}\sum_{i \in \Ical_{k_t}}  
  \{m(1,\bX_i; \hat{\bgb}_{1,k_t})  - m(0,\bX_i; \hat{\bgb}_{0,k_t}) \} \\ &+ 
  n_{k_s}^{-1}\sum_{i \in \Ical_{k_s}} \omega_{k_t,k_s}(\bX_i;\hat{\bgg}_{k_t,k_s})\frac{(-1)^{1-A_i} }{\pi_{k_s} (A_i,\bX_i; \hat{\bga}_{k_s})} \{Y_i - m(A_i,\bX_i; \hat{\bgb}_{A_i,k_t})\}\Bigg). 
\end{align*}
When the underlying OR model in the $k_s \in \tgs$ source site is the same as in the target population, the estimator $\widehat{\Delta}_{\tgt,k_s}$ is doubly robust in the following sense: either
(i) the OR model is consistent for all $k \in \{\tgt, k_s\}$, 
or (ii) the PS and density ratio models are consistent for the source site. Shifts in covariate distributions may induce heterogeneity in OR estimates across sites under misspecified OR models, even if the conditional distribution $Y \mid A, \bX$ is shared. 
To achieve robustness against misspecified OR models, it is important to use the same 
$\hat{\bgb}_{a,k_t}$ for $\hat{M}_{\tgt}$ and $\widehat{\delta}_{\tgt,k_s}$ so that we may rely on the correct PS and density ratio models for consistency according to the alternative representation
\begin{align*}
N_{\tgt}^{-1} & \sum_{k_t \in \tgt}n_{k_t}\Bigg\{ n_{k_s}^{-1}\sum_{i \in \Ical_{k_s}} \omega_{k_t,k_s}(\bX_i;\hat{\bgg}_{k_t,k_s})\frac{(-1)^{1-A_i} }{\pi_{k_s} (A_i,\bX_i; \hat{\bga}_{k_s})} Y_i \\
& + n_{k_t}^{-1}\sum_{i \in \Ical_{k_t}}  
  m(1,\bX_i; \hat{\bgb}_{1,k_t}) 
  - n_{k_s}^{-1}\sum_{i \in \Ical_{k_s}} \omega_{k_t,k_s}(\bX_i;\hat{\bgg}_{k_t,k_s})\frac{A_i }{\pi_{k_s} (1,\bX_i; \hat{\bga}_{k_s})} m(1,\bX_i; \hat{\bgb}_{1,k_t})\\
 & -  n_{k_t}^{-1}\sum_{i \in \Ical_{k_t}}  
  m(0,\bX_i; \hat{\bgb}_{0,k_t}) 
  + n_{k_s}^{-1}\sum_{i \in \Ical_{k_s}} \omega_{k_t,k_s}(\bX_i;\hat{\bgg}_{k_t,k_s})\frac{1-A_i }{\pi_{k_s} (0,\bX_i; \hat{\bga}_{k_s})} m(0,\bX_i; \hat{\bgb}_{0,k_t})\bigg\}. 
\end{align*}

To protect against negative transfer from source sites with biased TATE estimators, we combine information from each source site with the target sites through our adaptive aggregation step in Section \ref{ssec:method-aggre}.
\end{remark}

\subsection{Step 3: Adaptive Aggregation}\label{ssec:method-aggre}

In the final step, we obtain our FACE estimator by adaptively aggregating the initial TATE estimator $\widehat{\Delta}_{\tgt,\tgt}$ and the source site estimators $\widehat{\Delta}_{\tgt, k_s}$. 
Denote 
$\widehat{\delta}_{\tgt, \tgt} = N_{\tgt}^{-1}\sum_{k \in \tgt} n_{k} \widehat{\delta}_{\tgt, k}$.
The AC can estimate $\Delta_{\tgt}$ by taking a linear combination of the initial TATE estimator $\widehat{\Delta}_{\tgt,\tgt}$ and the source site estimators $\widehat{\Delta}_{\tgt, k_s}$, where the weights are estimated to make an optimal bias-variance tradeoff. The proposed FACE estimator can be viewed as an ``anchor and augmentation'' estimator, which weights the source site estimators $\widehat{\Delta}_{\tgt, k_s}$ by $\eta_{k_s}$, $k_s \in \tgs$ and the target estimator $\widehat{\Delta}_{\tgt, \tgt}$ by $(1-\sum_{k_s \in \tgs}\eta_{k_s} )$. FACE is given by
\begin{equation} \label{def:FACE}
    \FACE =  \widehat{\Delta}_{\tgt, \tgt} + \sum_{k_s \in \tgs} \eta_{k_s} \{\widehat{\Delta}_{\tgt, k_s}-\widehat{\Delta}_{\tgt, \tgt} \}
    = \widehat{\Delta}_{\tgt, \tgt} + \sum_{k_s \in \tgs} \eta_{k_s} \{\widehat{\delta}_{\tgt,k_s}-\widehat{\delta}_{\tgt, \tgt}\}, 
\end{equation}
which anchors on the initial TATE estimator $\widehat{\Delta}_{\tgt, \tgt}$ and is augmented with source site estimators $\widehat{\Delta}_{\tgt, k_s}$, with the weights $\{\eta_{k_s}, k_s \in \tgs\}$ to be estimated in a data-adaptive fashion to filter out potentially biased source site estimators. The second expression of $\FACE$ in \eqref{def:FACE} shows how the parameters from Steps 1 and 2 are used to construct the FACE estimator.

Moreover, the aggregation of the remaining unbiased source site augmentation terms should also minimize the estimation variance. Under the federated learning setting, the key to evaluating the variance of (\ref{def:FACE}) is to decompose it into contributions from separate sites so that they can be estimated within each site. For any subset of $\tgs$, $\tgs' \subseteq \tgs$, 
we consider the following decomposition
\begin{align}
  &\Var\left\{ \widehat{\Delta}_{\tgt, \tgt} + \sum_{k_s \in \tgs'} \eta_{k_s} (\widehat{\Delta}_{\tgt, k_s}-\widehat{\Delta}_{\tgt, \tgt})\right\} \notag \\
 & \approx \sum_{k_s \in \tgs'} \eta_{k_s}^2 \Var\left(\widehat{\delta}_{\tgt, k_s} \mid \Dscr_{\tgt}\right) \notag \\
 & +
 \sum_{k_t \in \tgt}\Var \left\{ \left(\frac{n_{k_t}}{N_{\tgt}},\frac{n_{k_t}-n_{k_t}\sum_{k_s \in \tgs'} \eta_{k_s}}{N_{\tgt}}, \sum_{k_s \in \tgs'} \eta_{k_s}\bd_{k_t,k_s}\trans\right) \left(\hat{M}_{\tgt}, \widehat{\delta}_{k_t}, \left(\bar{\bgps}_{k_t}\trans,
 \hat{\bgb}_{1,k_t}\trans,\hat{\bgb}_{0,k_t}\trans \right) \right)\trans \right\}, \label{eq:var-aggre}
\end{align}
where $\bd_{k_t,k_s}$ is the limit for $\hat{\bd}_{k_t,k_s}$, which is the partial derivative of $\widehat{\delta}_{\tgt,k_s}$ with respect to the broadcast estimators $\bar{\bgps}_{k_t}$,
 $\hat{\bgb}_{1,k_t}$ and $\hat{\bgb}_{0,k_t}$. 
The source site augmentation terms $\widehat{\delta}_{\tgt,k_s}$ involve both the source site $k_s \in \tgs$ data and estimated parameters $\left\{\bar{\bgps}_{k_t},
 \hat{\bgb}_{1,k_t},\hat{\bgb}_{0,k_t}\right\}$ from target sites $k_t \in \tgt$.  
 We characterize the uncertainty contributions from these two independent sources by $\widehat{\delta}_{\tgt,k_s}\mid \Dscr_{\tgt}$   
 and $\left(\bar{\bgps}_{k_t}\trans,
\hat{\bgb}_{1,k_t}\trans,\hat{\bgb}_{0,k_t}\trans\right)\bd_{k_t,k_s}$, respectively. 
We decouple the dependence of the source site augmentation terms $\widehat{\delta}_{\tgt,k_s}$ on the target sites 
by subtracting the first order approximation of the dependence $\left(\bar{\bgps}_{k_t}\trans,
 \hat{\bgb}_{1,k_t}\trans,\hat{\bgb}_{0,k_t}\trans\right)\bd_{k_t,k_s}$. 
The resulting $\widehat{\delta}_{\tgt,k_s} - \bd_{k_t,k_s}\trans\bar{\bgps}_{k_t}$ is asymptotically independent of the target sites. 

Since including information from non-informative source sites $\tgs\setminus \tgs^*$ may lead to biases, we adopt an adaptive combination strategy similar to the one given in \cite{cheng2021adaptive} for combining data from a randomized trial and an observation study. Here, we overcome the additional challenge of data sharing constraints, and we propose the following adaptive $L_1$ penalized optimal aggregation
\begin{align}\label{def:ada-aggre}
    \hat{\bge} &= \argmin_{\bge \in \R^{K}}
    \underbrace{N\left[ \sum_{k_s \in \tgs} \eta_{k_s}^2\frac{\hat{\sigma}_{k_s}^2}{n_{k_s}} 
    + \sum_{k_t \in \tgt} \hat{\bh}_{k_t}(\bge)\trans \frac{\hat{\Sigma}_{k_t}}{n_{k_t}} \hat{\bh}_{k_t}(\bge)
    \right]}_{\hat{L}(\bge)} + \lambda \sum_{k_s\in\tgs} |\eta_{k_s}| \left(\widehat{\delta}_{\tgt,k_s} - \widehat{\delta}_{\tgt,\tgt}\right)^2,     
\end{align}
where 
\[
\hat{\bh}_{k_t}(\bge)  =
  \left(\frac{n_{k_t}}{N_{\tgt}},\frac{n_{k_t}-n_{k_t}\sum_{k_s \in \tgs^*} \eta_{k_s}}{N_{\tgt}}, \sum_{k_s \in \tgs} \eta_{k_s}\hat{\bd}_{k_t,k_s}\trans\right)\trans,
\]
with $\hat{\Sigma}_{k_t}$ estimated from Step 1 and $\hat{\sigma}_{k_s}^2$ and $\hat{\bd}_{k_t,k_s}$ estimated from Step 2. 
The multiplicative $N$ factor is required to stabilize the loss. Choosing $\lambda \asymp N^\nu$ with $\nu \in (0,1/2)$, we achieve the following oracle property for selection
and aggregation: (i) biased source site augmentation terms have zero weights
with high probability; (ii) regularization on the weights for unbiased
source site augmentation terms is asymptotically negligible ($\ll N^{-1/2}$). Analogous to the phenomenon in meta-analysis, 
the estimation uncertainty of $\hat{\bge}$ has no asymptotic effect on the aggregated estimator. 

Using the variance estimator (stabilized by ``$N$'' factor likewise)
\begin{equation}\label{def:var}
    \hat{\Vcal} = N\left\{\sum_{k_s\in \tgs} \hat{\eta}_{k_s} \frac{\hat{\sigma}_{k_s}^2}{n_{k_s}}
    + \sum_{k_t \in \tgt} \hat{\bh}_{k_t}(\hat{\bge})\trans \frac{\hat{\Sigma}_{k_t}}{n_{k_t}} \hat{\bh}_{k_t}(\hat{\bge})\right\}
\end{equation}
and the $1-\alpha/2$ quantile for the standard normal distribution $\Zcal_{\alpha/2}$,
we construct the $(1-\alpha) \times 100\%$ confidence interval (CI)
\begin{equation}\label{def:CI}
    \hat{\Ccal}_{\alpha} = \left[\FACE - \sqrt{\hat{\Vcal}/N} \Zcal_{\alpha/2},
    \FACE + \sqrt{\hat{\Vcal}/N} \Zcal_{\alpha/2}\right].
\end{equation}
The full FACE workflow is summarized in Algorithm \ref{alg:FACE}.

\vspace{10 pt}

\begin{algorithm}[H]
\label{alg1}
\vspace{5 pt}
 \KwData{$J$ Target sites $k_t \in \tgt$, $K$ Source sites $k_s \in \tgs$, and a Leading AC}

 \For{Target sites $k_t \in \tgt$}{ Estimate $\bga_{k_t}$, $\bgb_{a,k_t}$ to calculate the initial TATE $\widehat{\Delta}_{\tgt,k_t}$, its augmentation $\widehat{\delta}_{\tgt,k_t}$,
  and the variance estimator $\hat{\Sigma}_{k_t}$, and transfer to the leading AC.
  Calculate $\bar{\bgps}_{k_t}$ and broadcast to source sites along with $\hat{\bgb}_{a,k_t}$. 

 }
 \For{Source sites $k_s \in \tgs$}{
     Estimate $\bgg_{k_t,k_s}$ and $\bga_{k_s}$
     to calculate the site-specific augmentation $\widehat{\delta}_{\tgt,k_s}$ and transfer to the leading AC. Calculate $\hat{\sigma}^2_{k_s}$, $\hat{\bd}_{k_t,k_s}$ and transfer to the leading AC.

 }
  \For{Leading AC}{
 Estimate $\bge$ by solving the penalized regression in (\ref{def:ada-aggre}). Construct the final global estimator as $\FACE$ by (\ref{def:FACE}). Calculate the global estimator variance by (\ref{def:var}) and construct 95\% CI.

 }

 \KwResult{Global TATE estimate, $\FACE$ and $95\%$ CI }
 \caption{FACE under generic model specifications}\label{alg:FACE}
\end{algorithm}

\begin{remark}
Our aggregation procedure is communication-efficient and privacy-protected, whereas aggregation procedures given in the current literature such as those in \cite{cheng2021adaptive} require sharing individual-level influence functions. 
 Equation \eqref{def:ada-aggre} is constructed using summary statistics, which provides a federated learning solution when individual-level data sharing is forbidden.
\end{remark}

\subsection{Cross-Validation and Tuning Parameters}
\label{ssec:method-tuning}
    To choose an optimal tuning parameter $\lambda$, we propose a sample splitting approach that does not require sharing individual-level data. In each site, the data is first split into training and validation datasets, keeping the same proportion within each site. In the training datasets, Algorithm \ref{alg1} is implemented to obtain the summary statistics ($\hat{\Sigma}_{k_t}$, $\hat{\bd}_{k_s}$, $\hat{\sigma}^2_{k_s}$, $\widehat{\delta}_{\tgt,k_s}$, and $\widehat{\delta}_{\tgt,\tgt}$) needed for Equation \eqref{def:ada-aggre}. The AC selects a grid of $\lambda$ values and calculates $\hat{\bge}(\lambda)$ by solving the penalized regression in (\ref{def:ada-aggre}).  In parallel, the validation datasets are used to obtain summary statistics denoted by  ($\tilde{\Sigma}_{k_t}$, $\tilde{\bd}_{k_tk_s}$, $\tilde{\sigma}^2_{k_s}$, $\tilde{\delta}_{\tgt,k_s}$ and $\tilde{\delta}_{\tgt,\tgt}$). These summary statistics are calculated using the validation datasets and plugging in the parameters estimated from the corresponding training datasets. The AC sets the value of the optimal tuning parameter, $\lambda_{\mathrm{opt}}$, to be the value corresponding to $\hat{\bge}$ that minimizes $Q(\hat{\bge})$ in the validation datasets, defined as
     $$Q(\hat{\bge}) = N^V\left[ \sum_{k_s \in \tgs} \hat{\eta}_{k_s}^2\frac{\tilde{\sigma}_{k_s}^2}{n_{k_s}^V} 
    + \sum_{k_t \in \tgt} \tilde{\bh}_{k_t}(\hat{\bge})\trans \frac{\tilde{\Sigma}_{k_t}}{n_{k_t}^V} \tilde{\bh}_{k_t}(\hat{\bge})
    \right],$$
    where $N^V$, $n_{k_s}^V$, and $n_{k_t}^V$ are the sample sizes for validation data from all sites,  source sites $k_s \in \tgs$, and target sites $k_t \in \tgt$, respectively. 

    \begin{remark}
      The upper and lower bounds on the grid of $\lambda$ values can be left unrestricted; in practice, we have found that searching between $0.01$ to $100$ to be sufficiently large to provide good finite sample performance. For increased stability to the choice of $\lambda$, we have implemented five-fold cross-validation, where we take $\lambda_{opt}$ to be the value corresponding to $\hat{\bge}$ that minimizes the average of $Q(\cdot)$ over five folds \citep{ChernozhukovEtal18DML}.
    \end{remark}

\section{Theoretical Guarantees}\label{sec:theory}
In this section, we provide the theoretical results for the FACE estimator.
We start with a high-level theory for a generic choice of models in
Section \ref{ssec:theory-gen}. 
Then, we discuss the efficiency gained from leveraging source 
sites in Section \ref{ssec:theory-RE}. 
In our asymptotic theory, $N$ is allowed to grow but the distribution for  $(Y,\bX\trans,A,R)\trans$ and 
$J+K$ are fixed.

\subsection{Theory for General FACE}\label{ssec:theory-gen}

To compress notation, we combine the broadcast parameters
and their asymptotic limits as
\begin{equation}
    \hat{\bgth}_{k_t} 
    = \left(\bar{\bgps}_{k_t}\trans,\hat{\bgb}_{1,k_t}\trans,\hat{\bgb}_{0,k_t}\trans\right)\trans, \;
    \bar{\bgth}_{k_t} 
    = \left(\E\{\bgps(\bX)\trans\mid R = k_t\},\bar{\bgb}_{1,k_t}\trans,\bar{\bgb}_{0,k_t}\trans\right)\trans. 
\end{equation}

\begin{assumption}{2}{}\label{assume:FACE}
For absolute constants $M, \varepsilon >0$,
\begin{enumerate}[label = (\alph*), ref = \ref{assume:FACE}(\alph*)]
\item\label{assume:FACE-if} (Regularity of estimators)
The estimators $\hat{M}_{\tgt}$, $\widehat{\delta}_{\tgt,k_t}$, $\hat{\bgb}_{a,k_t}$
and $\widehat{\delta}_{\tgt,k_s}$ admit the following asymptotically linear representations
\begin{align*}
 & \sqrt{N_{\tgt}}(\hat{M}_{\tgt} - \bar{M}_{\tgt,\tgt})
 = \frac{1}{\sqrt{N_{\tgt}}}\sum_{k_t \in \tgt}\sum_{i\in \Ical_{k_t}} \zeta_i + o_p(1),  \\
 & \sqrt{N_{\tgt}}(\widehat{\delta}_{\tgt,\tgt} - \bar{\delta}_{\tgt,\tgt})
 = \frac{1}{\sqrt{N_{\tgt}}}\sum_{k_t\in\tgt}\sum_{i\in \Ical_{k_t}} \xi_{i,\tgt} + o_p(1), \\
 & \sqrt{n_{k_s}}(\widehat{\delta}_{\tgt,k_s} - \bar{\delta}_{\tgt,k_s})
 = \frac{1}{\sqrt{n_{k_s}}}\sum_{i\in \Ical_{k_s}} \xi_{i,k_s}
 + \sqrt{n_{k_s}} \sum_{k_t \in \tgt}\bar{\bd}_{k_t,k_s}\trans \left(\hat{\bgth}_{k_t} - \bar{\bgth}_{k_t} \right)
 + o_p(1),\\
 &\sqrt{n_{k_t}}\left(\hat{\bgb}_{a,k_t} - \bar{\bgb}_{a,k_t}\right) =  \frac{1}{\sqrt{n_{k_t}}}
 \sum_{i \in \Ical_{k_t}} \bgu_{i,a} + o_p(1). 
\end{align*}
with bounded asymptotic limits $\bar{M}_{\tgt,\tgt}$, $\bar{\delta}_{\tgt,\tgt}$,
$\bar{\delta}_{\tgt,k_s}$, $\bar{\bd}_{k_t,k_s}$
and iid mean zero random variables $\zeta_i$, $\xi_{i,\tgt}$, $\xi_{i,k_s}$.

\item \label{assume:FACE-X} (Compact support)
The covariates $\bX$ and their functions $\bgps(\bX)$ in the density ratio are in compact sets
$\bX \in [-M,M]^p$ and $\bgps(\bX) \in [-M,M]^q$
almost surely. 

\item\label{assume:FACE-var} (Stable variance)
The variance of $\xi_{i,k_s}$
is in the set $[\varepsilon,M]$.
The variance-covariance matrix
$$
\Sigma_{k_t} = \Var\left\{\left(\zeta_i, \xi_{i,\tgt}, \bgps(\bX_i)\trans,\bgu_{i,1}\trans,\bgu_{i,0}\trans\right)\trans
\mid R=k_t\right\}
$$
has eigenvalues all in $[\varepsilon,M]$ for some positive constant $\varepsilon$ and $M$.

\item \label{assume:FACE-se} (Regularity of auxiliary estimators)
The estimators $\hat{\Sigma}_{k_t}$, $\hat{\sigma}^2_{k_s}$, $\hat{\bd}_{k_s}$
are $\sqrt{N}$-consistent
\begin{align*}
    & \sum_{k_t\in\tgt} \left\|\hat{\Sigma}_{k_t} - \Sigma_{k_t} \right\|   + \sum_{k_s\in\tgs} \left\{\left|\hat{\sigma}^2_{k_s} - \Var(\xi_{i,k_s}\mid R_i=k_s)\right|
+ \left\|\hat{\bd}_{k_s} - \bar{\bd}_{k_s}\right\|\right\}
= O_p\left(N^{-1/2}\right).
\end{align*}
\item\label{assume:FACE-dr} (Root-$N$ rate consistency)
For each target site $k_t \in \tgt$, at least one of the two models is correctly specified:
\begin{enumerate}[label = -\roman*, ref = \ref{assume:FACE-dr}-\roman*]
    \item \label{assume:FACE-ps}
    the PS model is consistently estimated: 
    $$
    \sup_{a=0,1}\sup_{\|\bx\|_{\infty} \le M}  \sum_{k_t \in \tgt}\left| \P(A=a\mid \bX=\bx, R = k_t) - \pi_k(a,\bx;\hat{\bga}_{k_t}) \right| = O_p\left(N^{-1/2}\right). 
    $$
    \item \label{assume:FACE-or}
    the OR model is consistently estimated: 
    $$
    \sup_{a=0,1}\sup_{\|\bx\|_{\infty} \le M} \sum_{k_t \in \tgt}\left| \E(Y\mid A=a, \bX=\bx, R = k_t) - m_{k_t}(a,\bx;\hat{\bgb}_{a,k_t}) \right| =  O_p\left(N^{-1/2}\right).
    $$
\end{enumerate}
\end{enumerate}
\end{assumption}

Assumptions \ref{assume:FACE-if} and \ref{assume:FACE-dr} are the typical regularity conditions and can be verified in two steps: 1) asymptotic normality of model estimators \citep{van2000asymptotic} and
2) local expansion of the estimators.
Assumption \ref{assume:FACE-var} regulates the scale of variability of the data,
which leads to a stable variance for $\FACE$.
Assumption \ref{assume:FACE-dr} ensures identification of the true TATE
by anchoring on $\widehat{\Delta}_{\tgt,\tgt}$  \citep{bang2005doubly}. Note that in the setting of multiple target sites, Assumption \ref{assume:FACE-dr} allows for each target site to have different correct model specifications for either the OR model or the PS model. In Supplement \ref{ssec:theory-real}, we provide a detailed set of conditions corresponding to the realization of logistic regression models to estimate nuisance models.

We now state the theory for the general FACE estimation.
\begin{theorem}\label{thm:FACE}
Under Assumptions \ref{assume:causal} and \ref{assume:FACE},
the FACE estimator is consistent and asymptotically normal
with consistent variance estimation $\hat{\Vcal}$,
$$
\sqrt{N/\hat{\Vcal}} \left(\FACE - \Delta_{\tgt}\right)
\leadsto \mathcal{N}(0,1).
$$
We use $\leadsto$ for convergence in distribution.
\end{theorem}
Theorem \ref{thm:FACE} implies that \eqref{def:CI} provides asymptotically honest coverage.
\begin{corollary}\label{cor:CI}
Under Assumptions \ref{assume:causal} and \ref{assume:FACE},
the coverage rate of
the confidence interval \eqref{def:CI} approaches the nominal level
asymptotically
$$
\lim_{N\to\infty}\P\left(\Delta_{\tgt} \in \hat{\Ccal}_{\alpha}\right) = 1-\alpha
$$
\end{corollary}
The proof of Theorem \ref{thm:FACE} and Corollary \ref{cor:CI} is given in Supplement \ref{assec:proof-FACE}. A key step in the proof of Theorem \ref{thm:FACE} is the analysis of
the $L_1$ penalized adaptive selection and aggregation \eqref{def:ada-aggre}.
We are able to establish the oracle property \citep{SCADoracle01}, i.e., the data-driven selection and aggregation through \eqref{def:ada-aggre} is asymptotically equivalent to the process with a priori selection and optimal aggregation. 
The problem is different from the typical penalized regression, so we develop a new proof strategy.
We first analyze the optimal combination with oracle selection, in which the biased augmentations are excluded.
For unbiased augmentations, $\widehat{\Delta}_{\tgt,k_s} - \widehat{\Delta}_{\tgt,\tgt}
= O_p\left(N^{-1/2}\right)$, the penalty term is asymptotically
negligible, i.e., $\lambda(\widehat{\Delta}_{\tgt,k_s} - \widehat{\Delta}_{\tgt,\tgt})^2 = o_p\left(N^{-1/2}\right)$ when $\lambda$ is chosen such that $\lambda \asymp N^\nu$ with $\nu \in (0,1/2)$.
Thus, the estimated combination converges to the asymptotic limit at the regular $N^{-1/2}$ rate.
Finally, we show that the estimated combination with oracle selection also solves the original problem with high probability.

\begin{remark}\label{remark:eta-reg}
 The proposed FACE estimator has CI with honest coverage of the true TATE if all the biases are detectable $|\bar{\delta}_{\tgt,k_s} - \bar{\delta}_{\tgt,\tgt}|\gg N^{-1/2}$. However, in accordance with the limit characterized by the information lower bound, it is not possible to detect source sites with small biases of $|\bar{\delta}_{\tgt,k_s} - \bar{\delta}_{\tgt,\tgt}|\lesssim N^{-1/2}$ order. Involving these sites in the final TATE will introduce non-negligible bias that cannot be corrected. With the presence of weakly biased $\hat{\delta}_{\tgt,k_s}$, 
the undetectable bias may induce a biased, non-regular FACE estimator, as well as undercoverage of the confidence interval.
While such an issue would not occur in the 
large $N$ and finite $K$ framework of our theory, 
we offer a possible remedy for regularity with regard to finite sample performance. 
In the aggregation \eqref{def:ada-aggre}, we may substitute the penalty factor with a truncated Wald statistic
$$
\hat{\bge} = \argmin_{\bge \in \R^{K}}
\hat{L}(\bge)
+ \lambda \sum_{k_s\in\tgs} 
|\eta_{k_s}| \left\{\left(\frac{\sqrt{N}\left|\hat{\delta}_{\tgt,k_s}-\hat{\delta}_{\tgt,\tgt}\right|}{\hat{\sigma}_{\delta,k_s}} - 1/\lambda\right) \vee 0\right\}, \,
N^{-1/2} \ll \lambda \ll 1, 
$$
where $\hat{\sigma}_{\delta,k_s}^2$ is the 
variance estimator for 
the $\hat{\delta}_{\tgt,k_s}-\hat{\delta}_{\tgt,\tgt}$, 
$$
\hat{\sigma}_{\delta,k_s}^2 =  N\left(\eta_{k_s}^2\frac{\hat{\sigma}_{k_s}^2}{n_{k_s}}
+ \sum_{k_t \in \tgt} \hat{\bh}_{k_t,k_s}\trans \frac{\hat{\Sigma}_{k_t}}{n_{k_t}} \hat{\bh}_{k_t,k_s}\right), \,
\hat{\bh}_{k_t,k_s}  = 
  \left(0,-\frac{n_{k_t}}{N_{\tgt}},\hat{\bd}_{k_t,k_s}\trans\right)\trans.
$$
Using the modified penalty factor that converges to its limit at $N^{-1/2}$ rate for $k_s \in \tgs^\dagger = \{k_s: \left|\bar{\delta}_{\tgt,k_s}-\bar{\delta}_{\tgt,\tgt}\right| \lesssim \lambda^{-1} N^{-1/2}\}$ and diverges to $+\infty$ elsewhere,
we will be able to establish $N^{-1/2}$ concentration of 
$\hat{\bge}$ at
$$
\Tilde{\bge} = \argmin_{\bge \in \tgs^\dagger}
L^*(\bge) + \lambda \sum_{k_s\in\tgs} 
|\eta_{k_s}| \left\{\left(\frac{\sqrt{N}\left|\bar{\delta}_{\tgt,k_s}-\bar{\delta}_{\tgt,\tgt}\right|}{\bar{\sigma}_{\delta,k_s}} - 1/\lambda\right) \vee 0\right\}
$$
over the support $\tgs^\dagger$. 
The approximation in Lemma \ref{lem:FACE} 
will hold, as the extra bias term decays to zero
\begin{align*}
    \sum_{k_s \in \tgs^\dagger}\left|(\hat{\eta}_{k_s}-\Tilde{\eta}_{k_s})
    \sqrt{N}(\bar{\delta}_{\tgt,k_s}-\bar{\delta}_{\tgt,\tgt})
    \right|
    = O_p \left(\bar{\delta}_{\tgt,k_s}-\bar{\delta}_{\tgt,\tgt}\right)
    = o_p(1). 
\end{align*}
Consequently, the modified FACE estimator would be asymptotically normal after removing the bias. 
The confidence interval would have reasonable coverage 
if the accumulated bias 
$\sum_{k_s \in \tgs^\dagger} \Tilde{\eta}_{k_s}(\bar{\delta}_{\tgt,k_s}-\bar{\delta}_{\tgt,\tgt})$ 
is smaller than the standard deviation. 
\end{remark}

\begin{remark}
    We may modify the CI to conservatively capture the uncertainty of $\FACE$. Denote the estimators needed for aggregation as $\hat{\bgth}$, including 
$\hat{M}_{\tgt}$, $\hat{\delta}_{\tgt,\tgt}$, $\hat{\delta}_{\tgt,k_s}$, 
$\hat{\Sigma}_{k_t}$, $\hat{\bh}_{k_t}$. 
We denote the process of getting $\FACE$ from $\hat{\bgth}$ as $\FACE=\Hscr(\hat{\bgth})$, which is continuous along $\bgth$ and it is deterministic given  $\hat{\bgth}$.
Suppose $\bgth_*$ is the asymptotic limit of $\hat{\bgth}$. 
Based on the standard asymptotic normality of $\sqrt{N}(\hat{\bgth}-\bgth_*)$, we may construct a standard confidence region
$\Cscr(\hat{\bgth},\alpha)$ such that 
$$
\liminf_{N\to \infty}\P\{ \bgth_* \in  \Cscr(\hat{\bgth},\alpha)\}
\ge 1-\alpha. 
$$
Mapping through $\Hscr$ will produce a confidence interval for $\Ascr(\bgth_*)$, 
$$
\liminf_{N\to \infty}\P\{\Hscr(\bgth_*) \in \Hscr \circ \Cscr(\hat{\bgth},\alpha)\}
\ge 1-\alpha. 
$$
This process will account for the uncertainty in site selection but may produce a conservative CI.
\end{remark}

\begin{remark}
For consistency of $\FACE$, we require that the PS or OR model is correct for the target sites but allow the models for the source sites and density ratio to be misspecified. To meaningfully leverage information from source sites for the TATE,
we would expect that many $k_s \in \tgs$ among the source sites (i) satisfy the ignorability condition \ref{assume:causal-unconfA} and  (ii) either the OR model $m(a)$ is correct,
or both the PS $\pi_{k_s}$ and the density ratio $\omega_{k_t,k_s}$ models are correct. For source sites satisfying the conditions above, their site-specific augmentations are unbiased and thus contribute to the efficiency improvement of $\FACE$.
\end{remark}

\begin{remark}
\label{remark:highD}
    When the oracle property can be achieved for estimating 
   sparse $\hat{\bgb}_{a,k}$, $\hat{\bgg}_{k_t,k_s}$ and $\hat{\bga}_{k}$ under the high-dimensional setting ($p$ grows with $n_k$) with minimal signal strength \citep{fan2001variable, zou2006adaptive, lv2009unified}, it ensures the exact identification of the non-zero elements in the coefficients, which reduces the high-dimensional setting to the low-dimensional setting studied in our paper. For the general high-dimensional setting without guarantee of the oracle property, 
    our FACE method can be extended under ideal situations with perfectly specified models through cross-fitting \citep{ChernozhukovEtal18DML}. Extensions beyond such an ideal setting require careful investigation beyond the approaches considered in the existing literature since General Neyman orthogonality requires all estimated models to be consistent for their target conditional expectations. As a fundamental issue in learning the TATE, the potentially incompatible source sites (i.e., source sites in $\tgs\setminus\tgs^*$) will invalidate such a consistency requirement.
\end{remark}

\subsection{Relative Efficiency}\label{ssec:theory-RE}

Notice that we recover the initial TATE estimator $\widehat{\Delta}_{\tgt,\tgt}$ from \eqref{def:FACE} 
if $\hat{\bge} = \mathbf{0}$. 
Since we are minimizing the post-aggregation variance, 
the optimal solution must be no worse than any alternative solutions. 
If there exists informative source sites in $\tgs'$, as defined in Assumption \ref{assume:S}, improvement in the efficiency of FACE compared to the target only estimator is guaranteed. 

\begin{assumption}{3}{}\label{assume:S}
For a nonempty set $\tgs' \subseteq \tgs$, 
\begin{enumerate}[label = (\alph*), ref = \ref{assume:S}(\alph*)]
\item \label{assume:S-dr} 
One of the following holds:
\begin{enumerate}[label = (\roman*), ref = \ref{assume:S-dr}(\roman*)]
\item \label{assume:S-or}Correct OR:
    the OR model is consistently estimated: 
    $$
    \sup_{a=0,1}\sup_{\|\bx\|_{\infty} \le M} \sum_{k_t \in \tgt}\left| \E(Y\mid A=a, \bX=\bx, R = k_t) - m_{k_t}(a,\bx;\hat{\bgb}_{a,k_t}) \right| =  O_p\left(N^{-1/2}\right); 
    $$
\item \label{assume:S-psdr} Consistent weighting: 
the PS and density ratio models are consistently estimated: 
\begin{align*}
\sup_{a=0,1}\sup_{\|\bx\|_{\infty} \le M} & \sum_{k_s \in \tgs'}\left| \P(A=a\mid \bX=\bx, R = k_s) - \pi_{k_s}(a,\bx;\hat{\bga}_{k_s}) \right| \\& + \sum_{k_t \in \tgt}\sum_{k_s\in\tgs'}\left| \frac{\P(R = k_t \mid \bX=\bx) \P(R = k_s)}{\P(R = k_s \mid \bX=\bx)\P(R = k_t)} - \omega_{k_t,k_s}(\bx;\hat{\bgg}_{k_t,k_s}) \right|
 =  O_p\left(N^{-1/2}\right). 
\end{align*}
\end{enumerate}
\item \label{assume:S-info} Informative source: 
Let $\bgvth = \left(\bgps(\bX)\trans, \bgu_{1}\trans,\bgu_{0}\trans \right)\trans$ be the combined influence function for broadcast estimators. For all $k_s \in \tgs'$
$$
\left| \Cov\left(\frac{\zeta+\xi_{\tgt}}{\P(R\in\tgt)},
 -\frac{\xi_{\tgt}}{\P(R\in\tgt)} + \sum_{k_t\in \tgt} \frac{\ind(R=k_t)}{\P(R=k_t)}
 \bgvth\trans \bar{\bd}_{k_t,k_s}
 \mid R \in \tgt\right)
\right| \ge \varepsilon. 
$$
\end{enumerate}
\end{assumption}
The two model consistency conditions in Assumption \ref{assume:S-dr} ensure the consistency 
of the doubly robust estimator $\widehat{\Delta}_{\tgt,k_s}$. 
Assumption \ref{assume:S-info} characterizes the informativeness of a source site $k_s$ such that the updated direction $\left(\widehat{\delta}_{\tgt,k_s}-\widehat{\delta}_{\tgt,\tgt}\right)$ is correlated with the initial $\widehat{\Delta}_{\tgt,\tgt}$. The covariance in the condition is likely to be negative with the opposite sign of $\xi_{\tgt}$.

\begin{proposition}\label{prop:no_worse}
Under the conditions of Theorem \ref{thm:FACE}, 
the asymptotic variance of $\FACE$ is no larger than that of $\widehat{\Delta}_{\tgt,\tgt}$. 
Moreover, if Assumption \ref{assume:S} holds, 
the asymptotic variance of $\FACE$ is strictly smaller than that of $\widehat{\Delta}_{\tgt,\tgt}$. 
\end{proposition}
The proof is given in Supplement \ref{assec:proof-RE}. Proposition \ref{prop:no_worse} offers a guarantee on the relative efficiency in general settings. 
As the exact efficiency gain may take different forms under general settings, we showcase the efficiency gain with a clear interpretation under a simple ideal setting in Supplement \ref{exact efficiency}.

\section{Simulation Studies}\label{sec:sim}
We study the finite sample performance of (i) the FACE estimator against four estimators: (ii) an estimator that leverages target data only (target-only), (iii)  a sample-size adjusted estimator (SS), (iv) an inverse-variance weighted estimator (IVW), and (v) an exponentially-tilted augmented inverse probability weighted estimator (Tilted-AIPW) that multiplies density-ratio weighted site-specific AIPW estimators and aggregates via SS. We examine the bias, root mean square error (RMSE), coverage probability of the $95\%$ CIs, and length of the $95\%$ CIs of these estimators across $500$ simulations.

\subsection{Data Generation}
We set $J+K = 10$ sites, with the first site as the target and $K=9$ source sites. We set the sample size $n_k = 200$, $k=1,...,10$.  In Supplement \ref{supp:sim}, we include results for $n_k = 400$, $k=1,...,10$ and vary the number of sites $J+K \in \{10,20,40\}$. To explore the effect of model misspecification, we vary the misspecification of the true OR, PS, and density ratio models. To allow for heterogeneity in the covariate distribution between sites, the covariates in each site $\bX_{kp}$ are generated from a skewed normal distribution, $\bX_{kp} \sim \mathcal{SN}(\bx; \kappa_{kp}, \phi^2_{kp}, \nu_{kp})$, where $k = 1,...,J+K$ indexes the sites and $p=1,...,10$ indexes the ten covariates, $\kappa_{kp}$ is the location parameter, $\phi_{kp}$ is the scale parameter, and $\nu_{kp}$ is the skewness parameter.  For all sites, we let ${\kappa}_{k\cdot} \in (0.10, 0.15)$ and $\phi_{k\cdot} = (1, ..., 1)$. For the target site, we set $\nu_{k\cdot} = 0$. For the source sites, we let $\nu_{k\cdot} \in [0,0.2]$ so that the exponential tilt model provides varying approximation quality for projecting the covariate distributions to the target site. 

The true potential outcomes are generated as 
\begin{align*}
 Y_k(a) & = [(\bX_k - \mu_1)\trans, (\bX_k^{\circ 2})\trans] (\bgb_{1a}\trans , \bgb_{2a}\trans)\trans + \Delta_k I(a = 1)+ \varepsilon_{k}, \quad \varepsilon_k \sim \mathcal{N}(0, 2\sqrt{5}) , \quad a = 0,1,
 \end{align*}
 where $\bX_k^{\circ 2}$ denotes $\bX_k$ squared element-wise, $\bgb_{11} = (0.4,.., 1.2)$, and $\bgb_{10} = (0.4,.., 1.2)$ with equally-spaced increments for a length $10$, and $\Delta_k$ is the ATE. We consider eight different settings where the level of sparsity varies, ranging from all source sites being informative to all being strongly non-informative. Table \ref{sparsity_table} describes in each setting how similar $\Delta_k$, $k \in \mathcal{S}$ are to the target $\Delta_\tgt = 3.0$. 

 \begin{table}[h!]
    \centering
    \scalebox{0.85}{
    \begin{tabular}{|c|l|l|}
    \hline
         {Sparsity} & {Description of Source Sites} & {Source ATEs $\Delta_k$, $k=2,...,10$}\\
         \hline
         {1} & {All source sites informative} & $\Delta_2,...,\Delta_{10} = 3.0$. \\
         \hline
         {2} & {One weakly non-informative source site} & $\Delta_2 = 3.2$, $\Delta_3,...,\Delta_{10} = 3.0$. \\
         \hline
         {3} & {Two weakly non-informative source sites} & $\Delta_2 = \Delta_3 = 3.4$, $\Delta_4,...,\Delta_{10}= 3.0$. \\
         \hline
         {4} & {Three moderately non-informative source sites} & $\Delta_2=\Delta_3=\Delta_4 = 3.6$, $\Delta_5,...,\Delta_{10} = 3.0$. \\
         \hline
         {5} & {Five moderately non-informative source sites} & $\Delta_2,...,\Delta_6 = 3.8$, $\Delta_7,...,\Delta_{10} = 3.0$. \\
         \hline
         {6} & {Seven moderately non-informative source sites} & $\Delta_2,...,\Delta_8 = 4.0$, $\Delta_9 = \Delta_{10} = 3.0$ \\
         \hline
         {7} & {Eight strongly non-informative source sites} & $\Delta_2,...,\Delta_9 = 4.5$, $\Delta_{10} = 3.0$ \\
         \hline
         {8} & {All source sites strongly non-informative} & $\Delta_2,...,\Delta_{10} = 5.0$. \\
         \hline
    \end{tabular}}
    \caption{{Eight levels of sparsity corresponding to the informativeness of source sites where the true TATE is $\Delta_\tau = 3.0$.}}
    \label{sparsity_table}
\end{table}
 
 The true PS model is generated as 
 $$ A_k \mid \bX=\bx \sim \text{Bernoulli}(\pi_k),  \quad \pi_k = \text{expit} (\bX_k\bga_{1k} + \bX_k^{\circ 2}\bga_{2k}),$$ 
 where for the target and source sites, $\bga_{1k} = (0.5, ..., -0.5)$,  with equally-spaced decrements for a length $10$ and $\bga_{2k} = (-0.5,0,...,0)$. For all sites, we fit linear regression models for the OR and logistic regression models for the PS, where we misspecify models by only including the linear terms of the covariates $\bX_k$. 

\subsection{Simulation Settings}
We consider the following settings. In Setting 1, we examine the scenario where the OR and PS models are correctly specified, but the density ratio models are misspecified. In Setting 2, we misspecify the OR while keeping the PS and density ratio models correctly specified. In Setting 3, we misspecify the PS but correctly specify the OR model and density ratio models.  In each setting, we choose the tuning parameter $\lambda$ by the distributed cross-validation procedure described in Section \ref{ssec:method-tuning}, where we split the simulated datasets in each site into two equally sized training and validation datasets and take the optimal $\lambda$ over five folds.

\subsection{Simulation Results}
\begin{figure}[H]
    \centering
    \includegraphics[width=0.8\textwidth]{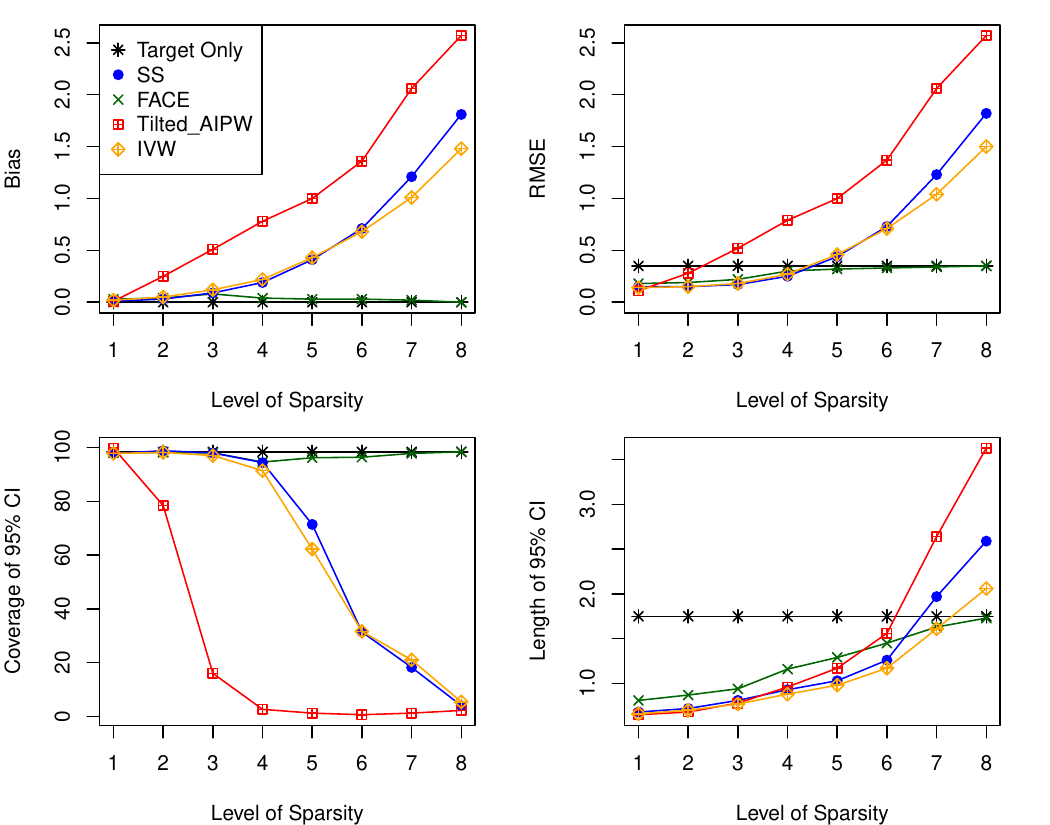}
    \caption{Setting 1. Bias, RMSE, coverage, and length of $95\%$ CI's of the target-only, SS, FACE, exponentially-tilted AIPW, and IVW estimators of the TATE across 500 simulations with misspecified density ratio models.} 
    \label{fig:misspec1}
\end{figure}

In Figure \ref{fig:misspec1}, we summarize the bias, RMSE, coverage, and length of the $95\%$ CIs of FACE and the four alternative estimators in Setting 1 where only the density ratio models are misspecified. Results for Setting 2 and Setting 3 with model misspecification are provided in Supplement \ref{supp:sim}. When all source sites are informative (sparsity level 1), all estimators perform well. When only one (level 2) or two (level 3) source sites are weakly non-informative, FACE, SS, and IVW perform well, with minimal bias, RMSE smaller than the target-only estimator, nominal coverage, and substantially shorter average CIs compared to the target-only. However, as the proportion and bias of non-informative source sites increase, only FACE shows relatively good robustness against negative transfer with minimal bias, RMSE below that of target-only, nominal coverage, and shorter average CIs compared to target-only. On the other hand, we observe large biases and poor coverage for the alternative estimators. Overall, the RMSE of FACE is lower than that of the target-only estimator and approaches the target-only as the proportion and bias of non-informative source sites increase. The coverage of FACE is close to the nominal $95\%$ across different levels of sparsity and the length of the $95\%$ CI of FACE is shorter than that of the target-only estimator when there are informative source sites.

\begin{figure}[H]
    \centering
    \includegraphics[width=0.6\textwidth]{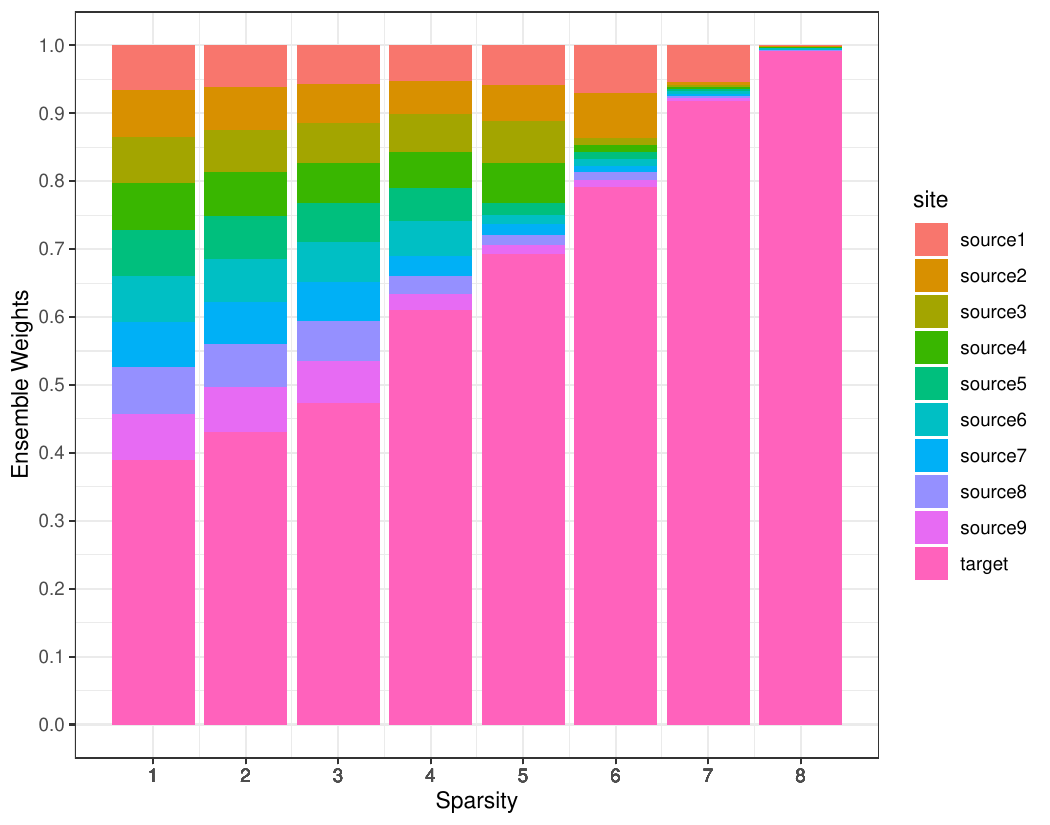}
    \caption{{FACE ensemble weights $\hat{\bge}$ for each site across eight sparsity levels}}
    \label{fig:eta}
\end{figure}

In Figure \ref{fig:eta}, the adaptivity of the FACE weights $\hat{\bge}$ can be seen across the eight sparsity levels. As the proportion of informative source sites decreases (sparsity level increases), the weight given to the target site increases. This adaptivity allows for reduced bias relative to other methods such as IVW or SS weighting, which have fixed weights that contribute to the large bias and low coverage when there are non-informative source sites.

\section{Comparative Effectiveness of COVID-19 Vaccines}\label{sec:data}

To illustrate FACE, we study the comparative effectiveness of BNT162b2 (Pfizer) versus mRNA-1273 (Moderna) for the prevention of COVID-19 outcomes in five VA sites. It is of interest to understand the real-world effectiveness of these vaccines, but head-to-head comparisons have been rare. A recent emulated target trial using the EHRs of US veterans showed that the 24-week risk of COVID-19 outcomes was low for patients who received either vaccine, but lower for veterans assigned to Moderna compared to Pfizer and potentially heterogeneous across patient populations \citep{dickerman2021}, suggesting that only reporting an overall ATE may be misleading for certain target populations. Utilizing FACE, we examine the TATE in a federated data setting where the target population of interest is one of five sites (North Atlantic, Southwest, Midwest, Continental, or Pacific) in the VA healthcare system. Our problem is more challenging than that of \cite{dickerman2021} or \cite{lin2022effectiveness} due to the federated data setting and the different target populations of interest that we are able to study.

Inclusion criteria included veteran status, at least 18 years of age by January 1, 2021, no previously documented COVID-19 infection, no previous COVID-19 vaccination, and documented two-dose COVID-19 vaccination with either Pfizer or Moderna between January 1 and March 24, 2021. For each eligible veteran, follow-up began on the day that the second dose of vaccine was received (baseline) and ended on the day of death, 120 or 180 days after
baseline, or the end of the study time period (September 24, 2021). The outcomes of interest were documented SARS-CoV-2 infection either 120 or 180 days after baseline and death with COVID-19 infection either 120 or 180 days after baseline. 

Among the $608,359$ eligible veterans, $293,137$ $(48.2\%)$ received Pfizer and $315,222$ $(51.8\%)$ received Moderna. Baseline characteristics among the two groups were similar within site. Across sites, there was heterogeneity in race (a larger proportion of Asians in the Pacific), and ethnicity (a larger Hispanic population in the Southwest and Pacific). Baseline characteristics in each of the five sites are summarized in Supplementary Tables \ref{supp:table1} and \ref{supp:table2}. All models were adjusted for age, sex, race, ethnicity, residence, and important comorbidities: chronic lung disease (including asthma, bronchitis, and chronic obstructive pulmonary disease), cardiovascular disease (including acute myocardial infarction, cardiomyopathy, coronary heart disease, heart failure, and peripheral vascular disease), hypertension, type 2 diabetes, chronic kidney disease, autoimmune diseases (including HIV infection, rheumatoid arthritis, etc.), and obesity (defined as body mass index of 30 or greater).

The raw event rates for documented COVID-19 infection within 180 days of receiving the second dose for Pfizer (Moderna) in the five sites were 2.81\% (1.93\%) in the North Atlantic, 3.58\% (3.23\%) in the Southwest, 2.25\% (2.08\%) in the Midwest, 2.97\% (2.36\%) in the Continental, and 2.80\% (1.43\%) in the Pacific. The raw event rates for death with COVID-19 infection within 180 days of receiving the second dose for Pfizer (Moderna) were 0.37\% (0.06\%) in the North Atlantic, 0.36\% (0.23\%) in the Southwest, 0.18\% (0.21\%) in the Midwest, 0.21\% (0.26\%) in the Continental, and 0.11\% (0.09\%) in the Pacific.

\begin{figure}[H]
        \centering
        \begin{subfigure}[b]{0.49\textwidth}
            \centering
            \includegraphics[width=\textwidth]{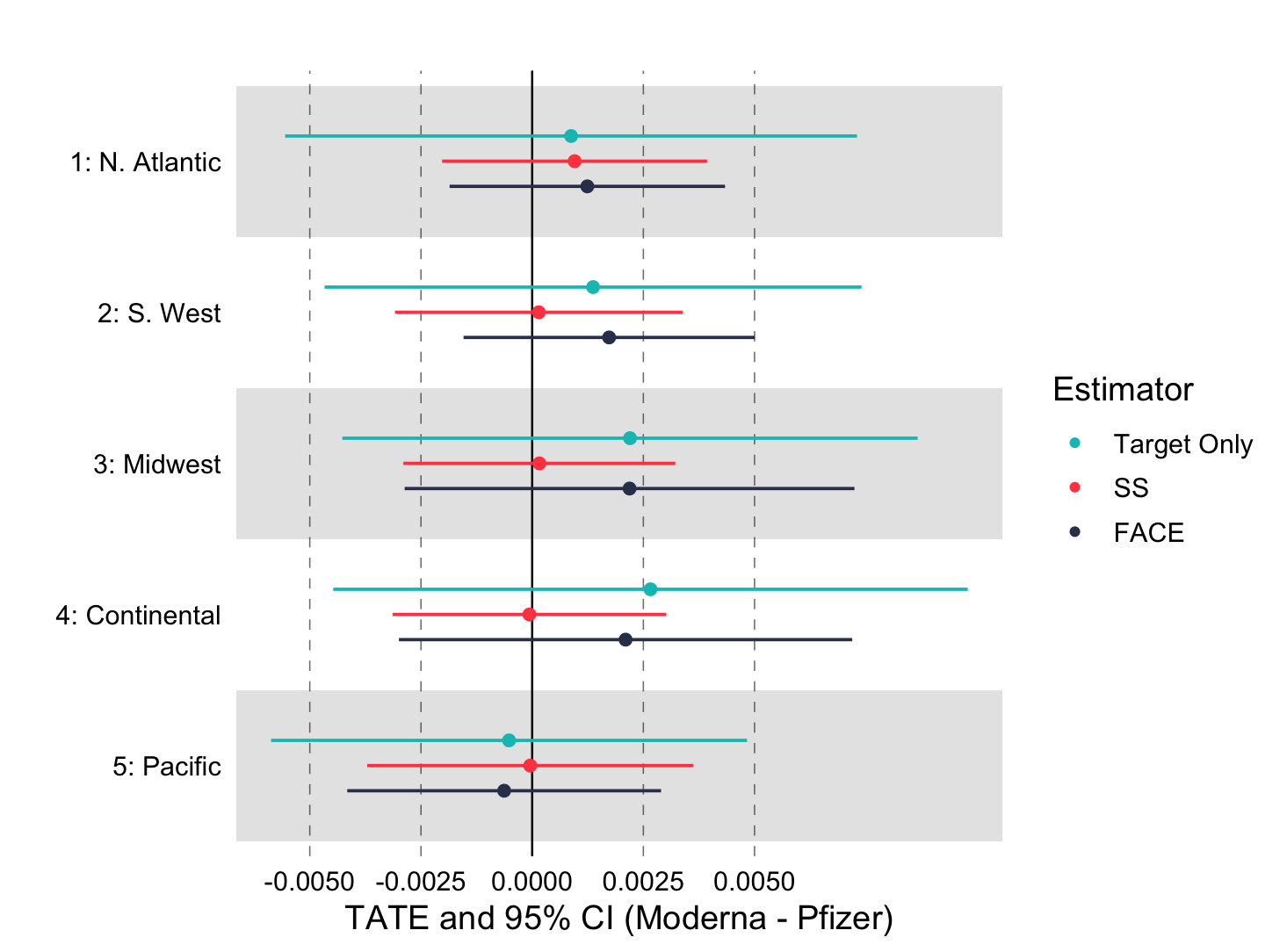}
            \caption[]%
           {{\small TATE for COVID-19 infection (120 days)}}    
           % \label{fig:a}
        \end{subfigure}
        \hfill
        \begin{subfigure}[b]{0.49\textwidth}  
            \centering 
            \includegraphics[width=\textwidth]{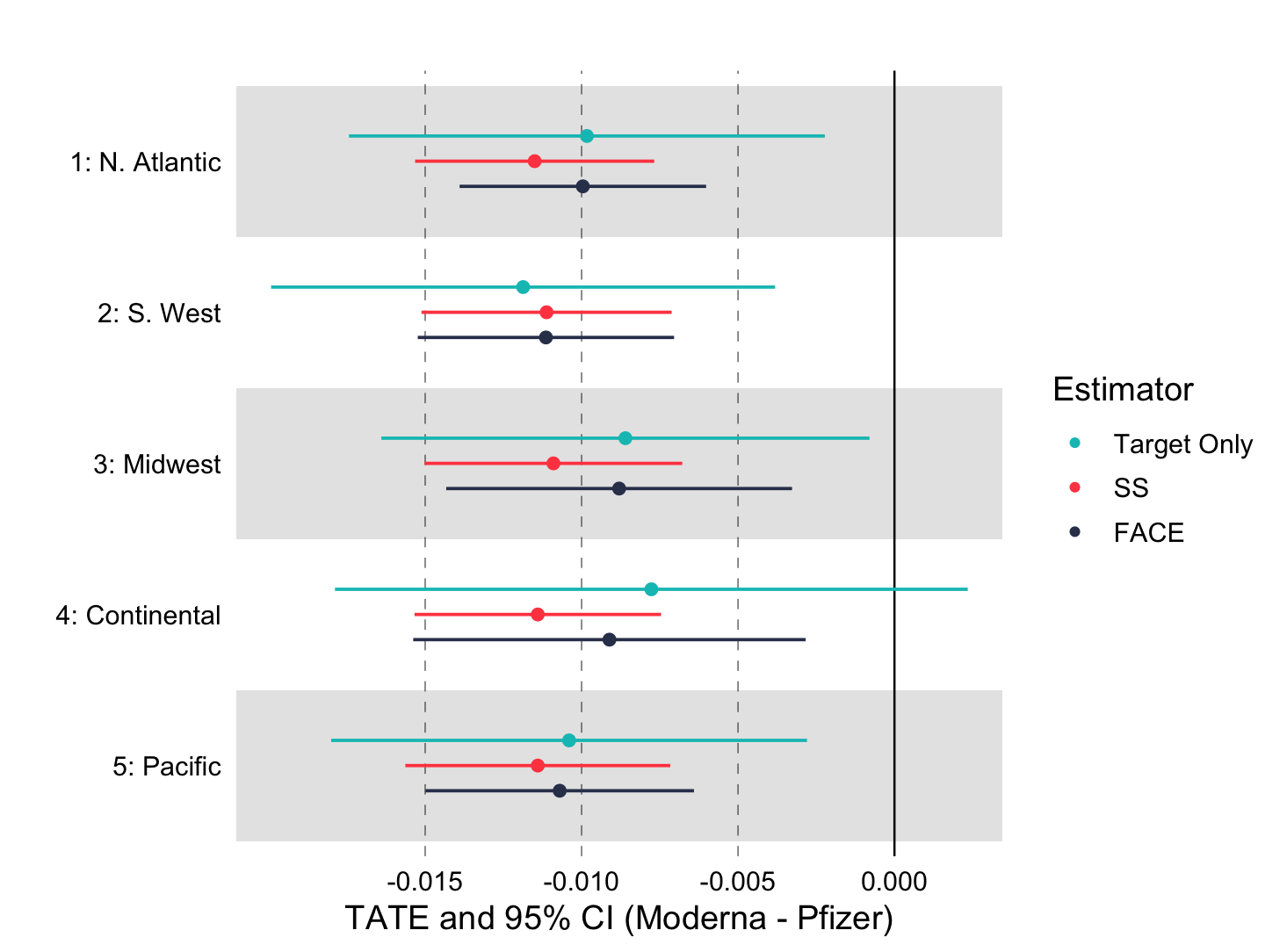}
            \caption[]%
           {{\small TATE for COVID-19 infection (180 days)}}    
           % \label{fig:b}
        \end{subfigure}
        \vskip\baselineskip
        \begin{subfigure}[b]{0.49\textwidth}   
            \centering 
            \includegraphics[width=\textwidth]{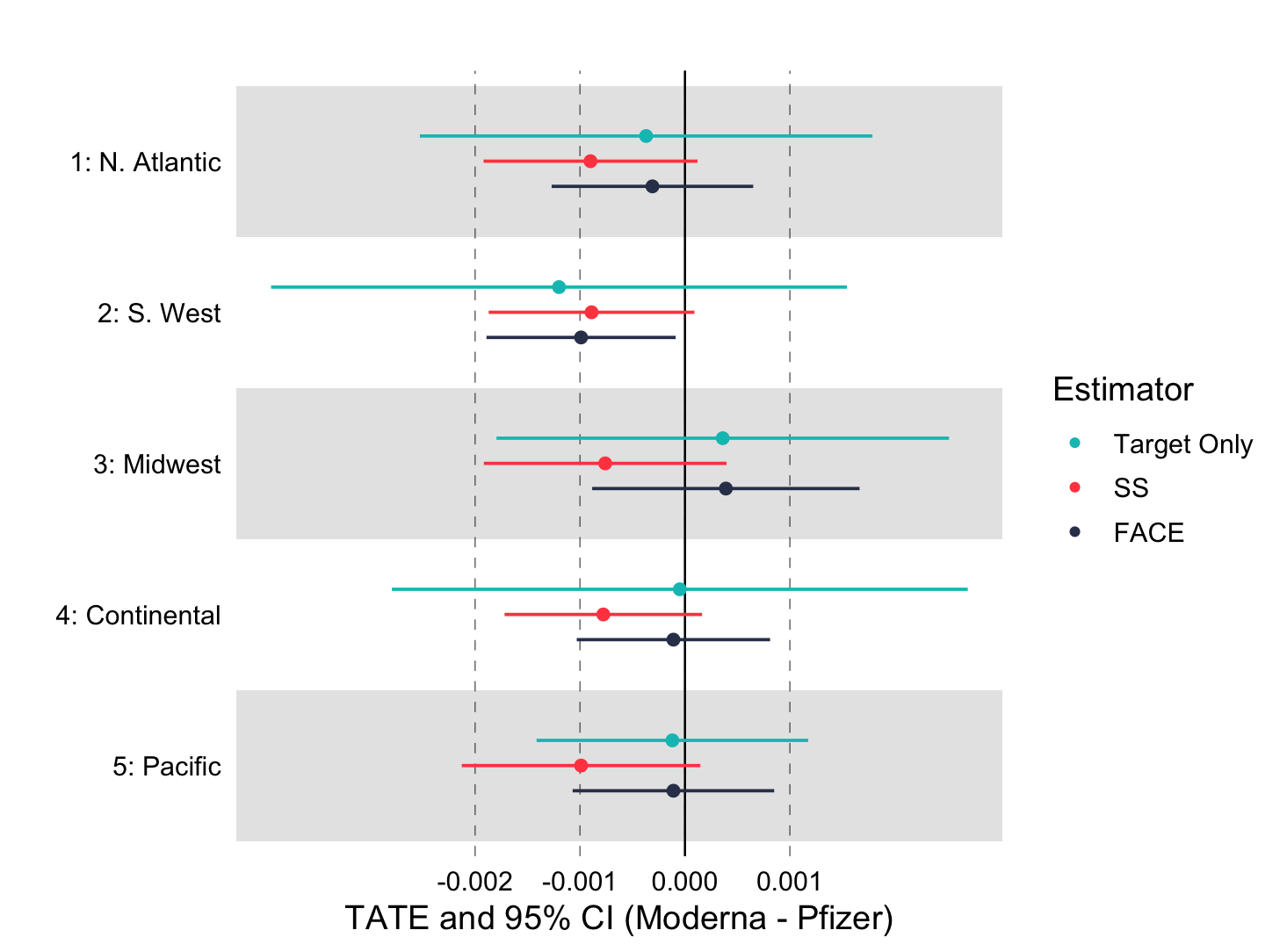}
            \caption[]%
            {{\small TATE for COVID-19 death (120 days)}}    
           % \label{fig:c}
        \end{subfigure}
        \hfill
        \begin{subfigure}[b]{0.49\textwidth}   
            \centering 
            \includegraphics[width=\textwidth]{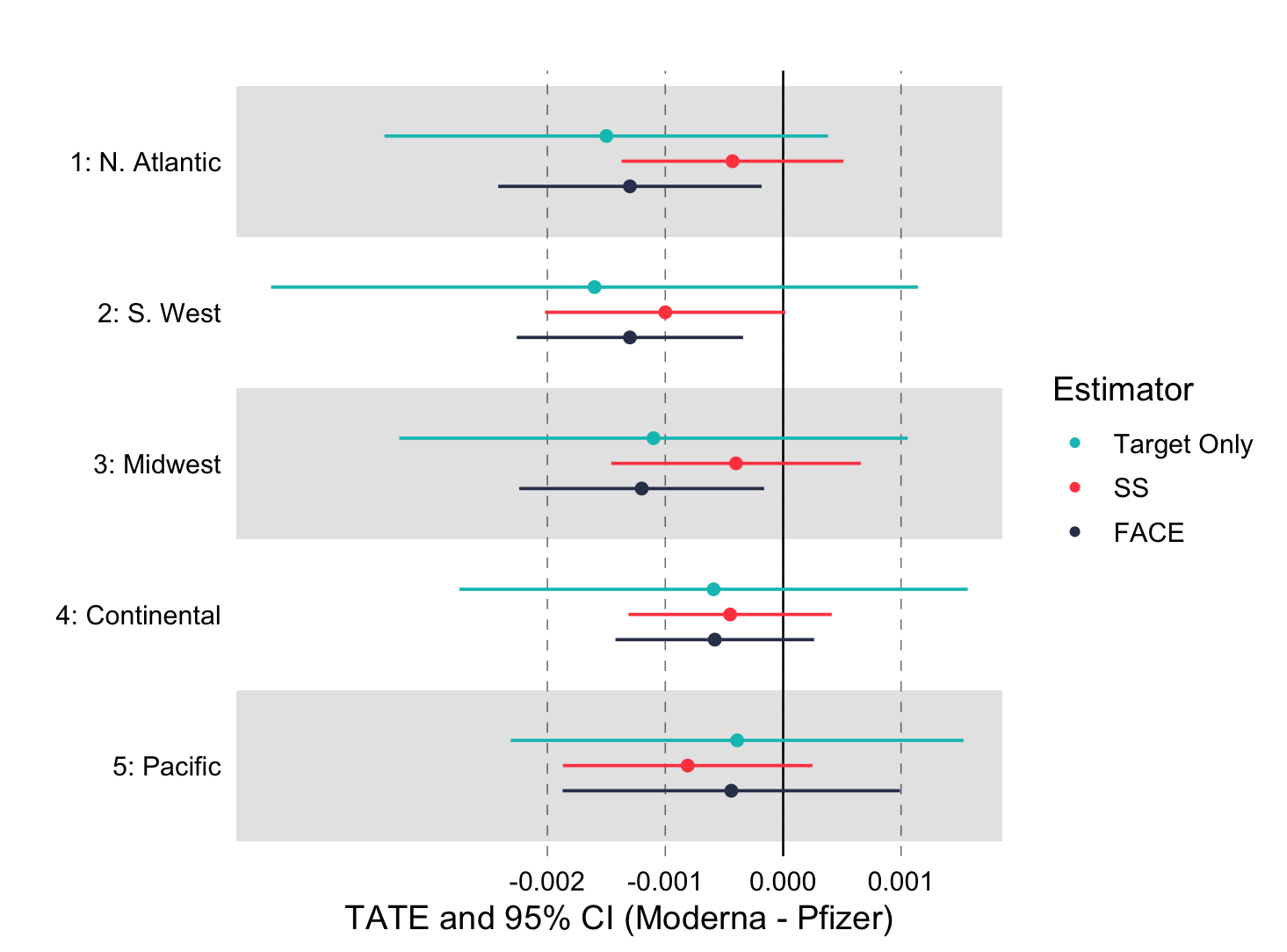}
             \caption[]%
           {{\small TATE for COVID-19 death (180 days)}}    
           % \label{fig:d}
        \end{subfigure}
        \caption[ ]
        {\small TATE estimates for the comparative effectiveness of Moderna vs. Pfizer vaccines for four outcomes} 
        \label{fig:allresults}

\end{figure}

Figure \ref{fig:allresults} shows the TATE estimates for the four outcomes of interest: (a) 120-day COVID-19 infection, (b) 180-day COVID-19 infection, (c) 120-day death with COVID-19 infection, and (d) 180-day death with COVID-19 infection. For each outcome, the target population is taken to be one of the five sites. Three estimators are compared along with their $95\%$ CIs: (i) target-only, (ii) a sample-size weighted estimator that leverages each site where $\eta_{k}$ is taken to be $n_k / N$ (SS), $k=1,...,5$, and (iii) the FACE estimator. Our results indicate that the FACE estimator tracks the target-only estimator more closely compared to the SS estimator. Compared to the target-only estimator, the FACE estimator has substantially tighter confidence intervals, resulting in qualitatively different conclusions in certain cases, e.g., 180-day COVID-19 infection in the Continental site, 120-day death with COVID-19 infection in the Southwest site, and 180-day death with COVID-19 infection in the Midwest, North Atlantic, and Southwest sites. Using FACE, our results show that veterans who received Moderna had an approximately $1\%$ lower rate of 180-day COVID-19 infection compared to Pfizer, and this difference appeared consistent across sites.

\section{Discussion}\label{sec:discuss}
In this paper, we have developed FACE to leverage heterogeneous data from multiple study sites to more precisely estimate treatment effects for a target population of interest. FACE accounts for heterogeneity in the distribution of covariates through a density ratio weighting approach and protects against distributional heterogeneity and model misspecification of the source sites through an adaptive integration strategy. It improves upon the precision of the target-only estimator by leveraging source population information without inducing bias. FACE is privacy-preserving and communication-efficient, requiring only one round of communication of aggregated summary statistics between sites. If individual-level data could be pooled together, our FACE estimation strategy could still be used, with no efficiency gain when both the outcome regression model and propensity score and density ratio models are correctly specified, but possible efficiency gain if the propensity score and density ratio models are misspecified since the estimation of the outcome regression model could be different. In addition to providing theoretical double robustness and efficiency guarantees, FACE does not rely on prior knowledge of model stability or correct model specification, which is a substantial improvement on current federated methods for causal inference \citep{xiong2021federated}. We also obtained promising results from a real-world analysis of COVID-19 outcomes for veterans assigned to either Pfizer or Moderna vaccines among five federated VA sites.

FACE can easily be generalized to the setting where some sites have RCT data. In such a setting, one could define the target population as the set of trial participants. When the RCT data is treated as the anchoring site, the target site PS model is known, so the target site estimator for the TATE is consistent, and the global adaptive estimator is likely to be more reliable. Our FACE framework can thus be viewed as a contribution to recent work on using observational studies to reduce the variance associated with treatment effect estimates from experimental studies \citep{athey2020combining}. For greater generalizability, participants for whom there is only observational data can be taken to be the target population. FACE can also be adapted to target different causal parameters of interest, such as the average treatment effect of the treated (ATT).

Our proposed FACE estimator is essentially a linear combination of efficient estimators from the target site and compatible source sites, and the aggregation step recovers the optimal linear combination satisfying $\eta_k = 0$ if $k\in\mathcal{S} \setminus \mathcal{S^*}$ and  $\bar{\bge}$ minimizes the variance of the final FACE estimator. Having proved the asymptotic convergence of $\hat{\bge}$ to $\bar{\bge}$ in Lemmas \ref{lem:aggre}, \ref{lem:eta}, and \ref{lem:FACE}, we expect the FACE estimator to attain the efficiency lower bound under the ideal setting where all source sites are compatible. A recent line of work characterizes the semi-parametric efficiency bound under a similar setting when datasets are aligned (i.e., share the same conditional outcome distribution), which confirms that FACE is efficient under the ideal setting \citep{li2023efficient}. In the more challenging setting where some sites are incompatible, there have been efforts to develop the optimal estimation rate for high-dimensional regression, but the efficiency bound is not applicable \citep{li2022transfer} in our setting. \citep{li2023data} discussed an estimator when fusing weakly aligned datasets; however, their estimator is not efficient. In addition, the existing literature does not address the key issues of data communication workflows under privacy constraints. Formal characterization of the efficiency lower bound in the complex setting of our study may be a compelling direction for future research.

Future work may also consider developing methods for estimands defined by subpopulations of interest. For example, the conditional average treatment effect (CATE) is an important estimand of real-world interest, particularly for understanding the benefits and dangers of treatments for underrepresented groups and fairness research. It may also be of interest to extend FACE to model-based treatment effect metrics defined by, for example, marginal structural models or structural nested models, although further elaboration on the causal interpretations would be needed. The major advantage of using the risk difference (e.g., ATE) as a metric is its model-free property. In contrast, relative risk metrics cannot characterize the population shift in a justifiable manner. Either the causal parameter is defined conditionally on $\bX$, thus totally independent of population shift \citep{vansteelandt2014structural, hou2023treatment}, or the causal parameter is defined marginally based on a single population, but such a model would no longer hold in another shifted population \citep{hernan2001marginal}.

An interesting extension is when no outcome or treatment information is observed for the target site. This may be the case when it is expensive, time-consuming, or otherwise challenging to collect information on target samples \citep{leek2010tackling, ling2022batch}. In such scenarios, it may be desirable to utilize data from source sites that include covariates, treatment, and outcome information. When there is substantial heterogeneity and outliers may exist, a strategy of integrative analysis is to identify a prevailing model, defined as the model satisfied by the majority of the sites. Identifying the prevailing model can be achieved via, for example, the majority rule  \citep{hastie2005robust}. \cite{guo2023robust} developed a theoretically justified robust inference for federated meta-learning (RIFL) framework to construct uniformly valid confidence intervals for the unknown prevailing model using multi-source data. Future work may seek to extend RIFL to more flexible target distributions.

\bibliographystyle{agsm}

\bibliography{ref}

@article{hernan2001marginal,
  title={Marginal structural models to estimate the joint causal effect of nonrandomized treatments},
  author={Hern{\'a}n, Miguel A and Brumback, Babette and Robins, James M},
  journal={Journal of the American Statistical Association},
  volume={96},
  number={454},
  pages={440--448},
  year={2001},
  publisher={Taylor \& Francis}
}

@article{vansteelandt2014structural,
  title={Structural Nested Models and G-estimation: The Partially Realized Promise},
  author={Vansteelandt, Stijn and Joffe, Marshall},
  journal={Statistical Science},
  pages={707--731},
  year={2014},
  publisher={JSTOR}
}

@article{hou2023treatment,
  title={Treatment effect estimation under additive hazards models with high-dimensional confounding},
  author={Hou, Jue and Bradic, Jelena and Xu, Ronghui},
  journal={Journal of the American Statistical Association},
  volume={118},
  number={541},
  pages={327--342},
  year={2023},
  publisher={Taylor \& Francis}
}

@article{lv2009unified,
  title={A unified approach to model selection and sparse recovery using regularized least squares},
  author={Lv, Jinchi and Fan, Yingying},
  journal={The Annals of Statistics},
  volume={37},
  number={6A},
  pages={3498},
  year={2009},
  publisher={Institute of Mathematical Statistics}
}

@article{leek2010tackling,
  title={Tackling the widespread and critical impact of batch effects in high-throughput data},
  author={Leek, Jeffrey T and Scharpf, Robert B and Bravo, H{\'e}ctor Corrada and Simcha, David and Langmead, Benjamin and Johnson, W Evan and Geman, Donald and Baggerly, Keith and Irizarry, Rafael A},
  journal={Nature Reviews Genetics},
  volume={11},
  number={10},
  pages={733--739},
  year={2010},
  publisher={Nature Publishing Group}
}

@article{ling2022batch,
  title={Batch effects removal for microbiome data via conditional quantile regression},
  author={Ling, Wodan and Lu, Jiuyao and Zhao, Ni and Lulla, Anju and Plantinga, Anna M and Fu, Weijia and Zhang, Angela and Liu, Hongjiao and Song, Hoseung and Li, Zhigang and others},
  journal={Nature Communications},
  volume={13},
  number={1},
  pages={1--14},
  year={2022},
  publisher={Nature Publishing Group}
}

@article{li2023data,
  title={Data fusion using weakly aligned sources},
  author={Li, Sijia and Gilbert, Peter B and Luedtke, Alex},
  journal={arXiv preprint arXiv:2308.14836},
  year={2023}
}

@article{li2023efficient,
  title={Efficient estimation under data fusion},
  author={Li, Sijia and Luedtke, Alex},
  journal={Biometrika},
  pages={asad007},
  year={2023},
  publisher={Oxford University Press}
}

@article{dehejia2021local,
  title={From local to global: External validity in a fertility natural experiment},
  author={Dehejia, Rajeev and Pop-Eleches, Cristian and Samii, Cyrus},
  journal={Journal of Business \& Economic Statistics},
  volume={39},
  number={1},
  pages={217--243},
  year={2021},
  publisher={Taylor \& Francis}
}

@article{li2022transfer,
  title={Transfer learning for high-dimensional linear regression: Prediction, estimation and minimax optimality},
  author={Li, Sai and Cai, T Tony and Li, Hongzhe},
  journal={Journal of the Royal Statistical Society Series B: Statistical Methodology},
  volume={84},
  number={1},
  pages={149--173},
  year={2022},
  publisher={Oxford University Press}
}

@article{fan2001variable,
  title={Variable selection via nonconcave penalized likelihood and its oracle properties},
  author={Fan, Jianqing and Li, Runze},
  journal={Journal of the American Statistical Association},
  volume={96},
  number={456},
  pages={1348--1360},
  year={2001},
  publisher={Taylor \& Francis}
}

@article{zou2006adaptive,
  title={The adaptive lasso and its oracle properties},
  author={Zou, Hui},
  journal={Journal of the American Statistical Association},
  volume={101},
  number={476},
  pages={1418--1429},
  year={2006},
  publisher={Taylor \& Francis}
}

@article{nie2021covariate,
  title={Covariate balancing sensitivity analysis for extrapolating randomized trials across locations},
  author={Nie, Xinkun and Imbens, Guido and Wager, Stefan},
  journal={arXiv preprint arXiv:2112.04723},
  year={2021}
}

@article{andrews2017weighting,
  title={Weighting for external validity},
  author={Andrews, Isaiah and Oster, Emily},
  year={2017},
  journal={National Bureau of Economic Research}
}

@article{nguyen2017sensitivity,
  title={Sensitivity analysis for an unobserved moderator in RCT-to-target-population generalization of treatment effects},
  author={Nguyen, Trang Quynh and Ebnesajjad, Cyrus and Cole, Stephen R and Stuart, Elizabeth A},
  journal={The Annals of Applied Statistics},
  pages={225--247},
  year={2017},
  publisher={JSTOR}
}

@article{hirshberg2019minimax,
  title={Minimax linear estimation of the retargeted mean},
  author={Hirshberg, David A and Maleki, Arian and Zubizarreta, Jose R},
  journal={arXiv preprint arXiv:1901.10296},
  year={2019}
}

@article{hirshberg2021augmented,
  title={Augmented minimax linear estimation},
  author={Hirshberg, David A and Wager, Stefan},
  journal={The Annals of Statistics},
  volume={49},
  number={6},
  pages={3206--3227},
  year={2021},
  publisher={Institute of Mathematical Statistics}
}

@article{cheng2021adaptive,
  title={Adaptive Combination of Randomized and Observational Data},
  author={Cheng, David and Cai, Tianxi},
  journal={arXiv preprint arXiv:2111.15012},
  year={2021}
}

@article{lin2022effectiveness,
  title={Effectiveness of Covid-19 Vaccines over a 9-Month Period in North Carolina},
  author={Lin, Dan-Yu and Gu, Yu and Wheeler, Bradford and Young, Hayley and Holloway, Shannon and Sunny, Shadia-Khan and Moore, Zack and Zeng, Donglin},
  journal={New England Journal of Medicine},
  year={2022},
  publisher={Mass Medical Soc}
}

@article{dickerman2021,
author = {Dickerman, Barbra A. and Gerlovin, Hanna and Madenci, Arin L. and Kurgansky, Katherine E. and Ferolito, Brian R. and Figueroa Muñiz, Michael J. and Gagnon, David R. and Gaziano, J. Michael and Cho, Kelly and Casas, Juan P. and Hernán, Miguel A.},
title = {Comparative Effectiveness of BNT162b2 and mRNA-1273 Vaccines in U.S. Veterans},
journal = {New England Journal of Medicine},
year = {2021},
doi = {10.1056/NEJMoa2115463}
}

@book{van2000asymptotic,
  title={Asymptotic Statistics},
  author={Van der Vaart, Aad W},
  volume={3},
  year={2000},
  publisher={Cambridge University Press}
}

@article{bang2005doubly,
  title={Doubly robust estimation in missing data and causal inference models},
  author={Bang, Heejung and Robins, James M},
  journal={Biometrics},
  volume={61},
  number={4},
  pages={962--973},
  year={2005},
  publisher={Wiley Online Library}
}

@misc{hernan2020causal,
  title={Causal inference: What if?},
  author={Hern{\'a}n, Miguel A and Robins, James M},
  year={2020},
  publisher={CRC Boca Raton, FL}
}

@article{athey2020combining,
  title={Combining experimental and observational data to estimate treatment effects on long term outcomes},
  author={Athey, Susan and Chetty, Raj and Imbens, Guido},
  journal={arXiv preprint arXiv:2006.09676},
  year={2020}
}

@article{vo2021federated,
  title={Federated Estimation of Causal Effects from Observational Data},
  author={Vo, Thanh Vinh and Hoang, Trong Nghia and Lee, Young and Leong, Tze-Yun},
  journal={arXiv preprint arXiv:2106.00456},
  year={2021}
}

@article{xiong2021federated,
  title={Federated Causal Inference in Heterogeneous Observational Data},
  author={Xiong, Ruoxuan and Koenecke, Allison and Powell, Michael and Shen, Zhu and Vogelstein, Joshua T and Athey, Susan},
  journal={arXiv preprint arXiv:2107.11732},
  year={2021}
}

@article{pan2009survey,
  title={A survey on transfer learning},
  author={Pan, Sinno Jialin and Yang, Qiang},
  journal={IEEE Transactions on Knowledge and Data Engineering},
  volume={22},
  number={10},
  pages={1345--1359},
  year={2009},
  publisher={IEEE}
}

@article{rubin1974estimating,
  title={Estimating causal effects of treatments in randomized and nonrandomized studies.},
  author={Rubin, Donald B},
  journal={Journal of Educational Psychology},
  volume={66},
  number={5},
  pages={688},
  year={1974},
  publisher={American Psychological Association}
}

@article{degtiar2023review,
  title={A review of generalizability and transportability},
  author={Degtiar, Irina and Rose, Sherri},
  journal={Annual Review of Statistics and Its Application},
  volume={10},
  pages={501--524},
  year={2023},
  publisher={Annual Reviews}
}

@article{stuart2011use,
  title={The use of propensity scores to assess the generalizability of results from randomized trials},
  author={Stuart, Elizabeth A and Cole, Stephen R and Bradshaw, Catherine P and Leaf, Philip J},
  journal={Journal of the Royal Statistical Society: Series A (Statistics in Society)},
  volume={174},
  number={2},
  pages={369--386},
  year={2011},
  publisher={Wiley Online Library}
}

@article{stuart2015assessing,
  title={Assessing the generalizability of randomized trial results to target populations},
  author={Stuart, Elizabeth A and Bradshaw, Catherine P and Leaf, Philip J},
  journal={Prevention Science},
  volume={16},
  number={3},
  pages={475--485},
  year={2015},
  publisher={Springer}
}

@article{stuart2018generalizability,
  title={Generalizability of randomized trial results to target populations: design and analysis possibilities},
  author={Stuart, Elizabeth A and Ackerman, Benjamin and Westreich, Daniel},
  journal={Research on Social Work Practice},
  volume={28},
  number={5},
  pages={532--537},
  year={2018},
  publisher={SAGE Publications Sage CA: Los Angeles, CA}
}

@article{hripcsak2016characterizing,
  title={Characterizing treatment pathways at scale using the OHDSI network},
  author={Hripcsak, George and Ryan, Patrick B and Duke, Jon D and Shah, Nigam H and Park, Rae Woong and Huser, Vojtech and Suchard, Marc A and Schuemie, Martijn J and DeFalco, Frank J and Perotte, Adler and others},
  journal={Proceedings of the National Academy of Sciences},
  volume={113},
  number={27},
  pages={7329--7336},
  year={2016},
  publisher={National Acad Sciences}
}

@article{neyman1923application,
  author = {Neyman, J.},
  journal = {Statistical Science},
  number = 5,
  pages = {463--480},
  title = {On the application of probability theory to agricultural experiments},
  volume = 5,
  year = {1923}
}

@article{dahabreh2020extending,
  title={Extending inferences from a randomized trial to a new target population},
  author={Dahabreh, Issa J and Robertson, Sarah E and Steingrimsson, Jon A and Stuart, Elizabeth A and Hernan, Miguel A},
  journal={Statistics in Medicine},
  volume={39},
  number={14},
  pages={1999--2014},
  year={2020},
  publisher={Wiley Online Library}
}

@article{josey2022calibration,
  title={A calibration approach to transportability and data-fusion with observational data},
  author={Josey, Kevin P and Yang, Fan and Ghosh, Debashis and Raghavan, Sridharan},
  journal={Statistics in Medicine},
  volume={41},
  number={23},
  pages={4511--4531},
  year={2022},
  publisher={Wiley Online Library}
}

@article{lee2023improving,
  title={Improving trial generalizability using observational studies},
  author={Lee, Dasom and Yang, Shu and Dong, Lin and Wang, Xiaofei and Zeng, Donglin and Cai, Jianwen},
  journal={Biometrics},
  volume={79},
  number={2},
  pages={1213--1225},
  year={2023},
  publisher={Wiley Online Library}
}

@article{qin1998inferences,
  title={Inferences for case-control and semiparametric two-sample density ratio models},
  author={Qin, Jing},
  journal={Biometrika},
  volume={85},
  number={3},
  pages={619--630},
  year={1998},
  publisher={Oxford University Press}
}

@article{huang2019distributed,
  title={A distributed one-step estimator},
  author={Huang, Cheng and Huo, Xiaoming},
  journal={Mathematical Programming},
  volume={174},
  number={1},
  pages={41--76},
  year={2019},
  publisher={Springer}
}

@article{duan20201fast,
     title = {A fast score test for generalized mixture models},
     author = {Duan, Rui and Ning, Yang and Wang, Shuang and Lindsay, Bruce and Carroll, Raymond and Chen, Yong},
     journal = {Biometrics},
     volume = {76},
     pages = {811-820},
     year = {2020b}
    }

@article{qin11,
     title = {Hypothesis testing in a mixture case-control model},
     author = {Qin, J. and Liang, K-Y.},
     journal = {Biometrics},
     volume = {67},
     pages = {182-193},
     year = {2011}
    }

@book{imbens2015causal,
  title={Causal inference in statistics, social, and biomedical sciences},
  author={Imbens, Guido W and Rubin, Donald B},
  year={2015},
  publisher={Cambridge University Press}
}

@article{brat2020international,
  title={International electronic health record-derived COVID-19 clinical course profiles: the 4CE Consortium},
  author={Brat, Gabriel A and Weber, Griffin M and Gehlenborg, Nils and Avillach, Paul and Palmer, Nathan P and Chiovato, Luca and Cimino, James and Waitman, Lemuel R and Omenn, Gilbert S and Malovini, Alberto and others},
  journal={medRxiv},
  year={2020},
  publisher={Cold Spring Harbor Laboratory Press}
}

@article{tan2020model,
  title={Model-assisted inference for treatment effects using regularized calibrated estimation with high-dimensional data},
  author={Tan, Zhiqiang and others},
  journal={Annals of Statistics},
  volume={48},
  number={2},
  pages={811--837},
  year={2020},
  publisher={Institute of Mathematical Statistics}
}

@article{lee2017communication,
  title={Communication-efficient sparse regression},
  author={Lee, Jason D and Liu, Qiang and Sun, Yuekai and Taylor, Jonathan E},
  journal={The Journal of Machine Learning Research},
  volume={18},
  number={1},
  pages={115--144},
  year={2017},
  publisher={JMLR. org}
}

@article{duan2020learninga,
    author = {Duan, Rui and Boland, Mary Regina and Liu, Zixuan and Liu, Yue and Chang, Howard H and Xu, Hua and Chu, Haitao and Schmid, Christopher H and Forrest, Christopher B and Holmes, John H and Schuemie, Martijn J and Berlin, Jesse A and Moore, Jason H and Chen, Yong},
    title = "{Learning from electronic health records across multiple sites: A communication-efficient and privacy-preserving distributed algorithm}",
    journal = {Journal of the American Medical Informatics Association},
    volume = {27},
    number = {3},
    pages = {376-385},
    year = {2019},
    month = {12},
    issn = {1527-974X},
    doi = {10.1093/jamia/ocz199}
}

@article{chen2006regression,
  title={Regression cubes with lossless compression and aggregation},
  author={Chen, Yixin and Dong, Guozhu and Han, Jiawei and Pei, Jian and Wah, Benjamin W and Wang, Jianyong},
  journal={IEEE Transactions on Knowledge and Data Engineering},
  volume={18},
  number={12},
  pages={1585--1599},
  year={2006},
  publisher={IEEE}
}

@article{robins1994estimation,
  title={Estimation of regression coefficients when some regressors are not always observed},
  author={Robins, James M and Rotnitzky, Andrea and Zhao, Lue Ping},
  journal={Journal of the American statistical Association},
  volume={89},
  number={427},
  pages={846--866},
  year={1994},
  publisher={Taylor \& Francis}
}

@article{weiss2016survey,
	title={A survey of transfer learning},
	author={Weiss, Karl and Khoshgoftaar, Taghi M and Wang, DingDing},
	journal={Journal of Big Data},
	volume={3},
	number={1},
	pages={1--40},
	year={2016},
	publisher={SpringerOpen}
}

@article{li2013statistical,
	title={Statistical inference in massive data sets},
	author={Li, Runze and Lin, Dennis KJ and Li, Bing},
	journal={Applied Stochastic Models in Business and Industry},
	volume={29},
	number={5},
	pages={399--409},
	year={2013},
	publisher={Wiley Online Library}
}

@article{duan2020odal,
	title={{ODAL}: A one-shot distributed algorithm to perform logistic regressions on electronic health records data from multiple clinical sites.},
	author={Duan, Rui and Boland, Mary Regina and Moore, Jason H and Chen, Yong},
	journal={Pacific Symposium on Biocomputing},
	pages={30--41},
	year={2020a},
	publisher={World Scientific}
}

@article{wang2019distributed,
	title={Distributed Inference for Linear Support Vector Machine},
	author={Wang, Xiaozhou and Yang, Zhuoyi and Chen, Xi and Liu, Weidong},
	journal={Journal of Machine Learning Research},
	volume={20},
	number={113},
	pages={1--41},
	year={2019}
}

@article{lian2017divide,
	title={Divide-and-conquer for debiased l 1-norm support vector machine in ultra-high dimensions},
	author={Lian, Heng and Fan, Zengyan},
	journal={The Journal of Machine Learning Research},
	volume={18},
	number={1},
	pages={6691--6716},
	year={2017},
	publisher={JMLR. org}
}

@article{chen2014split,
	title={A split-and-conquer approach for analysis of extraordinarily large data},
	author={Chen, Xueying and Xie, Min-ge},
	journal={Statistica Sinica},
	pages={1655--1684},
	year={2014},
	publisher={JSTOR}
}

@article{guo2023robust,
  title={Robust Inference for Federated Meta-Learning},
  author={Guo, Zijian and Li, Xiudi and Han, Larry and Cai, Tianxi},
  journal={arXiv preprint arXiv:2301.00718},
  year={2023}
}

@article{hastie2005robust,
  title={The robust beauty of majority rules in group decisions.},
  author={Hastie, Reid and Kameda, Tatsuya},
  journal={Psychological Review},
  volume={112},
  number={2},
  pages={494},
  year={2005},
  publisher={American Psychological Association}
}

@article{ChernozhukovEtal18DML,
author = {Chernozhukov, Victor and Chetverikov, Denis and
                Demirer, Mert and Duflo, Esther and Hansen, Christian
                and Newey, Whitney and Robins, James},
title = {Double/Debiased Machine Learning for Treatment and Structural Parameters},
journal = {The Econometrics Journal},
year = {2018},
volume = {21},
number = {1},
pages = {C1-C68},
doi = {0.1111/ectj.12097}}

@article{SCADoracle01,
author = {Jianqing Fan and Runze Li},
title = {Variable Selection via Nonconcave Penalized Likelihood and its Oracle Properties},
journal = {Journal of the American Statistical Association},
volume = {96},
number = {456},
pages = {1348-1360},
year  = {2001},
publisher = {Taylor & Francis},
doi = {10.1198/016214501753382273},

URL = {
        https://doi.org/10.1198/016214501753382273

},
eprint = {
        https://doi.org/10.1198/016214501753382273

}

}

\clearpage
\begin{center}
{\large\bf SUPPLEMENTARY MATERIAL}
\end{center}
The Supplementary Materials are divided into four sections. In Section \ref{supp:flowchart}, we illustrate the workflow of FACE to construct a global estimator in a federated data setting. In Section \ref{ssec:theory-real}, we provide a mild set of sufficient conditions for the necessary regularity conditions to hold in the special case with logistic regression models for the nuisance functions and illustrate FACE under logistic regression models. In Section \ref{supp:proofs}, we provide proofs for the theoretical results in Section 4 of the main paper. We also showcase the efficiency gain of FACE relative to the initial TATE estimator with an exact calculation under a simple ideal setting. In Section \ref{supp:sim}, we provide additional simulation results. In Section \ref{supp:realdata}, we provide supplementary results corresponding to the real data analysis.

\section{FACE Workflow}
\label{supp:flowchart}
\begin{figure}[H]
    \centering
    \includegraphics[scale=0.6]{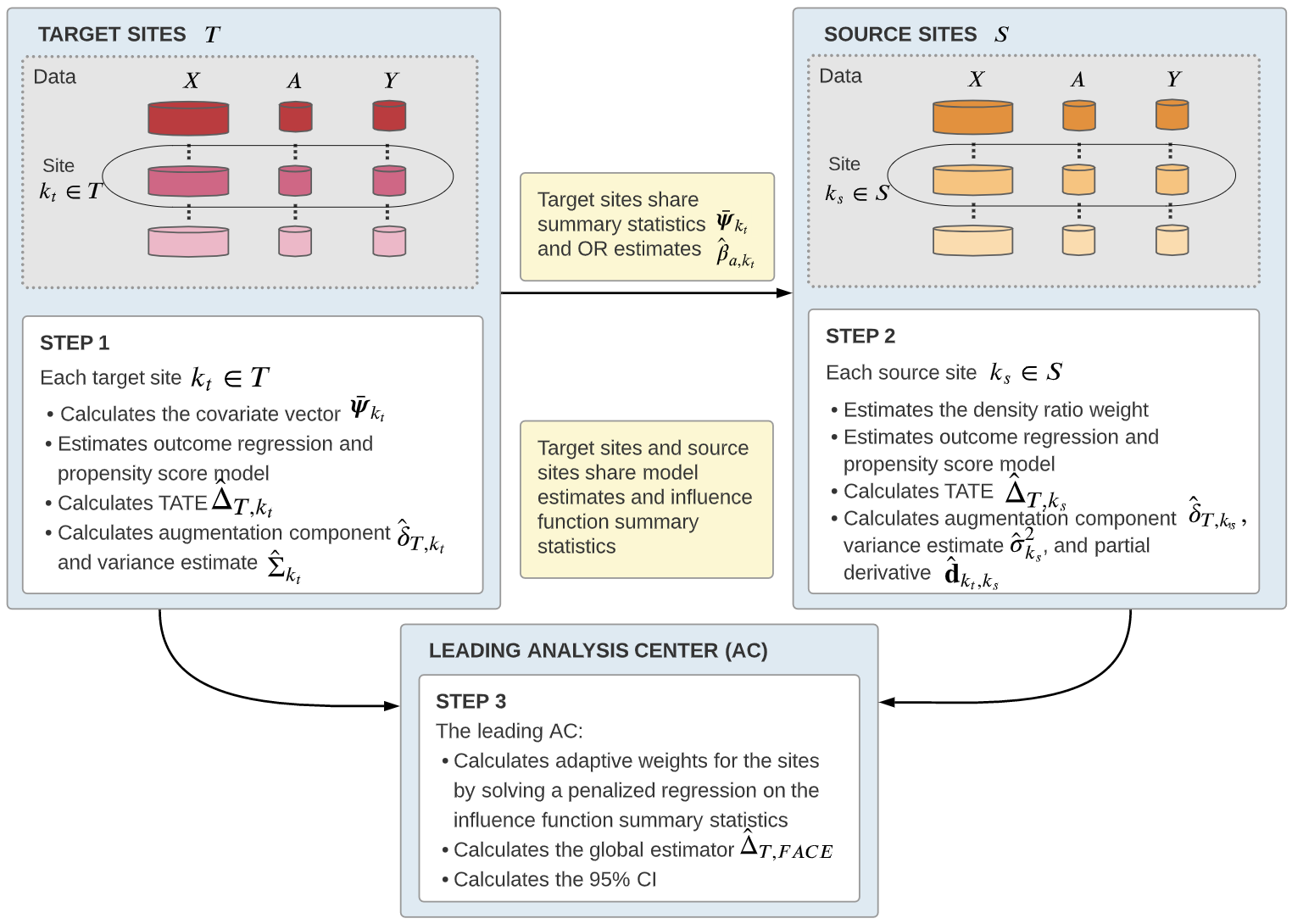}
    \caption{Workflow of FACE to construct a global estimator in a federated data setting }
    \label{fig:flowchart}
\end{figure}

%%%%%%%%%%%%%%%%%%%%%%%%%%%%%%%%%%%%%%%%%%%%%%%%%%%%%%%%%%%%%%%%%%%%%%%%%%%%%%%%%%%
\section{Special Case: FACE Under Logistic Regression Models}\label{ssec:theory-real}

As an example, we illustrate FACE under logistic regression models with $Y$ being binary,  $J+K = 5$ total sites, and $\tgt = \{1\}$ as the target site.
For notational ease, let $\bX$ be the vector of covariates with an intercept term.
We fit logistic regression models
with link $g(x) = 1/(1+e^{-x})$ and loss
$\ell(y,x) = \log(1+e^x)-yx$
for all PS and OR models. For simplicity, we let 
$\bgps(\bX) = \bX$. 

We denote the asymptotic parameters as
\begin{align*}
  \bar{\bga}_k &= \argmin_{\bga \in \R^p} \E\{\ell(A,\bga\trans\bX)\mid R=k\}, \\
  \bar{\bgb}_{a,k} &= \argmin_{\bga \in \R^p} \E\{\ell(Y,\bga\trans\bX)\mid A=a, R=k\}, \\
  \bar{\bgg}_{k_s} &= \argmin_{\bgg \in \R^q} \E\{\exp(\bgg\trans\bX)-\bgg\trans\E(\bX \mid R\in\tgt)
  \mid R = k_s\}.
\end{align*}
We give a mild set of sufficient conditions for Assumption \ref{assume:FACE}.
\begin{assumption}{S}{2}\label{assume:real}
For absolute constants $M,\varepsilon >0$,
\begin{enumerate}[label = (\alph*), ref = \ref{assume:real}(\alph*)]
\item\label{assume:real-X} (Design)
$\|\bX\|_\infty \le M$ almost surely, and all eigenvalues of $\E(\bX\bX\trans)$ are in $[\varepsilon,M]$.
\item \label{assume:real-inv} (Overlap)
For all $k=1,\dots,J+K$, $a=0,1$ and $i \in \Ical_k$,
$g(\bar{\bga}_k\trans\bX_i)$, $g'(\bar{\bgb}_{a,k}\trans\bX_i)$
and $\exp\{\bar{\bgg}_{k_s}\trans\bX_i\}$ are in
$[\varepsilon,1-\varepsilon]$ almost surely.
\item \label{assume:real-dr} (Double robustness) For each target site $k_t \in \tgt$, at least one of the two models is correctly specified:
\begin{enumerate}[label = -\roman*, ref = \ref{assume:real-dr}-\roman*]
    \item \label{assume:real-ps}
    the PS model is correct: $\P(A=1 \mid \bX, R = k_t) = g(\bar{\bga}_{k_t}\trans\bX);$
    \item \label{assume:real-or}
    the OR model is correct: $\E(Y \mid \bX,A=a , R = k_t) = g(\bar{\bgb}_{a,k_t}\trans\bX).$
\end{enumerate}
\end{enumerate}
\end{assumption}

In \underline{Step 1}, we calculate the mean covariate vector in the target site $k_t= \{1\}$ as
$\bar{\bgps}_{\tgt} = \frac{1}{n_1}\sum_{i \in \Ical_1} \bX_i$ and transfer it to sites 2 through 5.
Then, we estimate the models for $k_t = \{1\}$
$$
\hat{\bga}_1 =  \argmin_{\bga \in \R^{p+1}} \frac{1}{n_1}\sum_{i \in \Ical_1}
    \ell(A_i, \bga\trans\bX_i), \;
    \hat{\bgb}_{a,1}  = \argmin_{\bgb \in \R^{p+1}} \frac{1}{n_1}\sum_{i \in \Ical_1} I(A_i=a)\ell(Y_i, \bga\trans\bX_i).
$$
Using the estimated models, we obtain the initial estimator and its augmentation term
\begin{align*}
    \hat{M}_{\tgt} &= \frac{1}{n_1}\sum_{i \in \Ical_1} \left\{
    g\left(\hat{\bgb}_{1,1}\trans\bX_i\right) - g\left(\hat{\bgb}_{0,1}\trans\bX_i\right)
    \right\}, \\
    \widehat{\delta}_{\tgt,\tgt} &= \frac{1}{n_1}\sum_{i \in \Ical_1} \left[ \frac{A_i}{g\left(\hat{\bga}_1\trans\bX_i\right)}\left\{Y_i - g\left(\hat{\bgb}_{1,1}\trans\bX_i\right)\right\}
    -\frac{1-A_i}{g\left(-\hat{\bga}_1\trans\bX_i\right)}\left\{Y_i - g\left(\hat{\bgb}_{0,1}\trans\bX_i\right)\right\}
    \right]
\end{align*}
and $\widehat{\Delta}_{\tgt,\tgt} = \hat{M}_{\tgt} + \widehat{\delta}_{\tgt,\tgt}$. 
The variance covariance matrix ${\Sigma}_1$ can be estimated as $\hat{\Sigma}_1 = n_1^{-1} \sum_{i \in \Ical_1} \hat{\bU}_i\hat{\bU}_i\trans$ through the estimated influence functions, where $\hat{\bU}_i = (\hat{\zeta}_{i}, \hat{\xi}_{i,1}, \bgps(\bX_i)\trans, \hat{\bgu}_{1,i}, \hat{\bgu}_{0,i})\trans,$ and the exact form of $\hat{\xi}_{i,1}$, $\hat{\zeta}_i$ and $\hat{\bgu}_{a,i}$ are given in Supplement \ref{assec:proof-real}.

In \underline{Step 2}, we estimate the models for $k_s = \{2,\dots,5\}$
$$
\hat{\bga}_{k_s} =  \argmin_{\bga \in \R^{p+1}} n_{k_t}^{-1}\sum_{i \in \Ical_{k_s}}
    \ell(A_i, \bga\trans\bX_i), \;
\hat{\bgg}_{k_s} =  \argmin_{\bgg \in \R^{p+1}} n_{k_t}^{-1}\sum_{i \in \Ical_{k_s}} \exp(\bgg\trans\bX_i)-\bgg\trans \bar{\bgps}_{\tgt}.
$$
Using the estimated models, we obtain the site-specific augmentations
$$
    \widehat{\delta}_{\tgt,k_s} =  n_{k_s}^{-1}\sum_{i \in \Ical_{k_s}} e^{\hat{\bgg}_{k_s}\trans\bX_i}\left[ \frac{A_i}{g\left(\hat{\bga}_{k_s}\trans\bX_i\right)}\left\{Y_i - g\left(\hat{\bgb}_{1,1}\trans\bX_i\right)\right\}
    -\frac{1-A_i}{g\left(-\hat{\bga}_{k_s}\trans\bX_i\right)}\left\{Y_i - g\left(\hat{\bgb}_{0,1}\trans\bX_i\right)\right\}
    \right],
$$
along with the partial derivative of $\widehat{\delta}_{\tgt,k_s}$ with respect to $\bar{\bgps}_{\tgt}$, 
$\hat{\bd}_{k_s} =  (\hat{\bd}_{k_s,\psi}\trans,
\hat{\bd}_{k_s,\bgb_1}\trans,
\hat{\bd}_{k_s,\bgb_0}\trans)\trans$, as 
\begin{align*}
   \hat{\bd}_{k_s,\psi}
   =  & - \left\{n_{k_t}^{-1}\sum_{i \in \Ical_{k_s}}e^{\hat{\bgg}_{k_s}\trans\bX_i}\bX_i\bX_i\trans\right\}^{-1}n_{k_t}^{-1}\sum_{i \in \Ical_{k_s}} e^{\hat{\bgg}_{k_s}\trans\bX_i} \frac{(-1)^{1-A_i}}{g\left(\hat{\bga}_{k_s}\trans\bX_i\right)}\left\{Y_i - g\left(\hat{\bgb}_{A_i,k_s}\trans\bX_i\right)\right\}\bX_i,  \\
\hat{\bd}_{k_s,\bgb_a} &= 
  (-1)^a  n_{k_t}^{-1}\sum_{i \in \Ical_{k_s}} e^{\hat{\bgg}_{k_s}\trans\bX_i} \frac{\ind(A_i=a)}{g\left\{(-1)^{1-a}\hat{\bga}_{k_s}\trans\bX_i\right\}}g'\left(\hat{\bgb}_{A_i,k_s}\trans\bX_i\right)\bX_i. 
\end{align*}
The variance estimator $\hat{\sigma}^2_{k_s}$ can be calculated as $ \hat{\sigma}^2_{k_s} = n_{k_t}^{-1}\sum_{i \in \Ical_{k_s}}\hat{\xi}_{i,k_s}^2$ through the estimated influence function, where the form of $\hat{\xi}_{i,k_s}$ is given in Supplement \ref{assec:proof-real}.

In \underline{Step 3}, we use $\hat{\Sigma}_1$, $\hat{\bd}_{k_s}$, $\hat{\sigma}^2_{k_s}$, $\widehat{\delta}_{\tgt,k_s}$ and $\widehat{\delta}_{\tgt,\tgt}$
to solve the adaptive selection and aggregation \eqref{def:ada-aggre},
which leads to $\FACE$ and the confidence interval $\hat{\Ccal}_{\alpha}$.

After verifying that Assumptions \ref{assume:causal}  and \ref{assume:real} imply
the generic Assumption \ref{assume:FACE},
we can apply Theorem \ref{thm:FACE} in that realization.
\begin{corollary}\label{cor:real}
Under Assumptions \ref{assume:causal} and \ref{assume:real},
the FACE estimator is consistent and asymptotically normal
with consistent variance estimation $\hat{\Vcal}$,
$$
\sqrt{N/\hat{\Vcal}} \left(\FACE - \Delta_{\tgt}\right)
\leadsto \mathcal{N}(0,1).
$$
\end{corollary}

The proof is given in Supplement \ref{assec:proof-real}.

%%%%%%%%%%%%%%%%%%%%%%%%%%%%%%%%%%%%%%%%%%%%%%%%%%%%%%%%%%%%%%%%%%%%%%%%%%%%%%%%%%%
\section{Proofs}
\label{supp:proofs}
In this section, we provide proofs for the theoretical statements in the main text and supplement. 
In Sections \ref{assec:proof-DR} and \ref{assec:proof-agg}, we declare and prove the key preliminary results. We then use these results to prove 
Theorem \ref{thm:FACE} and Corollary \ref{cor:CI} in Section \ref{assec:proof-FACE}, 
Corollary \ref{cor:real} in Section
\ref{assec:proof-real}, 
Proposition \ref{prop:no_worse} 
in Section \ref{assec:proof-RE}
and Proposition \ref{prop:ideal}
in Section \ref{assec:proof-ideal}

%%%%%%%%%%%%%%%%%%%%%%%%%%%%%%%%%%%%%%%%%%%%%%%%%%%%%%%%%%%%%%%%%%%%%%%%%%%%%%%%%%%
\subsection{Double Robustness of $\hat{\Delta}_{\tgt,\tgt}$ and $\hat{\Delta}_{\tgt,k_s}$}\label{assec:proof-DR}
We first establish the consistency and asymptotic normality of the initial TATE estimator $\hat{\Delta}_{\tgt,\tgt}$
and source site TATE estimator $\hat{\Delta}_{\tgt,k_s}$. 

\begin{lemma}\label{lem:Dt}
Under Assumptions \ref{assume:causal}, \ref{assume:FACE-if}-\ref{assume:FACE-var} and \ref{assume:FACE-dr}, 
$$
\sqrt{N_{\tgt}}\left(\hat{\Delta}_{\tgt,\tgt} 
- \Delta_{\tgt}\right)
\leadsto \mathcal{N}(0, \sigma^2_{\tgt,\tgt})
$$
with asymptotic variance
$$
\sigma^2_{\tgt,\tgt} = 
\Var\left(\zeta+\xi_{\tgt} \mid R \in \tgt\right). 
$$
\end{lemma}

\begin{proof}[Proof of Lemma \ref{lem:Dt}]

From the influence function representation in Assumption \ref{assume:FACE-if}
$$
\hat{\Delta}_{\tgt,\tgt} -\bar{\Delta}_{\tgt,\tgt} = \frac{1}{N_{\tgt}} \sum_{k_t \in \tgt}\sum_{i \in \Ical_{k_t}}
\zeta_i+\xi_{i,\tgt} + o_p\left(N^{-1/2}\right),
$$
where $\bar{\Delta}_{\tgt,\tgt}$ is the asymptotic limit, and the stable variance in Assumption \ref{assume:FACE-var}
$$
\Var\left(\zeta+\xi_{\tgt} \mid R \in \tgt\right)
\in [2\varepsilon, 2M], 
$$
we have the asymptotic normality of $\hat{\Delta}_{\tgt,\tgt}$
$$
\sqrt{N_{\tgt}}\left(\hat{\Delta}_{\tgt,\tgt} 
- \bar{\Delta}_{\tgt,\tgt}\right)
\leadsto \mathcal{N}(0, \sigma^2_{\tgt,\tgt}). 
$$
Under the typical Assumptions \ref{assume:causal-cons}, \ref{assume:causal-posA}, \ref{assume:causal-unconfA} and \ref{assume:FACE-dr}, 
the doubly robust estimator $\hat{\Delta}_{\tgt,\tgt}$ converges to 
the TATE $\Delta_{\tgt}$ \citep{bang2005doubly}. 
Thus, we must have $ \bar{\Delta}_{\tgt,\tgt} = \Delta_{\tgt}$.

\end{proof}

\begin{lemma}\label{lem:Ds}
Under Assumptions \ref{assume:causal} and \ref{assume:FACE-if}-\ref{assume:FACE-var}, 
$$
\sqrt{n_{k_s}}\left(\hat{\Delta}_{\tgt,k_s} 
- \bar{\Delta}_{\tgt,k_s}\right)
\leadsto \mathcal{N}(0, \sigma^2_{\tgt,k_s})
$$
with $\bar{\Delta}_{\tgt,k_s} = 
\Delta_{\tgt} - \bar{\delta}_{\tgt,\tgt} +  \bar{\delta}_{\tgt,k_s}$ and
$$
\sigma^2_{\tgt,k_s} = 
\Var\left(\xi_{k_s}\mid R = k_s\right)
+ n_{k_s}\sum_{k_t \in \tgt} 
n_{k_t}^{-1} \Var\left\{\left(\bgps(\bX)\trans, \bgu_1\trans,\bgu_0\trans\right)\bar{\bd}_{k_t,k_s} \mid R = k_t\right\}. 
$$
Additionally under Assumption \ref{assume:S-dr}, 
$\bar{\Delta}_{\tgt,k_s}
 = \Delta_{\tgt}$ 
 for $k_s \in \tgs'$. 
\end{lemma}

\begin{proof}[Proof of Lemma \ref{lem:Ds}]
From the influence function representation in Assumption \ref{assume:FACE-if}
\begin{align*}
\hat{\Delta}_{\tgt,k_s} - \bar{\Delta}_{\tgt,k_s} &=  \sum_{k_t \in \tgt}\frac{1}{n_{k_t}}\sum_{i \in \Ical_{k_t}} \left\{
\frac{n_{k_t}}{N_{\tgt}}\zeta_i
+ \left(\bgps(\bX_i)\trans-\E\{\bgps(\bX)\mid R=k_t\}\trans, \bgu_{i,1}\trans,\bgu_{i,0}\trans\right)\bar{\bd}_{k_t,k_s}\right\} \\
& + \frac{1}{n_{k_s}}\sum_{i \in \Ical_{k_s}} \xi_{i,k_s}
+ o_p\left(N^{-1/2}\right)
\end{align*}
and the stable variance in Assumption \ref{assume:FACE-var} $\Var\left(\xi_{i,k_s} \mid R = k_s\right)
\in [\varepsilon, M]$ and
$$
\Var\left\{\frac{n_{k_t}}{N_{\tgt}}\zeta_i
+ \left(\bgps(\bX_i)\trans, \bgu_{i,1}\trans,\bgu_{i,0}\trans\right)\bar{\bd}_{k_t,k_s} \mid R = k_t\right\} 
\le M \left\{\P(R=k_t)^2+ \|\bar{\bd}_{k_t,k_s}\|_2^2\right\}, 
$$
we have the asymptotic normality of $\hat{\Delta}_{\tgt,k_s}$
$$
\sqrt{N_{\tgt}}\left(\hat{\Delta}_{\tgt,k_s} 
- \bar{\Delta}_{\tgt,k_s}\right)
\leadsto \mathcal{N}(0, \sigma^2_{\tgt,k_s}). 
$$

Similar to $\hat{\Delta}_{\tgt,\tgt}$, the source site estimator $\hat{\Delta}_{\tgt,k_s}$ is also doubly robust under Assumptions \ref{assume:causal}
and \ref{assume:S-dr}. 

When the OR model is consistently estimated under Assumption \ref{assume:S-or} (same as Assumption \ref{assume:FACE-or}) but the density ratio model and PS model may be mis-specified, we have through classical asymptotic analysis
\begin{align*}
    \hat{\Delta}_{\tgt,\tgt}
    &= 
    \sum_{k_t \in \tgt} \frac{n_{k_t}}{N_{\tgt}} \Bigg[
    \frac{1}{n_{k_t}}\sum_{i \in \Ical_{k_t}} \left\{m(1,\bX_i; \hat{\bgb}_{1,k_t}) - m(0,\bX_i; \hat{\bgb}_{1,k_t})\right\} \\
    &  \qquad +
    \frac{1}{n_{k_s}}\sum_{i \in \Ical_{k_s}} \omega_{k_t,k_s}(\bX_i;\hat{\bgg}_{k_t,k_s}) \frac{(-1)^{1-A_i}}{\pi_{k_s} (A_i,\bX_i; \hat{\bga}_{k_s})} 
  \{Y_i - m(A_i,\bX_i; \hat{\bgb}_{A_i,k_t})\} \Bigg]\\
  &= O_p\left(N^{-1/2}\right) + \underbrace{\sum_{k_t \in \tgt} \frac{\P(R = k_t)}{\P(R \in \tgt)}
  \E\{Y^{(1)} - Y^{(0)} \mid \bX_i, R = k_t\}}_{\displaystyle =\Delta_{\tgt}} \\
  &\qquad + \underbrace{\sum_{k_t \in \tgt} \frac{\P(R = k_t)}{\P(R \in \tgt)}\E\left[ \omega_{k_t,k_s}(\bX;\bar{\bgg}_{k_t,k_s})
  \frac{(-1)^{1-A}}{\pi_{k_s} (A,\bX; \bar{\bga}_{k_s})}
  \{Y - \E(Y\mid A,\bX)\}\mid R = k_s
  \right]}_{\displaystyle =0} \\
  &= O_p\left(N^{-1/2}\right) + \Delta_{\tgt}. 
\end{align*}
In the derivation, we utilized Assumption \ref{assume:causal-unconfA} to establish the ``$=0$'' by the identity
$$
\E(Y\mid A,\bX) = \E(Y\mid A,\bX, R = k_s). 
$$
Denote
$$
\omega^*_{k_t,k_s}(\bX) = \frac{\P(R = k_t \mid \bX=\bx) \P(R = k_s)}{\P(R = k_s \mid \bX=\bx)\P(R = k_t)}, 
$$
which produces the identity
$$
\E\{\omega^*_{k_t,k_s}(\bX) f(\bX) \mid R = k_s\}
= \E\{ f(\bX) \mid R = k_t\}. 
$$
When the PS and density ratio models are consistently estimated 
under Assumption \ref{assume:S-psdr} but the OR model may be mis-specified, 
we have through classical asymptotic analysis
\begin{align*}
   & \hat{\Delta}_{\tgt,\tgt}\\
    &= 
    \sum_{k_t \in \tgt} \frac{n_{k_t}}{N_{\tgt}} \Bigg[
    \frac{1}{n_{k_s}}\sum_{i \in \Ical_{k_s}} \omega_{k_t,k_s}(\bX_i;\hat{\bgg}_{k_t,k_s})\left\{ \frac{A_i}{\pi_{k_s} (1,\bX_i; \hat{\bga}_{k_s})}
    - \frac{1-A_i}{\pi_{k_s} (0,\bX_i; \hat{\bga}_{k_s})}\right\} Y_i \\
    & \qquad + \frac{1}{n_{k_t}}\sum_{i \in \Ical_{k_t}} m(1,\bX_i; \hat{\bgb}_{1,k_t})
    - \frac{1}{n_{k_s}}\sum_{i \in \Ical_{k_s}} \omega_{k_t,k_s}(\bX_i;\hat{\bgg}_{k_t,k_s}) \frac{A_i}{\pi_{k_s} (1,\bX_i; \hat{\bga}_{k_s})} 
  m(1,\bX_i; \hat{\bgb}_{1,k_t})\\
    & \qquad -  \frac{1}{n_{k_t}}\sum_{i \in \Ical_{k_t}}  m(0,\bX_i; \hat{\bgb}_{1,k_t}) + \frac{1}{n_{k_s}}\sum_{i \in \Ical_{k_s}} \omega_{k_t,k_s}(\bX_i;\hat{\bgg}_{k_t,k_s}) \frac{1-A_i}{\pi_{k_s} (0,\bX_i; \hat{\bga}_{k_s})} 
  m(0,\bX_i; \hat{\bgb}_{0,k_t}) \Bigg]\\
  &= O_p\left(N^{-1/2}\right) 
  + \sum_{k_t \in \tgt} \frac{\P(R = k_t)}{\P(R \in \tgt)} \Bigg(
  \E \left\{\omega^*_{k_t,k_s}(\bX)
  \frac{A}{\P(A=1\mid\bX,R=k_t)}Y
  \mid R = k_t\right\} \\
  & \qquad - 
  \E \left\{\omega^*_{k_t,k_s}(\bX) \frac{1-A}{\P(A=0\mid\bX,R=k_t)}Y
  \mid R = k_t\right\} \\
  & \qquad + \E\{m(1,\bX;\bar{\bgb}_{1,k_t}) -m(0,\bX;\bar{\bgb}_{0,k_t})\mid R = k_t\} \\
  & \qquad - \E[\omega^*_{k_t,k_s}(\bX)\{m(1,\bX;\bar{\bgb}_{1,k_t}) -m(0,\bX;\bar{\bgb}_{0,k_t})\}\mid R = k_s] \Bigg) \\
 &= \sum_{k_t \in \tgt} \frac{\P(R = k_t)}{\P(R \in \tgt)} 
  \E \left\{\omega^*_{k_t,k_s}(\bX)
  \E(Y^{(1)}\mid \bX)
  \mid R = k_t\right\}  - 
  \E \left\{\omega^*_{k_t,k_s}(\bX) \E(Y^{(0)}\mid \bX)
  \mid R = k_t\right\}\\
  & + O_p\left(N^{-1/2}\right)  \\
 &= \Delta_{\tgt} +  O_p\left(N^{-1/2}\right). 
\end{align*}
Therefore in either case $\bar{\Delta}_{\tgt,k_s} = \Delta_{\tgt}$. 
\end{proof}

%%%%%%%%%%%%%%%%%%%%%%%%%%%%%%%%%%%%%%%%%%%%%%%%%%%%%%%%%%%%%%%%%%%%%%%%%%%%%%%%%%%
\subsection{Optimal Aggregation}\label{assec:proof-agg}

We next consider the aggregation of the initial $\hat{\Delta}_{\tgt,\tgt}$ 
and the source site $\hat{\Delta}_{\tgt,k_s}$. 
Denote 
\begin{equation}\label{def:Lhat}
    \hat{L}(\bge) = N\left[ \sum_{k_s \in \tgs} \eta_{k_s}^2\frac{\hat{\sigma}_{k_s}^2}{n_{k_s}} 
    + \sum_{k_t \in \tgt} \hat{\bh}_{k_t}(\bge)\trans \frac{\hat{\Sigma}_{k_t}}{n_{k_t}} \hat{\bh}_{k_t}(\bge)
    \right]. 
\end{equation}
We define the oracle selection space for $\bge$ 
as 
\begin{equation}\label{def:RS}
\tgs^* = \{k_s \in \tgs: \bar{\Delta}_{\tgt,k_s} = \Delta_{\tgt}\}, \;
    \R^{\tgs^*} = \{\bge \in \R^{K}: 
    \eta_j = 0, \, \forall j \neq \tgs^*\},
\end{equation}
and the asymptotic loss function
\begin{align}
    L^*(\bge) &= 
     \sum_{k_s \in \tgs^*} \eta_{k_s}^2\Var(\xi_{k_s}\mid R=k_s)/\P(R=k_s)
    + \sum_{k_t \in \tgt}  \bh^*_{k_t}(\bge)\trans \Sigma_{k_t} \bh^*_{k_t}(\bge)
    / \P(R = k_t), \notag \\
  \bh^*_{k_t}(\bge)  &=
  \left(\P(R=k_t \mid R\in\tgt),\P(R=k_t \mid R\in\tgt)\left(1-\sum_{k_s \in \tgs^*} \eta_{k_s}\right), \sum_{k_s \in \tgs} \eta_{k_s}\bar{\bd}_{k_t,k_s}\trans\right)\trans   \label{def:loss-asym}. 
\end{align}
 Any combination $\bge \in \R^{\tgs^*}$ results in a 
consistent aggregated estimator for the TATE. 
The asymptotically optimal combination is 
\begin{equation}\label{def:eta-bar}
    \bar{\bge} = \argmin_{\bge \in \R^{\tgs^*}} 
    L^*(\bge). 
\end{equation}
In Lemma \ref{lem:aggre}, we establish the asymptotic 
distribution of the aggregated estimator with fixed $\bge \in \R^{\tgs^*}$. 
In Lemma \ref{lem:eta}, we show that the estimator $\hat{\bge}$ recovers the optimal $\bar{\bge}$. 
In Lemma \ref{lem:FACE}, we show that the uncertainty from $\hat{\bge}$ is negligible in estimating ${\Delta}_{\tgt}$ as $\FACE$. 

\begin{lemma}\label{lem:aggre}
Let $\hat{\Delta}(\bge) = \hat{\Delta}_{\tgt,\tgt} + 
\sum_{k_s\in \tgs'} \eta_{k_s}  \left(\hat{\Delta}_{\tgt,k_s} - \hat{\Delta}_{\tgt,\tgt}\right) $ be the aggregation with $\bge \in \R^{\tgs'}$. 
Under Assumptions \ref{assume:causal} and \ref{assume:FACE}, 
we have
$$
\sqrt{N} \left\{\hat{\Delta}(\bge) - \Delta_{\tgt} \right\} \leadsto \mathcal{N}\left(0, L^*(\bge)\right). 
$$
\end{lemma}

\begin{proof}[Proof of Lemma \ref{lem:aggre}]
By Lemma \ref{lem:Dt}, the initial estimator $\hat{\Delta}_{\tgt,\tgt}$ is consistent for $\Delta_{\tgt}$. 
According to the definition of $\tgs^*$ \eqref{def:RS},
$\hat{\Delta}_{\tgt,k_s}$ is consistent for $\Delta_{\tgt}$ for $k_s \in \tgs^*$. 
Thus, the weighted average $\hat{\Delta}(\bge)$ 
must also be consistent for $\Delta_{\tgt}$. 

Next, we establish the asymptotic normality of $\hat{\Delta}(\bge)$. From Assumption \ref{assume:FACE-if}, 
we have the influence function for $\hat{\Delta}(\bge)$
\begin{align*}
 & \hat{\Delta}(\bge)- \Delta_{\tgt} \\
  &=  o_p\left(N^{-1/2}\right) + \left(1-\sum_{k_s \in \tgs^*}\eta_{k_s}\right)\frac{1}{N_{\tgt}} \sum_{k_t \in \tgt}\sum_{i \in \Ical_{k_t}}
(\zeta_i+\xi_{i,\tgt})\\
& + \sum_{k_s \in \tgs^*}\eta_{k_s} \sum_{k_t \in \tgt}\frac{1}{n_{k_t}}\sum_{i \in \Ical_{k_t}} \left\{
\frac{n_{k_t}}{N_{\tgt}}\zeta_i
+ \left(\bgps(\bX_i)\trans-\E\{\bgps(\bX)\mid R=k_t\}\trans, \bgu_{i,1}\trans,\bgu_{i,0}\trans\right)\bar{\bd}_{k_t,k_s}\right\} \\
& + \sum_{k_s \in \tgs^*}\eta_{k_s} \frac{1}{n_{k_s}}\sum_{i \in \Ical_{k_s}} \xi_{i,k_s}\\
&=   o_p\left(N^{-1/2}\right) + \frac{1}{N}\sum_{k_s \in \tgs^*}\sum_{i \in \Ical_{k_s}} \frac{\eta_{k_s}\xi_{i,k_s}}{\P(R = k_s)}\\
& + \frac{1}{N} \sum_{k_t \in \tgt}\sum_{i \in \Ical_{k_t}} \Bigg\{\frac{\zeta_i+
\left(1-\sum_{k_s \in \tgs^*}\eta_{k_s}\right)\xi_{i,\tgt}}{\P(R \in \tgt)} \\ & \hspace{1in}
+ \frac{\left(\bgps(\bX_i)\trans-\E\{\bgps(\bX)\mid R = k_t\}, \bgu_{i,1}\trans,\bgu_{i,0}\trans\right)\bar{\bd}_{k_t,k_s}}{\P(R = k_t)}\Bigg\} . 
\end{align*}
We defined $L^*(\bge)$ to be precisely the variance of the influence function. To see this, we will show that $L^*(\bge)$ is the variance of $\left(1-\sum_{k\in \tgs}\eta_k\right)\hat{\Delta}_{\tgt,\tgt} + \sum_{k\in \tgs}\eta_k\hat{\Delta}_{\tgt,k}$ and use the influence function representation from Assumption  \ref{assume:FACE-if}. Denote $\eta_{\tgt} = 1-\sum_{k_s \in \tgs} \eta_{k_s}$ and define the asymptotic approximation of the aggregation under Assumption  \ref{assume:FACE-if} 
\begin{align*}
W(\bge) &= \frac{\eta_{\tgt}}{\sqrt{N}}\sum_{k_t \in \tgt} \sum_{i \in \Ical_{k_t}} \frac{N}{N_{\tgt}} \left(\zeta_i
+\xi_{i,\tgt}\right) \\
& + \sum_{k_s \in \tgs} 
\frac{\eta_{k_s}}{\sqrt{N}} \Bigg\{\sum_{k_t \in \tgt} \frac{N}{n_{k_t}}\sum_{i \in \mathcal{I}_{k_t}}\left\{\frac{n_{k_t}}{N_\tgt}\zeta_i + \left(\bgps(\bX_i)\trans-\E\{\bgps(\bX)\mid R=k_t\}\trans, \bgu_{i,1}\trans,\bgu_{i,0}\trans\right)\bar{\bd}_{k_t,k_s}\right\} \\
& \hspace{1in}
+ \frac{N}{n_{k_s}}\sum_{i \in \Ical_{k_s}} \xi_{i,k_s}  \Bigg\}\\
&=
\eta_{\tgt}\sqrt{N}(\hat{\Delta}_{\tgt,\tgt}-\bar{M}_{\tgt,\tgt} - \bar{\delta}_{\tgt,\tgt}) + \sum_{k_s\in \tgs}\eta_{k_s}\sqrt{N}(\hat{\Delta}_{\tgt,k_s} - \bar{M}_{\tgt,\tgt} - \bar{\delta}_{\tgt,k_s})
+ o_p(1). 
\end{align*}
where we have merged by site and individual indices to obtain the last line. By this alternative representation of $W(\bge)$, it is clear that its variance equals $L^*(\bge)$. 
Under Assumption \ref{assume:causal-posR} and \ref{assume:FACE-var}, $L^*(\bge)$ is stable
$$
   \frac{L^*(\bge)}{\|\bge\|_2^2 + \sum_{k_t\in\tgt}\|\bh^*_{k_t}(\bge)\|_2^2} \in [\varepsilon,M]. 
$$
Further, under Asssumptions \ref{assume:causal-posR} and \ref{assume:FACE-if}, we have
$$
\varepsilon \le \|\bh^*_{k_t}(\bge)\|_2^2 
\le 2 + \|\bge\|_1\left(1+\max_{k_s\in\tgs}\|\bar{\bd}_{k_t,k_s}\|_2\right) < \infty. 
$$
Hence for any bounded $\bge$, $L^*(\bge)$ is finite and nonzero, so we have 
$$
\sqrt{N} \left\{\hat{\Delta}(\bge) - \Delta_{\tgt} \right\} \leadsto \mathcal{N}\left(0, L^*(\bge)\right). 
$$
\end{proof}

\begin{lemma}\label{lem:eta}
Under Assumptions \ref{assume:causal} and \ref{assume:FACE}, we have
$$
\lim_{N \to \infty}\P(\hat{\bge} \in \R^{\tgs^*}) = 1, \; 
\|\hat{\bge} - \bar{\bge}\| = O_p\left(N^{-1/2}\right). 
$$
\end{lemma}

\begin{proof}[Proof of Lemma \ref{lem:eta}]
We define $\tilde{\bge}$ as the estimator under 
oracle selection
\begin{equation}\label{def:eta-tilde}
\tilde{\bge} =    \argmin_{\bge \in \R^{\tgs^*}}
    N\left[ \sum_{k_s \in \tgs} \eta_{k_s}^2\frac{\hat{\sigma}_{k_s}^2}{n_{k_s}} 
    + \sum_{k_t \in \tgt} \hat{\bh}_{k_t}(\bge)\trans \frac{\hat{\Sigma}_{k_t}}{n_{k_t}} \hat{\bh}_{k_t}(\bge)
    \right] + \lambda \sum_{k_s\in\tgs} |\eta_{k_s}| \left(\hat{\delta}_{\tgt,k_s} - \hat{\delta}_{\tgt,\tgt}\right)^2.
\end{equation}
We first show that $\|\tilde{\bge} - \bar{\bge}\| = O_p\left(N^{-1/2}\right)$. 
Then, we verify that $\tilde{\bge}$ satisfies the optimality condition, i.e.,  $\tilde{\bge} = \hat{\bge}$,  with high probability.
Note that $\hat{L}(\bge)$ and $L^*(\bge)$ are both quadratic functions of $\bge$, which can be expressed as
$$
L(\bge) = \bge^\top \hat{H} \bge + \hat{\bg}^\top \bge + \hat{c}
,\;L^*(\bge) = \bge^\top H \bge + \bg^\top \bge + c
$$
Using Assumptions \ref{assume:FACE-se} 
and the Chebyshev inequality under Assumptions \ref{assume:FACE-if} and  \ref{assume:FACE-var}, it is clear that $\hat{H}$, $\hat{\bg}$, and $\hat{c}$ 
are $\sqrt{N}$-consistent. Thus, $L(\bge) - L^*(\bge) \asymp (1+\|\bge\|^2)/\sqrt{N}$, since $H$, $\bg$ and $c$  are bounded under Assumptions \ref{assume:FACE-if} and  \ref{assume:FACE-var}. 

Under Assumptions \ref{assume:causal-posR} 
and \ref{assume:FACE-se}, 
we have the uniform approximation of the loss in a compact neighborhood of $\bar{\bge}$ of $S$
\begin{equation}\label{eq:approx-L}
\sup_{\|\bge - \bar{\bge}\| \le M} 
|\hat{L}(\bge)-L^*(\bge)|
 = O_p\left(N^{-1/2}\right). 
\end{equation}
By Lemmata \ref{lem:Dt} and \ref{lem:Ds}, 
we have for $k_s \in \tgs^*$
$$
\hat{\delta}_{\tgt,\tgt} - \hat{\delta}_{\tgt,k_s} = \hat{\Delta}_{\tgt,\tgt} - \hat{\Delta}_{\tgt,k_s}
= O_p\left(N^{-1/2}\right). 
$$
With $\lambda \lesssim N^{1/2}$, 
the penalty is small in the compact neighborhood of $\bar{\bge}$
\begin{equation}\label{eq:small-pen}
\sup_{\|\bge - \bar{\bge}\| \le M} \lambda \sum_{k_s\in\tgs} |\eta_{k_s}| \left(\hat{\delta}_{\tgt,k_s} - \hat{\delta}_{\tgt,\tgt}\right)^2
= O_p\left(N^{-1/2}\right). 
\end{equation}
Combining \eqref{eq:approx-L} and \eqref{eq:small-pen}, we have the approximation of the penalized loss 
$$
\sup_{\|\bge - \bar{\bge}\| \le M} 
\left|\hat{L}(\bge) + \lambda \sum_{k_s\in\tgs} |\eta_{k_s}| \left(\hat{\delta}_{\tgt,k_s} - \hat{\delta}_{\tgt,\tgt}\right)^2 -L^*(\bge)\right|
 = O_p\left(N^{-1/2}\right). 
$$
Following the convexity of $L^*(\bge)$ from Assumption \ref{assume:FACE-var}, 
we have 
$$
\|\tilde{\bge} - \bar{\bge}\| = O_p\left(N^{-1/2}\right). 
$$

The optimality condition of the original problem \eqref{def:FACE} is
$$
\frac{\partial}{\partial \eta_{k_s}} \hat{L}
= - \mathrm{sign}(\eta_{k_s}) \lambda \left(\hat{\delta}_{\tgt,k_s} - \hat{\delta}_{\tgt,\tgt}\right)^2, \,  \eta_{k_s} \neq 0; \;
\left|\frac{\partial}{\partial \eta_{k_s}} \hat{L}
\right| \le \lambda \left(\hat{\delta}_{\tgt,k_s} - \hat{\delta}_{\tgt,\tgt}\right)^2, \,  \eta_{k_s} = 0. 
$$
For $j \in \tgs^*$, the conditions are shared with \eqref{def:eta-tilde}, so $\tilde{\bge}$ must satisfy them. 
To establish the optimality of $\tilde{\bge}$ 
for \eqref{def:FACE}, it suffices to show 
\begin{equation}\label{eq:opt-zero}
\left|\frac{\partial}{\partial \eta_{k_s}} \hat{L}
\right| \le \lambda \left(\hat{\delta}_{\tgt,k_s} - \hat{\delta}_{\tgt,\tgt}\right)^2, \, 
k_s \in \tgs \setminus \tgs^*. 
\end{equation}
By the definition of $\tgs^*$, 
we have for biased sites
$$
\bar{\delta}_{\tgt,k_s} - \bar{\delta}_{\tgt,\tgt}
= \bar{\Delta}_{\tgt,k_s} - \bar{\Delta}_{\tgt,\tgt} \neq 0. 
$$
By Lemmata \ref{lem:Dt} and \ref{lem:Ds}, 
we have for $k_s \in \tgs \setminus \tgs^*$
$$
\hat{\delta}_{\tgt,k_s} - \hat{\delta}_{\tgt,\tgt} = \bar{\Delta}_{\tgt,k_s} - \bar{\Delta}_{\tgt,\tgt} + O_p\left(N^{-1/2}\right)
$$
bounded away from zero. 
With $\lambda \to \infty$, the penalty for biased sites diverges for $k_s \in \tgs \setminus \tgs^*$
\begin{equation}\label{eq:opt-zero-R}
\lambda \left(\hat{\delta}_{\tgt,k_s} - \hat{\delta}_{\tgt,\tgt}\right)^2 \to \infty. 
\end{equation}
Under Assumptions \ref{assume:causal-posR}, \ref{assume:FACE-var} and \ref{assume:FACE-se}, 
the derivative is tight
\begin{equation}\label{eq:opt-zero-L}
\frac{\partial}{\partial \eta_{k_s}} \hat{L}
= \frac{\partial}{\partial \eta_{k_s}} L^*
+ O_p\left(N^{-1/2}\right)
= O_p(1). 
\end{equation}
Combining \eqref{eq:opt-zero-R} and \eqref{eq:opt-zero-L}, 
we must have \eqref{eq:opt-zero} with high probability. 
This implies that $\hat{\bge}$ satisfies precisely the optimality condition with high probability. Therefore, we must have $\hat{\bge}= \tilde{\bge}$ according to the convexity of the problem with high probability. 

\end{proof}

\begin{lemma}\label{lem:FACE}
Under Assumptions \ref{assume:causal} and \ref{assume:FACE}, 
$$
\sqrt{N}\left\{ \hat{\Delta}(\bar{\bge}) - \FACE\right\} = o_p(1). 
$$
\end{lemma}

\begin{proof}[Proof of Lemma \ref{lem:FACE}]
We decompose the difference into informative source sites $k_s \in \tgs^*$ and biased source sites $k_s \in \tgs\setminus\tgs^*$
\begin{align*}
   \sqrt{N}\left\{ \hat{\Delta}(\bar{\bge}) - \FACE\right\} &=
   \sum_{k_s \in \tgs^*} (\bar{\eta}_{k_s}-\hat{\eta}_{k_s})
   \sqrt{N}\left(\hat{\Delta}_{\tgt,k_s} - \hat{\Delta}_{\tgt,\tgt}\right) \\
& + 
    \sum_{k_s \in \tgs\setminus\tgs^*}(\bar{\eta}_{k_s}-\hat{\eta}_{k_s})
   \sqrt{N}\left(\hat{\Delta}_{\tgt,k_s} - \hat{\Delta}_{\tgt,\tgt}\right). 
\end{align*}
By the definition of $\tgs^*$ \eqref{def:RS}
and the conclusions of Lemmata \ref{lem:Dt} and \ref{lem:Ds}, 
we have the tightness of terms for $k_s \in \tgs^*$
$$
\sqrt{N}\left(\hat{\Delta}_{\tgt,k_s} - \hat{\Delta}_{\tgt,\tgt}\right) = O_p\left(N^{-1/2}\right). 
$$
Applying the conclusion of Lemma \ref{lem:eta}, we have
for $k_s \in \tgs^*$
$$
(\bar{\eta}_{k_s}-\hat{\eta}_{k_s})\sqrt{N}\left(\hat{\Delta}_{\tgt,k_s} - \hat{\Delta}_{\tgt,\tgt}\right) = O_p\left(N^{-1}\right)
= o_p(1)
$$
and for $k_s \in \tgs \setminus \tgs^*$
$$
(\bar{\eta}_{k_s}-\hat{\eta}_{k_s})\sqrt{N}\left(\hat{\Delta}_{\tgt,k_s} - \hat{\Delta}_{\tgt,\tgt}\right) = 0
$$
with large probability. 
Therefore, we have obtained
$$
\sqrt{N}\left\{ \hat{\Delta}(\bar{\bge}) - \FACE\right\} = o_p(1). 
$$

\end{proof}

%%%%%%%%%%%%%%%%%%%%%%%%%%%%%%%%%%%%%%%%%%%%%%%%%%%%%%%%%%%%%%%%%%%%%%%%%%%%%%%%%%%
\subsection{Proof of Theorem \ref{thm:FACE} and Corollary \ref{cor:CI}}\label{assec:proof-FACE}

Applying Lemmata \ref{lem:aggre} and \ref{lem:FACE}, we have the asymptotic normality 
of $\FACE$, 
$$
\sqrt{N}\left(\FACE - \Delta_{\tgt}\right) \leadsto \mathcal{N}\left(0, L^*(\bar{\bge})\right). 
$$
Using the consistency of $\hat{\bge}$ for $\bar{\bge}$ and locally uniform convergence of $\hat{L}$ for $L^*$ (see \eqref{def:Lhat}-\eqref{def:eta-bar} for the definitions), we have the consistency of the variance estimator
$$
\hat{\Vcal} = \hat{L}(\hat{\bge}) = L^*(\bar{\bge}) + O_p\left(N^{-1/2}\right). 
$$
By the continuous mapping theorem, we have
$$
\sqrt{N/\hat{\Vcal}}\left(\FACE - \Delta_{\tgt}\right) \leadsto \mathcal{N}\left(0, 1\right). 
$$
The coverage probability in Corollary \ref{cor:CI} immediately follows. 

%%%%%%%%%%%%%%%%%%%%%%%%%%%%%%%%%%%%%%%%%%%%%%%%%%%%%%%%%%%%%%%%%%%%%%%%%%%%%%%%%%%
\subsection{Proof of Corollary \ref{cor:real}}\label{assec:proof-real}

In Supplement \ref{ssec:theory-real}, we noted that the variance covariance matrix for the target site, $\hat{\Sigma}_{1}$ can be calculated as as $\hat{\Sigma}_1 = \frac{1}{n_{\tgt}^2} \sum_{i \in \Ical_1} \hat{\bU}_i\hat{\bU}_i\trans$ through the estimated influence functions, where $\hat{\bU}_i = (\hat{\zeta}_{i}, \hat{\xi}_{i,1}, \bgps(\bX_i)\trans)\trans$. Here, we provide the exact form for $\hat{\xi}_{i,1}$ and $\hat{\zeta}_i$.
\begin{align*}
\hat{\bgu}_{i,1} &= 
\left\{\frac{1}{n_{\tgt}}\sum_{i \in \Ical_1} g'\left(\hat{\bga}_1\trans\bX_j\right)\bX_j\bX_j\trans\right\}^{-1}\bX_i\left\{A_i - g\left(\hat{\bga}_1\trans\bX_i\right)\right\},  \\
\hat{\bgu}_{i,0} &= 
\left\{\frac{1}{n_{\tgt}}\sum_{j \in \Ical_1}(1-A_j) g'\left(\hat{\bgb}_{0,1}\trans\bX_j\right)\bX_j\bX_j\trans\right\}^{-1}
     \bX_i (1-A_i)\left\{Y_i - g\left(\hat{\bgb}_{0,1}\trans\bX_i\right)\right\},\\
    \hat{\xi}_{i,1} &= \frac{A_i}{g(\hat{\bga}_1\trans\bX_i)} \{Y_i - g(\hat{\bgb}_{1,i}\trans\bX_i)\}
    -\frac{1-A_i}{g(-\hat{\bga}_1\trans\bX_i)} \{Y_i - g(\hat{\bgb}_{0,i}\trans\bX_i)\} \\
    & - \left[\frac{1}{n_{\tgt}}\sum_{j \in \Ical_1} e^{-(-1)^{A_j} \hat{\bga}_1\trans\bX_j} \left\{Y_j - g\left(\hat{\bgb}_{A_j,1}\trans\bX_j\right)\right\} \bX_j\trans\right]
    \left\{\frac{1}{n_{\tgt}}\sum_{i \in \Ical_1} g'\left(\hat{\bga}_1\trans\bX_j\right)\bX_j\bX_j\trans\right\}^{-1} \\
    & \hspace{3em}  \bX_i\left\{A_i - g\left(\hat{\bga}_1\trans\bX_i\right)\right\} \\
    & - \left\{\frac{1}{n_{\tgt}}\sum_{j \in \Ical_1} \frac{A_j}{g\left(\hat{\bga}_1\trans\bX_j\right)}g'\left(\hat{\bgb}_{1,1}\trans\bX_j\right)\bX_j\trans \right\}
    \hat{\bgu}_{i,1} \\
    & + \left\{\frac{1}{n_{\tgt}}\sum_{j \in \Ical_1} \frac{1-A_j}{g\left(-\hat{\bga}_1\trans\bX_j\right)}g'\left(\hat{\bgb}_{0,1}\trans\bX_j\right)\bX_j\trans \right\}
    \hat{\bgu}_{i,0}, \\
    \hat{\zeta}_{i} &= g\left(\hat{\bgb}_{1,1}\trans\bX_i\right) - g\left(\hat{\bgb}_{0,1}\trans\bX_i\right) + \left\{\frac{1}{n_{\tgt}}\sum_{j \in \Ical_1} g'\left(\hat{\bgb}_{1,1}\trans\bX_j\right)\bX_j\trans \right\}
    \hat{\bgu}_{i,1} \\
    & - \left\{\frac{1}{n_{\tgt}}\sum_{j \in \Ical_1}g'\left(\hat{\bgb}_{0,1}\trans\bX_j\right)\bX_j\trans \right\}
    \hat{\bgu}_{i,0}, \\
    \hat{\bU}_i &= (\hat{\zeta}_{i}, \hat{\xi}_{i,1}, \bgps(\bX_i)\trans,\hat{\bgu}_{i,1}\trans,\hat{\bgu}_{i,0}\trans)\trans.
\end{align*}

For source sites, the variance estimator $\hat{\sigma}_k^2$ can be calculated as $\hat{\sigma}_k^2 = \frac{1}{n_k} \sum_{i \in \Ical_k} \hat{\xi}_{i,k}^2$, where $\hat{\xi}_{i,k}$ is
\begin{align*}
    \hat{\xi}_{i,k} &= e^{\hat{\bgg}_k\trans\bX_i}\left[\frac{A_i}{g(\hat{\bga}_k\trans\bX_i)} \{Y_i - g(\hat{\bgb}_{1,i}\trans\bX_i)\}
    -\frac{1-A_i}{g(-\hat{\bga}_k\trans\bX_i)} \{Y_i - g(\hat{\bgb}_{0,i}\trans\bX_i)\} \right]\\
    & - \left[\frac{1}{n_k}\sum_{j \in \Ical_k} e^{\left(\hat{\bgg}_k-(-1)^{A_j} \hat{\bga}_k\right)\trans\bX_j} \left\{Y_j - g\left(\hat{\bgb}_{A_j,k}\trans\bX_j\right)\right\} \bX_j\trans\right]
    \left\{\frac{1}{n_k}\sum_{i \in \Ical_k} g'\left(\hat{\bga}_k\trans\bX_j\right)\bX_j\bX_j\trans\right\}^{-1} \\
    & \hspace{3em}  \bX_i\left\{A_i - g\left(\hat{\bga}_k\trans\bX_i\right)\right\} \\
     & +\hat{\bd}_{k,\psi}\trans\left(e^{\hat{\bgg}_k\trans\bX_i}\bX_i-\bar{\bgps}_{\tgt}\right).
\end{align*}

As Assumption \ref{assume:FACE} is satisfied, the FACE estimator is consistent and asymptotically normal
with consistent variance estimation $\hat{\Vcal}$,
$$
\sqrt{N/\hat{\Vcal}} \left(\FACE - \Delta_{\tgt}\right)
\leadsto \mathcal{N}(0,1).
$$

%%%%%%%%%%%%%%%%%%%%%%%%%%%%%%%%%%%%%%%%%%%%%%%%%%%%%%%%%%%%%%%%%%%%%%%%%%%%%%%%%%%
\subsection{Proof of Proposition \ref{prop:no_worse}}\label{assec:proof-RE}

Since the initial estimator $\hat{\Delta}_{\tgt,\tgt}$ corresponds to $\hat{\Delta}(\mathbf{0})$, the asymptotic variance
of $\sqrt{N}(\hat{\Delta}_{\tgt,\tgt} -\Delta_{\tgt})$ can be expressed as $L^*(\mathbf{0})$
by Lemma \ref{lem:aggre}. 
By Lemmata \ref{lem:aggre} and \ref{lem:FACE}, the asymptotic variance of $\sqrt{N}(\FACE-\Delta_{\tgt})$ is $L^*(\bar{\bge})$. 
By the definition of $\bar{\eta}$ as the minimum, 
we must have $L^*(\bar{\bge}) \le L^*(\mathbf{0})$. 
Thus, we have shown the non-inferiority of $\FACE$. 

To show that $L^*(\bar{\bge})$ is strictly smaller than $L^*(\mathbf{0})$, 
it suffices to find another $\check{\bge}$, an upper bound for $L^*(\bar{\bge})$ by the definition of $\bar{\bge}$, such that
\begin{equation}\label{eq:check-eta}
L^*(\bar{\bge}) \le L^*(\check{\bge}) <  L^*(\mathbf{0}). 
\end{equation}
Without loss of generality, we consider the simplified problem with one source site $k_* \in \tgs'$,
$$
\check{\Delta}(\eta) = \hat{\Delta}_{\tgt,\tgt} + \eta \left(\hat{\Delta}_{\tgt,k_*}-\hat{\Delta}_{\tgt,\tgt}\right). 
$$
Under Assumption \ref{assume:S-dr}, the TATE estimator of the site $\hat{\Delta}_{\tgt,k_*}$
is consistent for $\Delta_{\tgt}$  and asymptotically normal by Lemma \ref{lem:Ds}. 
Thus, $\check{\Delta}(\eta)$ is also consistent for $\Delta_{\tgt}$ and asymptotically normal
with any $\eta$. 
The optimal $\eta$ is given by the projection
$$
\eta_* = \frac{N\Cov\left(\hat{\Delta}_{\tgt,\tgt}, \hat{\Delta}_{\tgt,k_*}-\hat{\Delta}_{\tgt,\tgt}\right)}{N\Var\left( \hat{\Delta}_{\tgt,k_*}-\hat{\Delta}_{\tgt,\tgt}\right)}. 
$$
We can construct $\check{\bge}$ to be $\eta_*$ for site-$k_*$
and zero elsewhere such that $\hat{\Delta}(\check{\bge}) = \check{\Delta}(\eta_*)$. 
As long as $\Cov\left(\hat{\Delta}_{\tgt,\tgt}, \hat{\Delta}_{\tgt,k_*}-\hat{\Delta}_{\tgt,\tgt}\right) \neq 0$, the resulting estimator is different from the initial estimator $\check{\bge} \neq \mathbf{0} \Rightarrow \hat{\Delta}(\check{\bge}) \neq \hat{\Delta}_{\tgt,\tgt}$. 
Under Assumption \ref{assume:causal-posR} and \ref{assume:FACE-if}, 
the asymptotic covariance between $\sqrt{N}\hat{\Delta}_{\tgt,\tgt}$ and  $\sqrt{N}\left(\hat{\Delta}_{\tgt,k_*}-\hat{\Delta}_{\tgt,\tgt}\right)$ takes the form
$$
\Cov\left(\frac{\zeta+\xi_{\tgt}}{\P(R\in\tgt)},
 -\frac{\xi_{\tgt}}{\P(R\in\tgt)} + \sum_{k_t\in \tgt} \frac{\ind(R=k_t)}{\P(R=k_t)}
 \left(\bgps(\bX)\trans, \bgu_{1}\trans,\bgu_0\trans \right)\bar{\bd}_{k_t,k_*}
 \mid R \in \tgt\right).
$$
which is bounded away from zero by Assumption \ref{assume:S-info}. 
Thus, we have found the suitable $\check{\bge}$ that separates the asymptotic variance 
of $\FACE$ and $\hat{\Delta}_{\tgt,\tgt}$ through \eqref{eq:check-eta}. 

%%%%%%%%%%%%%%%%%%%%%%%%%%%%%%%%%%%%%%%%%%%%%%%%%%%%%%%%%%%%%%%%%%%%%%%%%%%%%%%%%%%
\subsection{Proof of Proposition \ref{prop:ideal}}\label{assec:proof-ideal}

Under the ideal setting of Assumption \ref{assume:ideal}, the influence functions of the doubly robust $\hat{\Delta}_{\tgt,\tgt}$ and $\hat{\Delta}_{\tgt,2}$ 
admit much simpler forms \citep{robins1994estimation} as a result of Neyman Orthogonality \citep{ChernozhukovEtal18DML}, 
\begin{align*}
   \hat{\Delta}_{\tgt,\tgt} - \Delta_{\tgt} &= o_p\left(N^{-1/2}\right)
   + \frac{1}{n_{\tgt}}\sum_{i \in \Ical_1} \Bigg[m(1,X_i;\bar{\bgb}_1) - m(0,X_i;\bar{\bgb}_0)
   - \Delta_{\tgt}\\
   & \qquad + \frac{A_i \{Y_i - m(1,X_i;\bar{\bgb}_1)\}}{\pi(1,\bX_i;\bar{\bga}_1)} - \frac{(1-A_i) \{Y_i - m(0,X_i;\bar{\bgb}_0)\}}{\pi(0,\bX_i;\bar{\bga}_1)} \Bigg]\\
  \hat{\Delta}_{\tgt,2} -  \Delta_{\tgt} &= o_p\left(N^{-1/2}\right)
   + \frac{1}{n_{\tgt}}\sum_{i \in \Ical_1} \left[m(1,X_i;\bar{\bgb}_1) - m(0,X_i;\bar{\bgb}_0)
   - \Delta_{\tgt}\right]\\
   & + \frac{1}{n_{\tgs}}\sum_{i\in \Ical_2}\omega_{1,2}(\bX_i;\bar{\bgg}_{1,2}) \Bigg[  \frac{A_i \{Y_i - m(1,X_i;\bar{\bgb}_1)\}}{\pi(1,\bX_i;\bar{\bga}_2)} - \frac{(1-A_i) \{Y_i - m(0,X_i;\bar{\bgb}_0)\}}{\pi(0,\bX_i;\bar{\bga}_2)} \Bigg].
\end{align*}
The asymptotic variance of the aggregation $\sqrt{N}\left\{(1-\eta)\hat{\Delta}_{\tgt,\tgt}+\eta \hat{\Delta}_{\tgt,2} - \Delta_{\tgt}  \right\}$ takes the form
$$
L^*(\eta) = \frac{N}{n_{\tgt}}\Vcal^2_{m} +  \frac{N}{n_{\tgt}}(1-\eta)^2\Vcal^2_{\tgt} + \eta^2\frac{N}{n_{\tgs}}\Vcal^2_{\tgs}. 
$$
Minimizing the quadratic function of $\eta$ give the optimal solution
$$
\bar{\eta} = \frac{n_{\tgs}\Vcal^2_{\tgt}}{n_{\tgs}\Vcal^2_{\tgt}+n_{\tgt}\Vcal^2_{\tgs}}. 
$$
We obtain the relative efficiency through 
$$
\frac{L^*(0) }{L^*(\bar{\eta}) } = 
\frac{\Vcal^2_{m}/n_{\tgt}+\Vcal^2_{\tgt}/n_{\tgt}}
{\Vcal^2_{m}/n_{\tgt}+\Vcal^2_{\tgt}\Vcal^2_{\tgs}/(n_{\tgt}\Vcal^2_{\tgs} + n_{\tgs}\Vcal^2_{\tgt})}
= 
1 + \frac{\Vcal_{\tgt}^4}{\Vcal_m^2\Vcal_{\tgt}^2
+ n_{\tgt}\left(\Vcal_m^2+\Vcal_{\tgt}^2\right)\Vcal_{\tgs}^2/n_{\tgs}}. 
$$

%%%%%%%%%%%%%%%%%%%%%%%%%%%%%%%%%%%%%%%%%%%%%%%%%%%%%%%%%%%%%%%%%%%%%%%%%%%%%%%%%%%
\subsection{Exact Efficiency Gain in an Ideal Setting}
\label{exact efficiency}
Recall that Proposition \ref{prop:no_worse} offers a guarantee on the efficiency gain of FACE relative to the initial TATE estimator.
When models are correctly specified, we have an explicit form for the oracle optimal combination $\bar{\eta}$ and the improvement in estimation efficiency for the TATE. 
\begin{assumption}{S}{3}\label{assume:ideal}
The PS, OR, and density ratio models
are consistently estimated at $\sqrt{N}$ rate:
\begin{align*}
\sup_{a=0,1}\sup_{\|\bx\|_{\infty} \le M} & \sum_{k = 1}^K\left| \P(A=a\mid \bX=\bx, R = k) - \pi_k(a,\bx;\hat{\bga}_k) \right| \\
& + 
\sum_{k_t \in \tgt}\left| \E(Y\mid A=a, \bX=\bx, R = k_t) - m_{k_t}(a,\bx;\hat{\bgb}_{a,k_t}) \right| \\
& + \sum_{k_t \in \tgt}\sum_{k_s\in\tgs}\left| \frac{\P(R = k_t \mid \bX=\bx) \P(R = k_s)}{\P(R = k_s \mid \bX=\bx)\P(R = k_t)} - \omega_{k_t,k_s}(\bx;\hat{\bgg}_{k_t,k_s}) \right|
 =  O_p\left(N^{-1/2}\right). 
\end{align*}
\end{assumption}

\begin{proposition}{}{}\label{prop:ideal}
Suppose $\tgt = \{1\}$ and $\tgs = \{2\}$. 
Denote 
\begin{gather}
     \Vcal^2_{m} =  \Var\left\{m(1,\bX;\bar{\bgb}_1)-
     m(0,\bX;\bar{\bgb}_0)-\Delta_{\tgt}\mid R=1\right\}, \notag \\
    \Vcal^2_{\tgt} =  \Var\left[\frac{(-1)^{1-A}}{\pi(A,\bX;\bar{\bga}_1)}\left\{Y - m(A,\bX;\bar{\bgb}_A)\right\}\mid R=1\right], \notag \\
     \Vcal^2_{\tgs} =  \Var\left[\omega_{1,2}(\bX;\bar{\bgg}_{1,2})\frac{(-1)^{1-A}}{\pi(A,\bX;\bar{\bga}_2)}\left\{Y - m(A,\bX;\bar{\bgb}_a)\right\}\mid R=2\right].   
\end{gather}
Under Assumptions \ref{assume:causal}-\ref{assume:ideal}, 
the optimal combination asymptotically approaches
$$
\bar{\eta} = \frac{n_{\tgs}\Vcal^2_{\tgt}}{n_{\tgs}\Vcal^2_{\tgt}+n_{\tgt}\Vcal^2_{\tgs}}. 
$$
The ratio of the asymptotic variance of the initial TATE estimator to that of FACE is  
$$
1 + \frac{\Vcal_{\tgt}^4}{\Vcal_m^2\Vcal_{\tgt}^2
+ n_{\tgt}\left(\Vcal_m^2+\Vcal_{\tgt}^2\right)\Vcal_{\tgs}^2/n_{\tgs}}, 
$$
\end{proposition}
which shows that FACE is at least as efficient as the initial TATE estimator.
Resulting from independence under the ideal setting, 
the weights $\{1-\bar{\eta}, \bar{\eta}\}$ coincide with the inverse variance weights for $\{\widehat{\delta}_{\tgt,1}, \widehat{\delta}_{\tgt,2}\}$.
According to Proposition \ref{prop:ideal}, the relative efficiency of FACE is monotone increasing in $n_{\tgs}/\Vcal_{\tgs}^2$. 
When $n_{\tgs}$ increases, the relative efficiency approaches 
$1+\Vcal_{\tgt}^2/\Vcal_m^2$. In that case, the asymptotic variance of FACE approaches 
\begin{align*}
\Vcal_m^2 / n_{\tgt} = & \Var\left\{ m(1,\bX;\bar{\bgb}_1)-
     m(0,\bX;\bar{\bgb}_0)\right\}/ n_{\tgt} \\
     = &  \Var\left\{ \frac{1}{n_{\tgt}} \sum_{i\in \Ical_{\tgt}} \E(Y_i^{(1)} - Y_i^{(0)}\mid \bX_i)\right\},
\end{align*}
which is the estimation variance of the TATE when one knows the true treatment effects.
Larger source sites will lead to better estimation of the individual treatment effect model approaching the oracle  $\E(Y_i^{(1)} - Y_i^{(0)}\mid \bX_i)$. 
The limiting $\Vcal_m^2 / n_{\tgt}$ represents the uncertainty from 
averaging individual treatment effects over the target sites for TATE, which is necessary if researchers are agnostic about the relationship in the population distribution of $\bX$ across sites.
Under the ideal setting, the two components in the initial TATE estimator, outcome regression $\hat{M}_{\tgt}$ and augmentation $\widehat{\delta}_{\tgt,\tgt}$, are independent.  
The FACE estimator includes the source site data to improve the augmentation component, leading to a reduction in its asymptotic variance.

%%%%%%%%%%%%%%%%%%%%%%%%%%%%%%%%%%%%%%%%%%%%%%%%%%%%%%%%%%%%%%%%%%%%%%%%%%%%%%%%%%%
\section{Additional Simulation Studies}
\label{supp:sim}

\subsection{High-Dimensional Setting}
We have provided additional numerical experiments to showcase the generalization of our FACE estimation strategy. In the first simulation, we set the target site to be of size $n_T = 400$ and nine source sites to be $n_k = 200$, $k=1,...,9$. In each site, we generate $p = 200$ covariates, where only the first $5$ covariates are non-null for the outcome regression and propensity score models. We consider five different settings where the level of sparsity varies, corresponding to how similar the ATEs in the source sites are to the target ATE of $3.0$.
    
    \begin{table}[h!]
    \centering
    \begin{tabular}{|c|l|}
    \hline
         {Level of sparsity} & {Description of source sites} \\
         \hline
         {1} & {All source sites have true ATEs of 3.0} \\
         \hline
         {2} & {Two source sites have true ATEs of 3.4} \\
         \hline
         {3} & {Five source sites have true ATEs of 3.8} \\
         \hline
           {4} & {Seven source sites have true ATEs of 4.0} \\
         \hline
         {5} & {Eight source sites have true ATEs of 4.5} \\
         \hline
    \end{tabular}
    \caption{Five levels of sparsity corresponding to how similar the source site ATEs are to the target site ATE of $3.0$.}
\end{table}

We examine the bias, RMSE, coverage, and length of $95\%$ CIs of FACE, as well as of four other estimators: target-only, SS, exponentially-tilted AIPW, and IVW across $300$ simulations, with correctly specified OR and PS models and misspecified density ratio models. Across the different sparsity levels, we observe a negative transfer phenomenon for the alternative approaches, while FACE shows relatively good robustness against negative transfer. The RMSE of FACE is lower than that of the target-only estimator and approaches the target-only as the number of useful source sites decreases. The coverage of FACE is close to the nominal $95\%$ across different levels of sparsity and the length of the $95\%$ CI of FACE is much shorter than that of the target-only estimator when there are many useful source sites, and approaches the target-only as the source sites become less informative.

\begin{figure}[H]
    \centering
    \includegraphics[width=\textwidth]{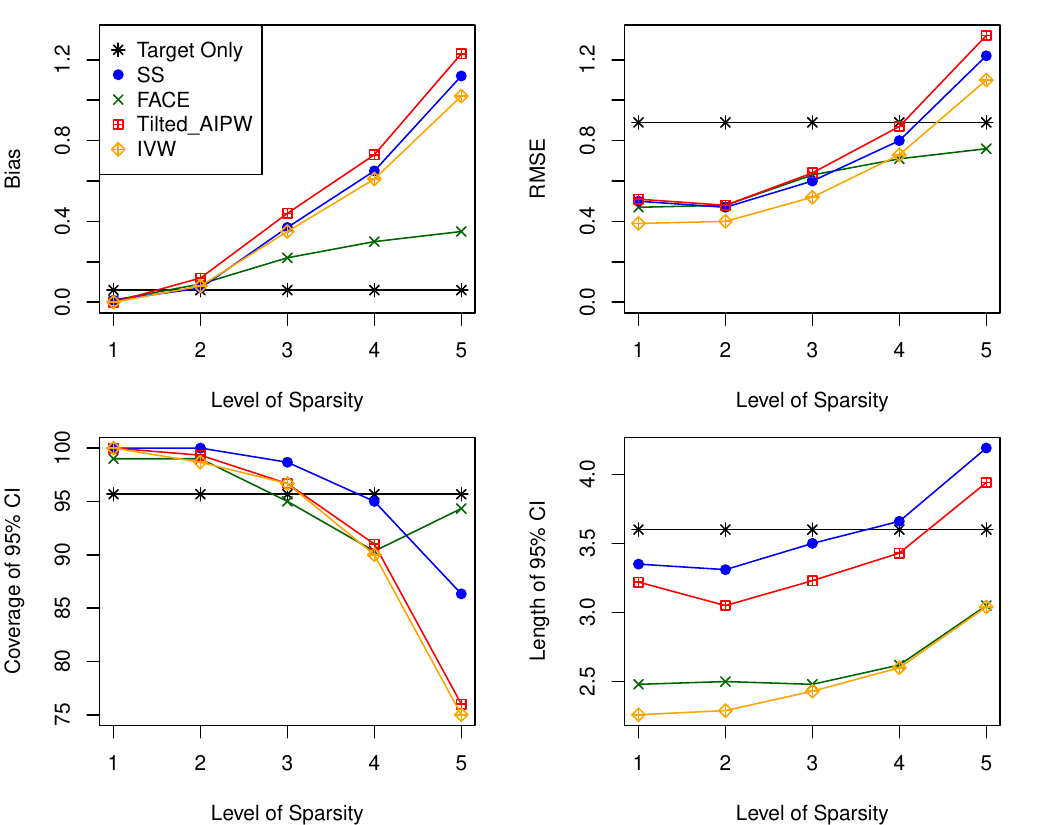}
    \caption{High-dimensional setting with target site size of $n_T = 400$ and source site sizes of $n_k = 200$, $k=1,...,9$ and $p=200$ covariates. Bias, RMSE, coverage, and length of $95\%$ CIs across 300 simulations with $10$ sites. Correctly specified OR and PS models and misspecified density ratio models. Estimators for comparison include the target-only, SS, FACE, exponentially-tilted AIPW, and IVW estimators of the TATE.} 
\end{figure}

\subsection{Model Misspecification}
We now present simulation results for Setting 2 and Setting 3 as described in the main text. Recall that in Setting 2, we misspecify the PS but correctly specify the OR model and density ratio models.

 \begin{figure}[H]
    \centering
    \includegraphics[width=\textwidth]{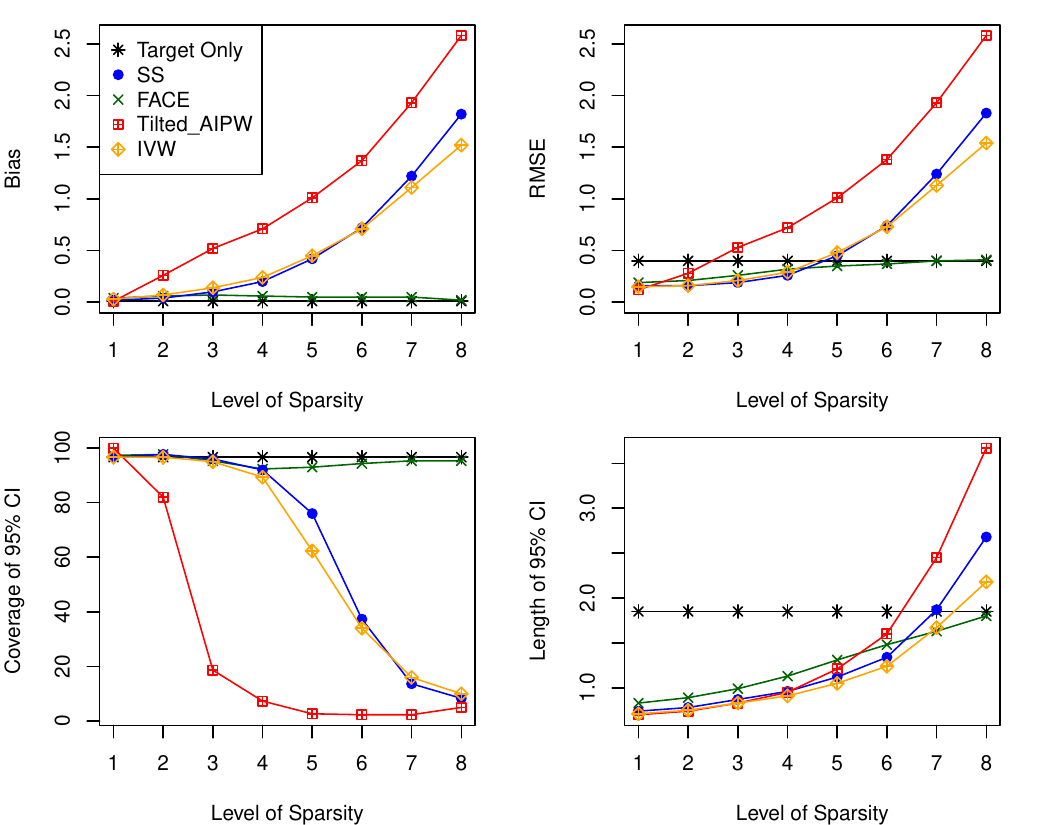}
    \caption{{Setting 2. Misspecified propensity score models. Bias, RMSE, coverage, and length of $95\%$ CIs of the target-only, SS, FACE, exponentially-tilted AIPW, and IVW estimators of the TATE across 300 simulations with $10$ sites of sample size $200$ and $p=10$ covariates.  } }
\end{figure}

\clearpage
In Setting 3, we misspecify the OR while keeping the PS and density ratio models correctly specified. 

 \begin{figure}[H]
    \centering
    \includegraphics[width=\textwidth]{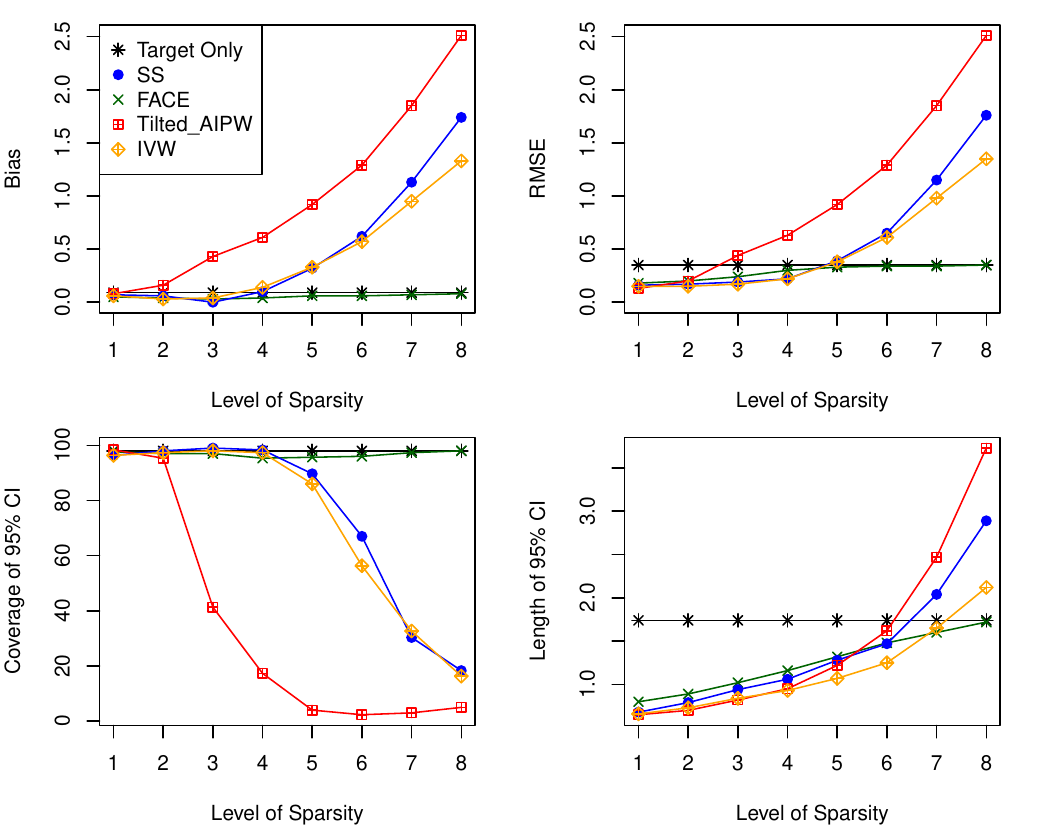}
    \caption{{Setting 3. Misspecified outcome regression models. Bias, RMSE, coverage, and length of $95\%$ CIs of the target-only, SS, FACE, exponentially-tilted AIPW, and IVW estimators of the TATE across 300 simulations with $10$ sites of sample size $200$ and $p=10$ covariates.  } }
\end{figure}

\subsection{Varying Sample Size}
To examine the effect of larger sample sizes, we set $n_k = 400$, $k=1,...,10$ for all sites and run the data generating mechanism of Setting I of the main text. 

\begin{figure}[H]
    \centering
    \includegraphics[width=\textwidth]{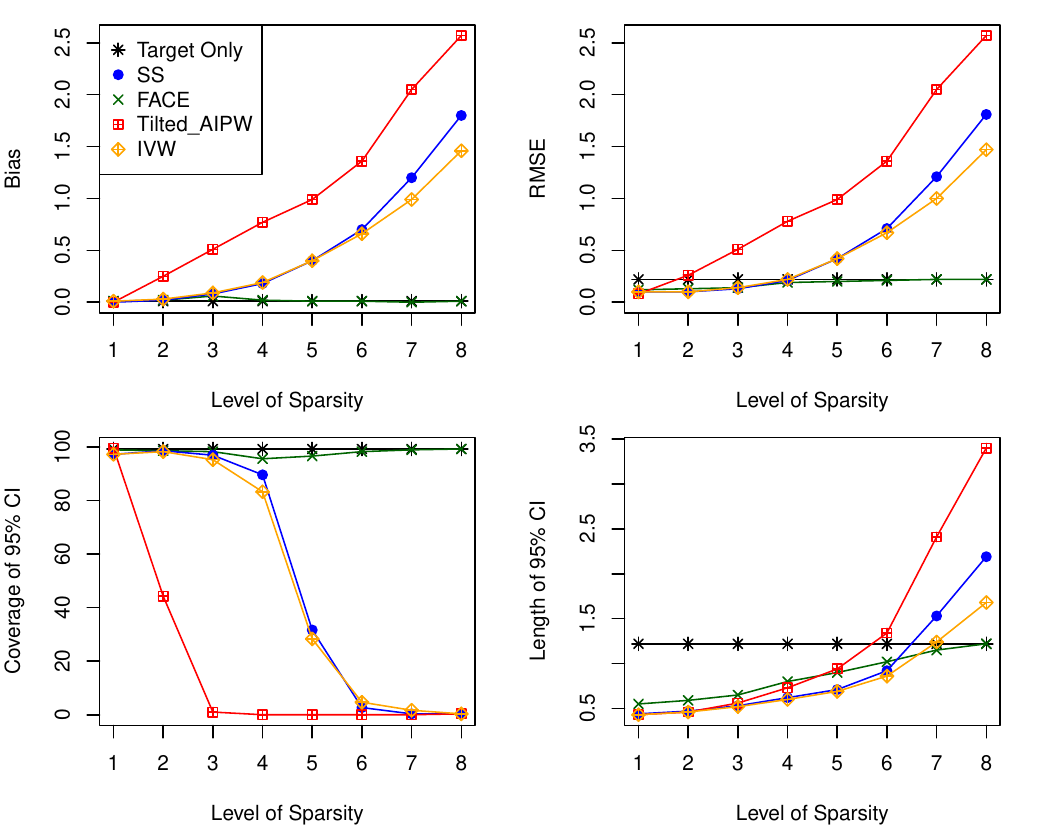}
    \caption{Setting 1. Misspecified density ratio models. Bias, RMSE, coverage, and length of $95\%$ CIs of the target-only, SS, FACE, exponentially-tilted AIPW, and IVW estimators of the TATE across 300 simulations with $10$ sites of sample size $400$ and $p=10$ covariates.}
\end{figure}

\begin{figure}[H]
    \centering
    \includegraphics[width=0.7\textwidth]{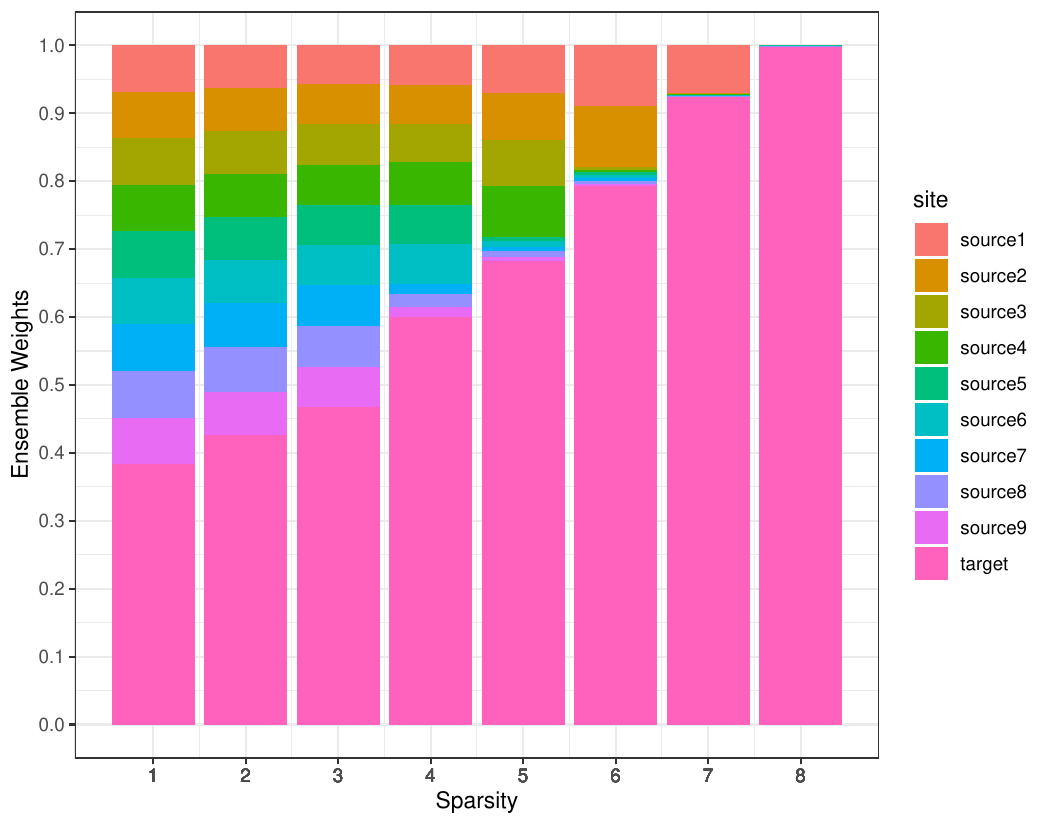}
    \caption{{FACE ensemble weights $\hat{\bge}$ for each site across eight sparsity levels}}
\end{figure}

\subsection{Varying Number of Source Sites}
To examine the effect of increasing the number of source sites $K$, we set $n_k = 200$ for all sites and run the data generating mechanism of Setting I of the main text. We fix the number of non-informative source sites, so that five source sites are moderately non-informative, with a true ATE of $3.8$, whereas the true target ATE is $3.0$. We examine the bias, RMSE, coverage, and length of the $95\%$ CIs when $J+K \in \{10,20,40\}$ . FACE displays minimal bias, smaller RMSE, and shorter average length of confidence intervals relative to the target-only estimator, with nominal coverage when $J+K \in \{10,20\}$ and slightly below nominal coverage when $J+K = 40$. The SS, exponentially-tilted AIPW, and IVW estimators have considerable bias and poor coverage, even when the proportion of informative source sites is high ($J+K = 40$).

\begin{table}[ht]
\centering
\begin{tabular}{l|crrrrr}
  \hline
  Estimator & Number of Sites & Bias & RMSE & Coverage & Length \\
  \hline
   & 10 & 0.00 & 0.35 & 98.40 & 1.75 \\
  Target-Only & 20 & 0.01 & 0.33 & 99.67 & 1.73 \\
   & 40 & 0.01 & 0.34 & 99.00 & 1.75 \\
   \hline
  & 10 & 0.03 & 0.32 & 96.20 & 1.29 \\
  FACE & 20 & 0.00 & 0.31 & 93.33 & 1.13 \\
   & 40 & 0.01 & 0.33 & 88.00 & 1.03 \\
   \hline
  & 10 & 0.41 & 0.44 & 71.40 & 1.03 \\
  SS & 20 & 0.20 & 0.24 & 80.67 & 0.66 \\
   & 40 & 0.11 & 0.15 & 85.67 & 0.44 \\
   \hline
   & 10 & 1.00 & 1.00 & 1.20 & 1.17 \\
  Tilted AIPW & 20 & 0.82 & 0.83 & 0.00 & 0.63 \\
   & 40 & 0.74 & 0.74 & 0.00 & 0.38 \\
   \hline
   & 10 & 0.43 & 0.46 & 62.20 & 0.98 \\
  IVW & 20 & 0.26 & 0.28 & 64.67 & 0.63 \\
   & 40 & 0.16 & 0.20 & 66.00 & 0.42 \\
  \hline
\end{tabular}
\caption{Setting 1 with $n_k = 200$ for all sites and $p=10$ covariates. Bias, RMSE, coverage, and length of $95\%$ CIs across 300 simulations varying the number of sites $J+K \in \{10,20,40\}$ and fixing five source sites to be moderately non-informative with a true ATE of $3.8$, whereas the true target ATE is $3.0$. Correctly specified OR and PS models and misspecified density ratio models. Estimators for comparison include the target-only, SS, FACE, exponentially-tilted AIPW, and IVW estimators of the TATE.}
\end{table}

\clearpage

%%%%%%%%%%%%%%%%%%%%%%%%%%%%%%%%%%%%%%%%%%%%%%%%%%%%%%%%%%%%%%%%%%%%%%%%%%%%%%%%%%%
\section{Additional COVID-19 Real Data Analyses}
\label{supp:realdata}
\begin{figure}[H]
        \centering
        \begin{subfigure}[b]{0.49\textwidth}
            \centering
            \includegraphics[width=\textwidth]{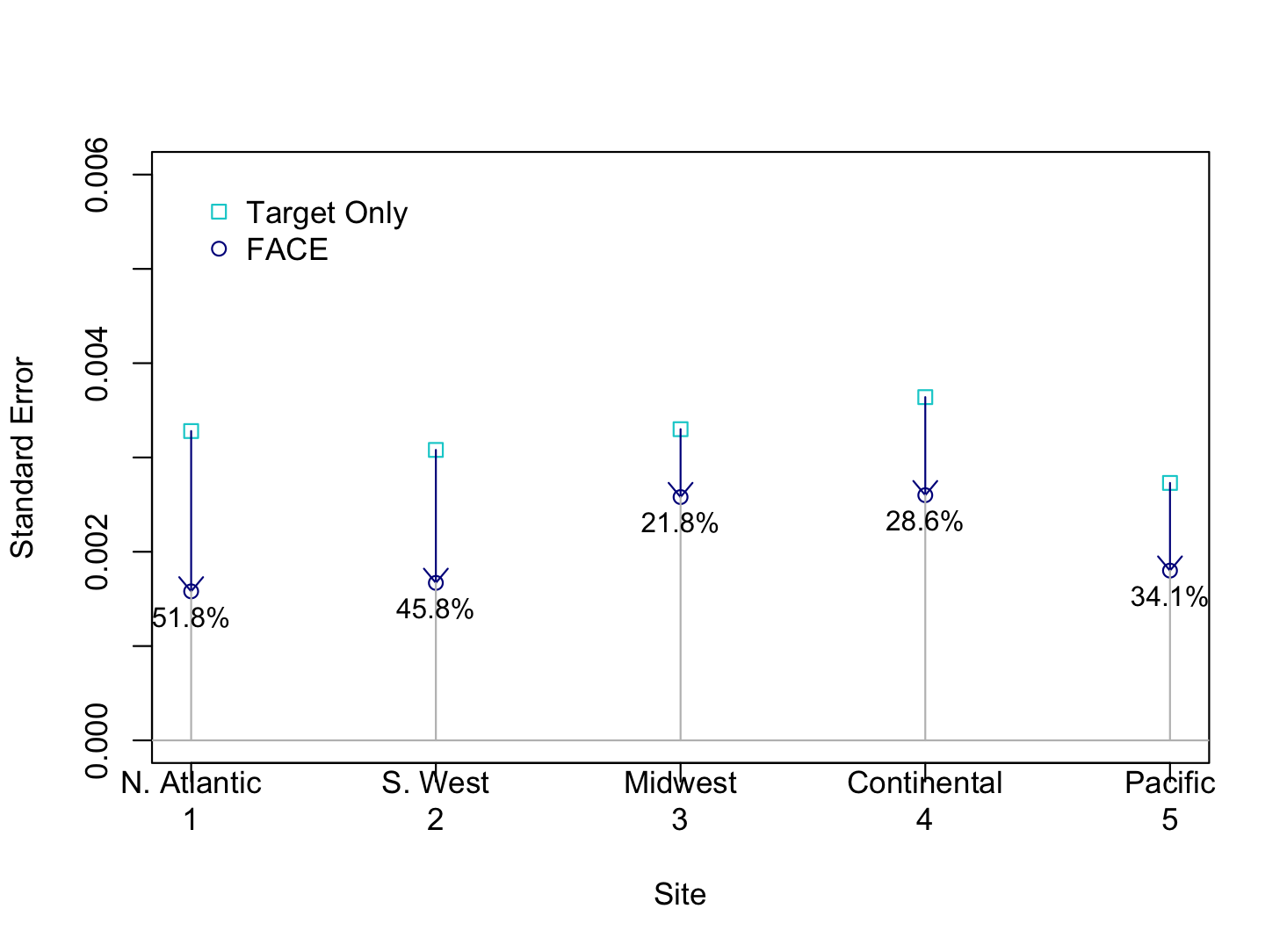}
            \caption[]
           {{\small COVID-19 infection (120 days) }}    
           % \label{fig:a}
        \end{subfigure}
        \hfill
        \begin{subfigure}[b]{0.49\textwidth}  
            \centering 
            \includegraphics[width=\textwidth]{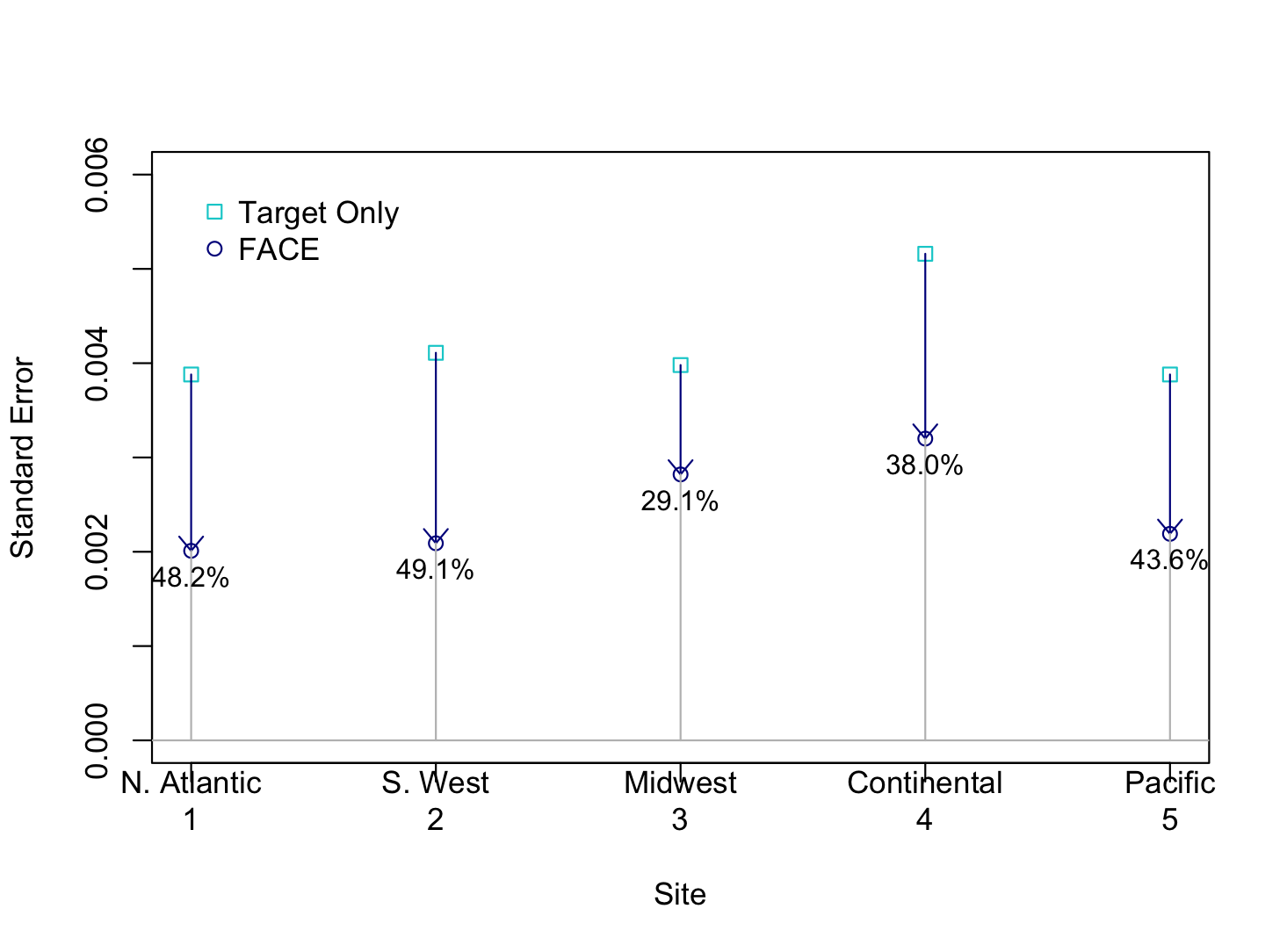}
            \caption[]%
           {{\small COVID-19 infection (180 days)}}    
           % \label{fig:b}
        \end{subfigure}
        % \vskip\baselineskip
        \begin{subfigure}[b]{0.49\textwidth}   
            \centering 
            \includegraphics[width=\textwidth]{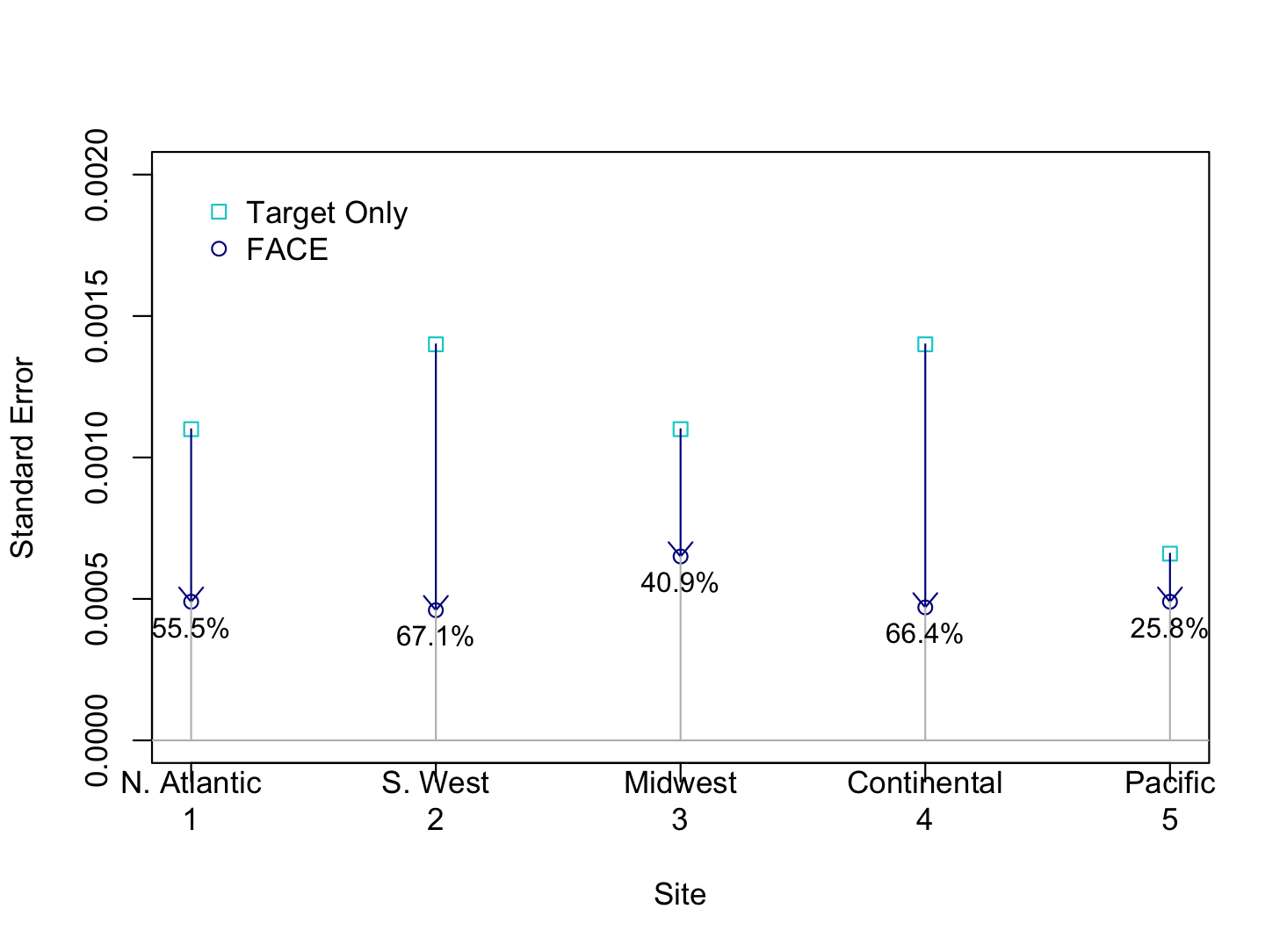}
            \caption[]%
            {{\small COVID-19 death (120 days)}}    
           % \label{fig:c}
        \end{subfigure}
        \hfill
        \begin{subfigure}[b]{0.49\textwidth}   
            \centering 
            \includegraphics[width=\textwidth]{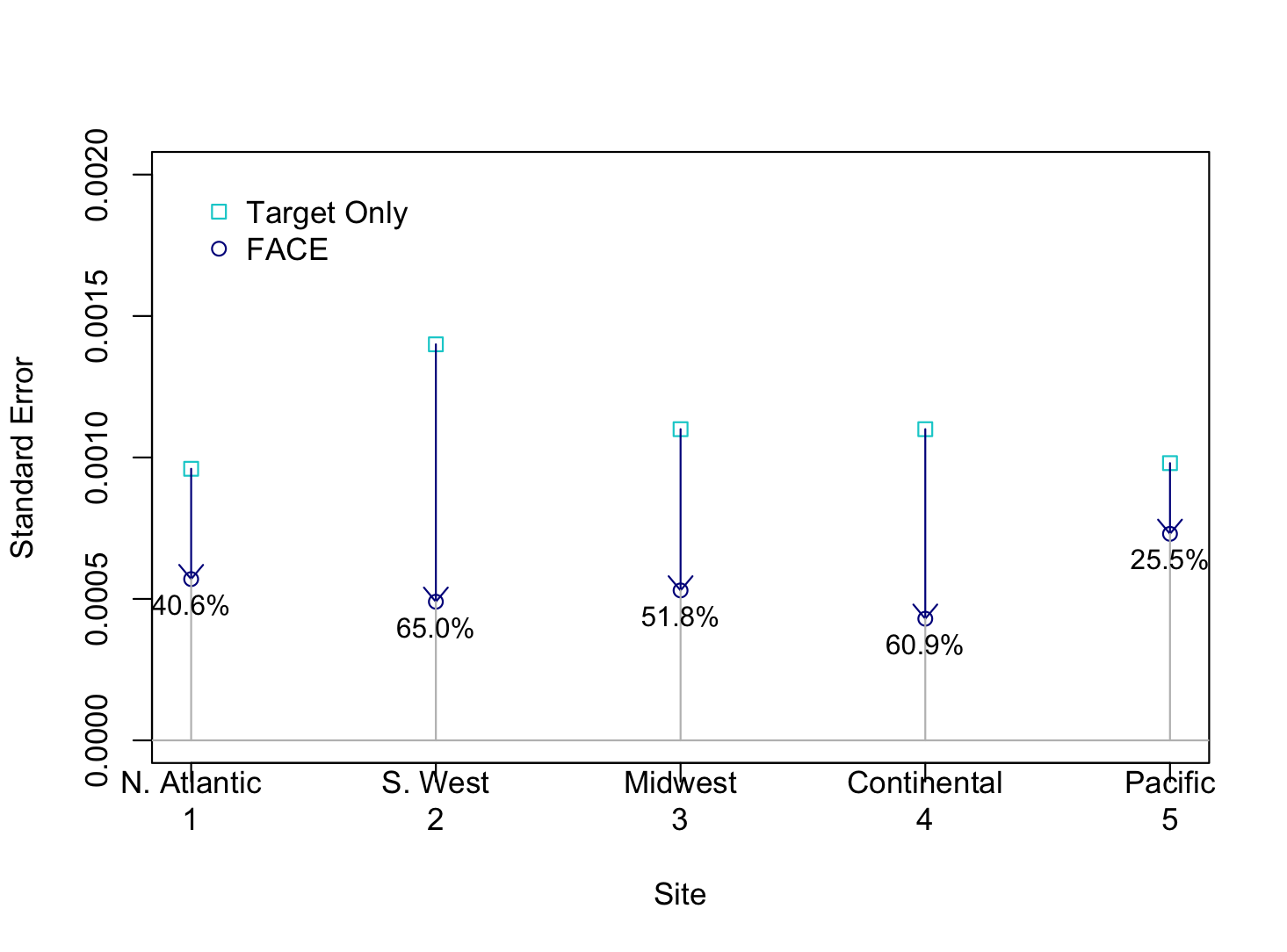}
             \caption[]%
           {{\small COVID-19 death (180 days)}}    
           % \label{fig:d}
        \end{subfigure}
        \caption[ ]
        {\small Gain in efficiency for TATE estimate using FACE vs Target Only estimator. For each site, the percent reduction in SE is calculated for each of the four outcomes}
        \label{fig:stdreduction}

\end{figure}

Figure \ref{fig:stdreduction}  visualizes the efficiency gain in using FACE compared to the Target Only estimator. For each of the four outcomes of interest, FACE meaningfully reduces the standard error of the TATE estimate for each target site, with the percentage reduction ranging from $25.5\%$ to $67.1\%$. 

\begin{table}[H]
\caption{Baseline characteristics of veterans in each of five VA sites}
\footnotesize
\begin{tabular}{lrrrrr}
\hline
  & & & Site & & \\
  \cmidrule(lr){2-6}
  & 1 & 2 & 3 & 4 & 5 \\
  & North Atlantic & Southwest & Midwest & Continental & Pacific\\

 & $(n_1=143,076)$ & $(n_2=128,792)$ & $(n_3=123,228)$ & $(n_4=93,822)$ & $(n_5=119,441)$ \\
\hline
Age (years) &  &  &  &  & \\
\hline
  \hspace{3mm}18-49 & 12,264 (8.6\%) & 10,064 (7.8\%) & 9,753 (7.9\%) & 9,807 (10.5\%) & 12,936 (10.8\%)\\

  \hspace{3mm}50-59 & 16,862 (11.8\%) & 16,906 (13.1\%) & 13,299 (10.8\%) & 13,146 (14.0\%) & 13,348 (11.2\%)\\

  \hspace{3mm}60-69 & 35,709 (25.0\%) & 35,092 (27.2\%) & 29,943 (24.3\%) & 24,670 (26.3\%) & 27,906 (23.4\%)\\

  \hspace{3mm}70-79 & 59,765 (41.8\%) & 50,839 (39.5\%) & 54,588 (44.3\%) & 36,230 (38.6\%) & 49,522 (41.5\%)\\

  \hspace{3mm}80 or older & 18,476 (12.9\%) & 15,891 (12.3\%) & 15,645 (12.7\%) & 9,969 (10.6\%) & 15,729 (13.2\%)\\
\hline
Sex &  &  &  &  & \\
\hline
  \hspace{3mm}Female & 11,752 (8.2\%) & 11,821 (9.2\%) & 8,829 (7.2\%) & 9,314 (9.9\%) & 9,897 (8.3\%)\\

  \hspace{3mm}Male & 131,324 (91.8\%) & 116,971 (90.8\%) & 114,399 (92.8\%) & 84,508 (90.1\%) & 109,544 (91.7\%)\\
\hline
Race &  &  &  &  & \\
\hline
  \hspace{3mm}Asian & 745 (0.5\%) & 391 (0.3\%) & 388 (0.3\%) & 535 (0.6\%) & 5,062 (4.2\%)\\

 \hspace{3mm}Black & 38,146 (26.7\%) & 34,064 (26.4\%) & 20,720 (16.8\%) & 24,182 (25.8\%) & 15,016 (12.6\%)\\
 
 \hspace{3mm}White & 96,890 (67.7\%) & 86,404 (67.1\%) & 94,769 (76.9\%) & 61,471 (65.5\%) & 82,750 (69.3\%)\\
 
  \hspace{3mm}Other & 7,295 (5.1\%) & 7,933 (6.2\%) & 7,351 (6.0\%) & 7,634 (8.1\%) & 16,613 (13.9\%)\\

\hline
Ethnicity &  &  &  &  & \\
\hline
  \hspace{3mm}Hispanic & 5,862 (4.1\%) & 16,768 (13.0\%) & 2,661 (2.2\%) & 9,127 (9.7\%) & 13,938 (11.7\%)\\

  \hspace{3mm} Not Hispanic & 137,214 (95.9\%) & 112,024 (87.0\%) & 120,567 (97.8\%) & 84,695 (90.3\%) & 105,503 (88.3\%)\\
\hline
Urbanicity &  &  &  &  & \\
\hline
  \hspace{3mm} Rural & 31,216 (21.8\%) & 25,223 (19.6\%) & 36,551 (29.7\%) & 21,932 (23.4\%) & 20,133 (16.9\%)\\

  \hspace{3mm} Urban & 111,860 (78.2\%) & 103,569 (80.4\%) & 86,677 (70.3\%) & 71,890 (76.6\%) & 99,308 (83.1\%)\\
\hline
Comorbidities & & & & & \\
\hline
\hspace{3mm}CLD* & 43,186 (30.2\%) & 39,267 (30.5\%) & 41,912 (34.0\%) & 27,124 (28.9\%) & 30,780 (25.8\%)\\

\hspace{3mm}CVD** & 40,565 (28.4\%) & 36,167 (28.1\%) & 38,512 (31.3\%) & 25,097 (26.7\%) & 28,999 (24.3\%)\\

\hspace{3mm}Hypertension & 104,775 (73.2\%) & 97,584 (75.8\%) & 92,355 (74.9\%) & 68,454 (73.0\%) & 79,986 (67.0\%)\\

\hspace{3mm}T2D  & 56,641 (39.6\%) & 52,356 (40.7\%) & 49,660 (40.3\%) & 38,585 (41.1\%) & 42,170 (35.3\%)\\

\hspace{3mm}CKD & 25,631 (17.9\%) & 24,029 (18.7\%) & 25,261 (20.5\%) & 17,396 (18.5\%) & 20,169 (16.9\%)\\

\hspace{3mm}Autoimmune$^\dag$ & 49,135 (34.3\%) & 46,313 (36.0\%) & 45,952 (37.3\%) & 30,392 (32.4\%) & 38,870 (32.5\%)\\

\hspace{3mm}Obesity$^\ddag$ & 39,626 (27.7\%) & 37,438 (29.1\%) & 36,465 (29.6\%) & 26,526 (28.3\%) & 31,330 (26.2\%)\\
\hline
\end{tabular}
\label{supp:table1}
* Chronic lung diseases (CLD) included asthma, bronchitis, and chronic obstructive pulmonary disease. \\
** Cardiovascular disease (CVD) included acute myocardial infarction, cardiomyopathy, coronary heart
disease, heart failure, and peripheral vascular disease. \\
$\dag$ Autoimmune diseases included HIV infection, rheumatoid arthritis, etc. \\
$\ddag$ Obesity was defined as a body-mass index of $30$ or greater.
\end{table}

\begin{landscape}
\begin{table}[H]
\caption{Baseline characteristics for veterans in each of the five sites in each vaccine group}
\tiny
\centering
\begin{tabular}{lrrrrrrrrrr}
\hline
   & Site 1:& North Atlantic & Site 2:& Southwest & Site 3:& Midwest & Site 4:& Continental & Site 5:& Pacific \\
  \cmidrule(lr){2-3} \cmidrule(lr){4-5} \cmidrule(lr){6-7} \cmidrule(lr){8-9} \cmidrule(lr){10-11} \\
  & Pfizer & Moderna & Pfizer & Moderna & Pfizer & Moderna & Pfizer & Moderna & Pfizer & Moderna\\
 & $(n=69,903)$ & $(n=73,173)$ & $(n=60,492)$ & $(n=68,300)$ & $(n=57,853)$ & $(n=65,375)$ & $(n=47,391)$ & $(n=46,431)$ & $(n=57,498)$ & $(n=61,943)$\\
\hline
Age (years) &  &  &  &  &  &  &  &  &  & \\
\hline
  \hspace{3mm}18-49 & 6,920 (9.9\%) & 5,344 (7.3\%) & 5,381 (8.9\%) & 4,683 (6.9\%) & 5,082 (8.8\%) & 4,671 (7.1\%) & 5,449 (11.5\%) & 4,358 (9.4\%) & 7,070 (12.3\%) & 5,866 (9.5\%)\\

  \hspace{3mm}50-59 & 9,180 (13.1\%) & 7,682 (10.5\%) & 8,407 (13.9\%) & 8,499 (12.4\%) & 6,131 (10.6\%) & 7,168 (11.0\%) & 7,207 (15.2\%) & 5,939 (12.8\%) & 6,968 (12.1\%) & 6,380 (10.3\%)\\

  \hspace{3mm}60-69 & 18,442 (26.4\%) & 17,267 (23.6\%) & 16,371 (27.1\%) & 18,721 (27.4\%) & 13,716 (23.7\%) & 16,227 (24.8\%) & 12,513 (26.4\%) & 12,157 (26.2\%) & 13,427 (23.4\%) & 14,479 (23.4\%)\\

  \hspace{3mm}70-79 & 27,601 (39.5\%) & 32,164 (44.0\%) & 23,196 (38.3\%) & 27,643 (40.5\%) & 25,967 (44.9\%) & 28,621 (43.8\%) & 17,919 (37.8\%) & 18,311 (39.4\%) & 22,990 (40.0\%) & 26,532 (42.8\%)\\

  \hspace{3mm}80 or older & 7,760 (11.1\%) & 10,716 (14.6\%) & 7,137 (11.8\%) & 8,754 (12.8\%) & 6,957 (12.0\%) & 8,688 (13.3\%) & 4,303 (9.1\%) & 5,666 (12.2\%) & 7,043 (12.2\%) & 8,686 (14.0\%)\\
\hline \\
Sex &  &  &  &  &  &  &  &  &  & \\
\hline
  \hspace{3mm}Female & 6,379 (9.1\%) & 5,373 (7.3\%) & 6,120 (10.1\%) & 5,701 (8.3\%) & 4,193 (7.2\%) & 4,636 (7.1\%) & 5,155 (10.9\%) & 4,159 (9.0\%) & 5,154 (9.0\%) & 4,743 (7.7\%)\\

  \hspace{3mm}Male & 63,524 (90.9\%) & 67,800 (92.7\%) & 54,372 (89.9\%) & 62,599 (91.7\%) & 53,660 (92.8\%) & 60,739 (92.9\%) & 42,236 (89.1\%) & 42,272 (91.0\%) & 52,344 (91.0\%) & 57,200 (92.3\%)\\
\hline \\
Race &  &  &  &  &  &  &  &  &  & \\
\hline
  \hspace{3mm}Asian & 479 (0.7\%) & 266 (0.4\%) & 224 (0.4\%) & 167 (0.2\%) & 196 (0.3\%) & 192 (0.3\%) & 323 (0.7\%) & 212 (0.5\%) & 2,270 (3.9\%) & 2,792 (4.5\%)\\

  \hspace{3mm}Black & 23,632 (33.8\%) & 14,514 (19.8\%) & 16,304 (27.0\%) & 17,760 (26.0\%) & 11,511 (19.9\%) & 9,209 (14.1\%) & 14,866 (31.4\%) & 9,316 (20.1\%) & 8,172 (14.2\%) & 6,844 (11.0\%)\\
  
 \hspace{3mm}White & 42,228 (60.4\%) & 54,662 (74.7\%) & 40,040 (66.2\%) & 46,364 (67.9\%) & 42,516 (73.5\%) & 52,253 (79.9\%) & 28,221 (59.5\%) & 33,250 (71.6\%) & 39,163 (68.1\%) & 43,587 (70.4\%)\\

  \hspace{3mm}Other & 3,564 (5.1\%) & 3,731 (5.1\%) & 3,924 (6.5\%) & 4,009 (5.9\%) & 3,630 (6.3\%) & 3,721 (5.7\%) & 3,981 (8.4\%) & 3,653 (7.9\%) & 7,893 (13.7\%) & 8,720 (14.1\%)\\
\hline \\
Ethnicity &  &  &  &  &  &  &  &  &  & \\
\hline
  \hspace{3mm}Hispanic & 2,929 (4.2\%) & 2,933 (4.0\%) & 5,951 (9.8\%) & 10,817 (15.8\%) & 1,531 (2.6\%) & 1,130 (1.7\%) & 5,062 (10.7\%) & 4,065 (8.8\%) & 6,615 (11.5\%) & 7,323 (11.8\%)\\

  \hspace{3mm}Not Hispanic & 66,974 (95.8\%) & 70,240 (96.0\%) & 54,541 (90.2\%) & 57,483 (84.2\%) & 56,322 (97.4\%) & 64,245 (98.3\%) & 42,329 (89.3\%) & 42,366 (91.2\%) & 50,883 (88.5\%) & 54,620 (88.2\%)\\
\hline \\
Urbanicity &  &  &  &  &  &  &  &  &  & \\
\hline
  \hspace{3mm}Rural & 11,546 (16.5\%) & 19,670 (26.9\%) & 11,701 (19.3\%) & 13,522 (19.8\%) & 12,442 (21.5\%) & 24,109 (36.9\%) & 8,598 (18.1\%) & 13,334 (28.7\%) & 8,538 (14.8\%) & 11,595 (18.7\%)\\

  \hspace{3mm}Urban & 58,357 (83.5\%) & 53,503 (73.1\%) & 48,791 (80.7\%) & 54,778 (80.2\%) & 45,411 (78.5\%) & 41,266 (63.1\%) & 38,793 (81.9\%) & 33,097 (71.3\%) & 48,960 (85.2\%) & 50,348 (81.3\%)\\
\hline \\
Comorbidities & & & & & & & & & & \\
\hline
\hspace{3mm}CLD & 19,423 (27.8\%) & 23,763 (32.5\%) & 18,356 (30.3\%) & 20,911 (30.6\%) & 18,253 (31.6\%) & 23,659 (36.2\%) & 13,031 (27.5\%) & 14,093 (30.4\%) & 14,598 (25.4\%) & 16,182 (26.1\%)\\

\hspace{3mm}CVD & 18,573 (26.6\%) & 21,992 (30.1\%) & 16,902 (27.9\%) & 19,265 (28.2\%) & 17,335 (30.0\%) & 21,177 (32.4\%) & 12,546 (26.5\%) & 12,551 (27.0\%) & 13,742 (23.9\%) & 15,257 (24.6\%)\\

\hspace{3mm}Hypertension & 49,985 (71.5\%) & 54,790 (74.9\%) & 45,094 (74.5\%) & 52,490 (76.9\%) & 42,622 (73.7\%) & 49,733 (76.1\%) & 34,362 (72.5\%) & 34,092 (73.4\%) & 37,453 (65.1\%) & 42,533 (68.7\%)\\

\hspace{3mm}T2D & 26,872 (38.4\%) & 29,769 (40.7\%) & 23,884 (39.5\%) & 28,472 (41.7\%) & 22,770 (39.4\%) & 26,890 (41.1\%) & 19,549 (41.3\%) & 19,036 (41.0\%) & 19,841 (34.5\%) & 22,329 (36.0\%)\\

\hspace{3mm}CKD & 12,241 (17.5\%) & 13,390 (18.3\%) & 11,287 (18.7\%) & 12,742 (18.7\%) & 11,197 (19.4\%) & 14,064 (21.5\%) & 8,665 (18.3\%) & 8,731 (18.8\%) & 9,542 (16.6\%) & 10,627 (17.2\%)\\

\hspace{3mm}Autoimmune & 22,431 (32.1\%) & 26,704 (36.5\%) & 21,898 (36.2\%) & 24,415 (35.7\%) & 21,260 (36.7\%) & 24,692 (37.8\%) & 14,912 (31.5\%) & 15,480 (33.3\%) & 18,228 (31.7\%) & 20,642 (33.3\%)\\

\hspace{3mm}Obesity & 18,799 (26.9\%) & 20,827 (28.5\%) & 18,406 (30.4\%) & 19,032 (27.9\%) & 16,731 (28.9\%) & 19,734 (30.2\%) & 13,168 (27.8\%) & 13,358 (28.8\%) & 15,190 (26.4\%) & 16,140 (26.1\%)\\
\hline
\end{tabular}
\label{supp:table2}
\end{table}

\end{landscape}

\end{document}